\documentclass[12pt]{article}
\usepackage{amsfonts,amsmath,amssymb,epsf,stmaryrd,mathrsfs}
\usepackage{extarrows}
\usepackage{graphics}
\usepackage{ifpdf}
\ifpdf
\usepackage{epstopdf}
\usepackage{hyperref}
\else
\usepackage[hypertex]{hyperref}
\fi

\newcounter{thm}

\newcommand\sect[1]{\section{#1}\setcounter{equation}0\setcounter{thm}0}

\newcommand\be            {\begin{equation}}
\newcommand\bea           {\begin{equation}\begin{array}l\displaystyle}
\newcommand\bearll        {\begin{array}{ll}\displaystyle}
\newcommand\ee            {\end{equation}}
\newcommand\eear          {\end{array}}
\newcommand\enl           {\\[1em]\displaystyle}
\newcommand\etb           {&\!\! \displaystyle}
\newcommand\labl[1]       {\label{#1}\ee}
\newcommand\nxt{\noindent\raisebox{.08em}{\rule{.44em}{.44em}}%
\hspace{.4em}}

\newcommand\arxiv[2]      {\href{http://arXiv.org/abs/#1}{#2}}
\newcommand\doi[2]        {\href{http://dx.doi.org/#1}{#2}}
\newcommand\httpurl[2]    {\href{http://#1}{#2}}

\newcommand\cir           {\,{\circ}\,}
\newcommand\eps           {\varepsilon}
\newcommand\Hol           {\mathrm{Hol}}
\newcommand\Hom           {\mathrm{Hom}}
\newcommand\id            {{\rm id}}
\newcommand\In            {\,{\in}\,}

\newcommand\one           {{\bf1}}
\newcommand\oti           {\,{\otimes}\,}

\newcommand\ti            {\,{\times}\,}
\newcommand{\Tvec}        {\mathit{TV}}
\newcommand{\WD}          {\mathit{WD}}

\newcommand\Cb            {\mathbb{C}}
\newcommand\Nb            {\mathbb{N}}
\newcommand\Rb            {\mathbb{R}}
\newcommand\Zb            {\mathbb{Z}}

\newcommand\Ac            {\mathcal{A}}
\newcommand\Bc            {\mathcal{B}}

\newcommand\Gc            {\mathcal{G}}
\newcommand\Hc            {\mathcal{H}}
\newcommand\Ic            {\mathcal{I}}
\newcommand\Jc            {\mathcal{J}}
\newcommand\Kc            {\mathcal{K}}

\newcommand\Oc            {\mathcal{O}}

\newcommand\Vc            {\mathcal{V}}

\newcommand\gt[1]         {\mathfrak{#1}}
\newcommand\sfk           {{\mathsf k}}
\newcommand\sfi           {{\mathsf i}}
\newcommand\sfd           {{\mathsf d}}
\newcommand\sfT           {{\mathsf T}}

\newcommand\qq            {\begin{eqnarray}}
\newcommand\qqq           {\end{eqnarray}}

\newcommand\til           {\widetilde}
\newcommand\bx            {\textrm{b}}
\newcommand\ggt           {\gt{g}}
\newcommand\ggtk          {\widehat\ggt_\sfk}
\newcommand\erm           {\textrm{e}}

\def\Zc{\mathcal{Z}}
\def\ggt{\gt{g}}
\def\too{\rightarrow}
\def\vv{\textrm{v}}
\def\Gx{\textrm{G}}
\def\Ker{\textrm{ker}}
\def\Im{\textrm{im}}

\def\a{\alpha}
\def\G{\Gamma}
\def\g{\gamma}
\def\ep{\epsilon}
\def\vep{\varepsilon}
\def\D{\Delta}
\def\d{\delta}
\def\La{\Lambda}
\def\la{\lambda}
\def\Si{\Sigma}
\def\si{\sigma}
\def\t{\tau}
\def\Om{\Omega}
\def\om{\omega}
\def\Th{\Theta}
\def\th{\theta}

\def\p{\partial}
\def\det{{\rm det}}
\def\vol{{\rm vol}}
\def\con{\lrcorner\,}
\def\ovl{\overline}
\def\unl{\underline}

\begin{document}

\thispagestyle{empty}
\def\thefootnote{\fnsymbol{footnote}}
\begin{flushright}
KCL-MTH-08-07\\
%0808.XXXX [hep-th]
\end{flushright}
\vskip 5.0em
\begin{center}\LARGE
Gerbe-holonomy for surfaces with defect networks
\end{center} \vskip
4em
\begin{center}\large
  Ingo Runkel\footnote{Email: {\tt ingo.runkel@kcl.ac.uk}}
  and
  Rafa\l ~R.\ Suszek\footnote{Email: {\tt rafal.suzek@kcl.ac.uk}}
\end{center}
\begin{center}
  Department of Mathematics, King's College London \\
  Strand, London WC2R 2LS, United Kingdom
\end{center}
\vskip 1em
\begin{center}
  August 2008
\end{center}

\vskip 4em
\begin{abstract}
We define the sigma-model action for world-sheets with embedded
defect networks in the presence of a three-form field strength. We
derive the defect gluing condition for the sigma-model fields and
their derivatives, and use it to distinguish between conformal and
topological defects. As an example, we treat the WZW model with
defects labelled by elements of the centre $\,Z(\Gx)\,$ of the
target Lie group $\,\Gx$;\ comparing the holonomy for different
defect networks gives rise to a 3-cocycle on $\,Z(\Gx)$.\ Next, we
describe the factorisation properties of two-dimensional quantum
field theories in the presence of defects and compare the
correlators for different defect networks in the quantum WZW model.
This, again, results in a 3-cocycle on $\,Z(\Gx)$.\ We observe that
the cocycles obtained in the classical and in the quantum
computation are cohomologous.
\end{abstract}

\setcounter{footnote}{0}
\def\thefootnote{\arabic{footnote}}

\newpage

\tableofcontents

\vspace{1em}

\sect{Introduction}\label{sec:intro}

In this paper, we consider two-dimensional sigma models
\be
S[X;\g] ~=~
\int_\Si\,\sfd\si^1\wedge\sfd\si^2\,\sqrt{\det\g}\,\bigl(\g^{-1}
\bigr)^{ab}\,G_{\mu \nu}(X)\,\p_a X^\mu\,\p_b X^\nu ~+~S_{\rm
top}[X]
\labl{eq:sigma-action}
for maps $\,X\,$ from a world-sheet $\,\Si\,$ with metric $\,\g\,$
to a target space $\,M\,$ with metric $\,G$. The field variable
$\,X\,$ is allowed to be discontinuous across lines on the
world-sheet. We shall refer to such lines of discontinuity as {\em
defects}. The most familiar setting in which defects occur is
provided by orbifold models, where the field has to be periodic only
up to the action of the group of automorphisms of the target space.
However, defect conditions much more general than a jump of the
field by a target-space automorphism are possible. One of the main
results of this paper is the formulation of the topological term
$\,S_{\rm top}[X]\,$ in the sigma-model action for world-sheets with
defects, and, in particular, for situations in which the defect
lines meet to form {\em defect junctions}. By varying the
sigma-model action, we obtain the gluing condition to be imposed on
the embedding field $\,X\,$ and its derivatives at the defect. This
allows us to analyse the world-sheet symmetries and, in particular,
to distinguish between conformal and topological defects.

Circular defect lines can be treated by thinking of them as boundary
conditions of a folded model \cite{Wong:1994pa}, but for defect
junctions this is no longer possible. One can therefore expect that
the study of defect junctions yields interesting information that
cannot be obtained through the analysis of boundary conditions of
the sigma model or some folded version thereof. We illustrate this
on the example of the WZW model.
\medskip

The topological term $\,S_{\rm top}[X]\,$ of the sigma-model action
can be understood as the logarithm of a ${\rm U}(1)$-valued holonomy
associated to an embedding of the world-sheet $\,\Si\,$ in the
target space $\,M$.\ The holonomy is computed in terms of the gauge
potential $\,B\,$ of a 3-form field strength $\,H\,$ on $\,M$.\
Typically, the gauge potential cannot be defined globally and exists
only patch-wise, which then leads to additional 1-forms $\,A\,$ and
functions $\,g\,$ on two- and three-fold overlaps of these patches,
respectively. These forms and functions enter the formulation of the
world-sheet holonomy \cite{Alvarez:1984es}. It was realised in
\cite{Gawedzki:1987ak} that the correct structure to capture the
data composed of $\,B,A\,$ and $\,g\,$ on the target space is a
class in the third (real) Deligne hypercohomology, which was
subsequently identified in
\cite{Brylinski:1993ab,Murray:1994db,Murray:1999ew} with a geometric
object called a {\em gerbe with connection}. In section
\ref{sec:gerbes-local-data}, we give a brief recollection of the
bits of the theory of gerbes that we shall need, and in section
\ref{sec:hol-no-defect}, we review the holonomy formula for
world-sheets with empty boundary and no defects. The notion of
holonomy was generalised to world-sheets with boundaries in
\cite{Gawedzki:1999bq,Kapustin:1999di,Carey:2002,Gawedzki:2002se,Gawedzki:2004tu}.
The boundary gets mapped to a D-brane which is a submanifold of the
target space that supports a (global) curvature 2-form and a
gerbe-twisted gauge bundle. The latter is described by a so-called
gerbe module \cite{Carey:2002,Gawedzki:2002se}. The holonomy in the
presence of circular defect lines was first formulated in
\cite{Fuchs:2007fw}. In this case, the defect circles get mapped to
a submanifold $\,Q\subset M\ti M$,\ termed a {\em bi-brane} in
\cite{Fuchs:2007fw}. The bi-brane world-volume is equipped with a
curvature 2-form and a gerbe bimodule, hence the name. We review
this construction and the necessary background for gerbe bimodules
in sections \ref{sec:bibrane-def} and \ref{sec:circle-hol}.

In sections \ref{sec:inter-bi-branes} and \ref{sec:holo-net}, we
extend the validity of the holonomy formula further to allow for
defect junctions. There is, again, a corresponding target-space
notion, which we call an {\em inter-bi-brane}. An inter-bi-brane
consists of a collection $\,T=\bigsqcup_{n\ge 1}\, T_n\,$ of
submanifolds $\,T_n\subset M\ti M\ti\cdots\ti M\,$ of $n$ copies of
$\,M$,\ where $n$ refers to the number of defect lines meeting at a
junction. Each $\,T_n\,$ is equipped with a twisted scalar field.

It turns out to be convenient not to restrict $\,Q\,$ and $\,T_n\,$
to be submanifolds of products of $\,M$,\ but, instead, to allow
arbitrary manifolds endowed with projections to $\,M\,$ and $\,Q$.\
We shall use this point of view in section \ref{sec:holo}.
\medskip

Defects in sigma models have also been investigated in the quantised
theory. Most of the known results apply to the conformal r\'egime,
e.g., for free theories
\cite{Bachas:2001vj,Fuchs:2007tx,Bachas:2007td}, for the WZW model
\cite{Bachas:2004sy,Alekseev:2007in,Schweigert:2007wd}, or for
rational conformal field theories in general
\cite{Petkova:2000ip,Quella:2002ct,Frohlich:2004ef,Quella:2006de}.
The first systematic treatment of CFT correlators with defect
junctions appeared in \cite{Frohlich:2006ch}. Properties of defects
were also studied in supersymmetric theories (see
\cite{Brunner:2007qu,Brunner:2008fa} for recent results), and in
classical and quantised integrable models in 1+1 dimensions (see,
e.g., \cite{Bajnok:2007jg,Corrigan:2008yv} and the references
therein).
\medskip

\begin{figure}[bt]

$$
 \raisebox{-25pt}{\begin{picture}(160,50)
  \put(0,0){\scalebox{0.40}{\includegraphics{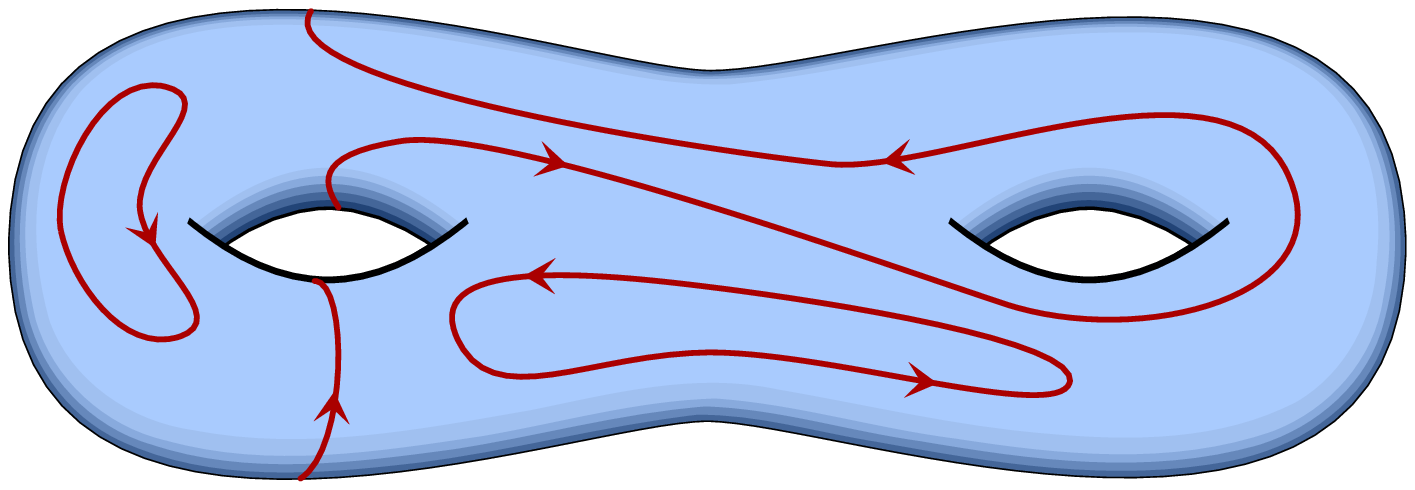}}}
  \end{picture}}
  ~=~ \sum_{\phi}~
 \raisebox{-25pt}{\begin{picture}(85,50)
  \put(0,0){\scalebox{0.40}{\includegraphics{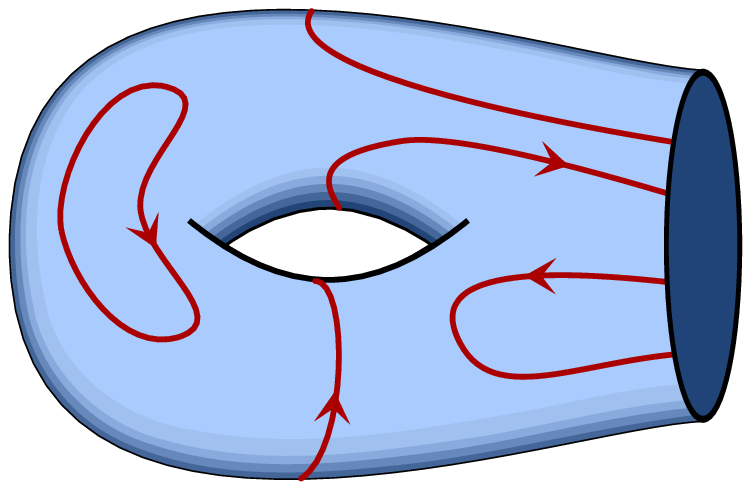}}}
  \end{picture}}
  \,|\phi\rangle ~\langle \phi|\,
 \raisebox{-25pt}{\begin{picture}(110,5)
  \put(0,0){\scalebox{0.40}{\includegraphics{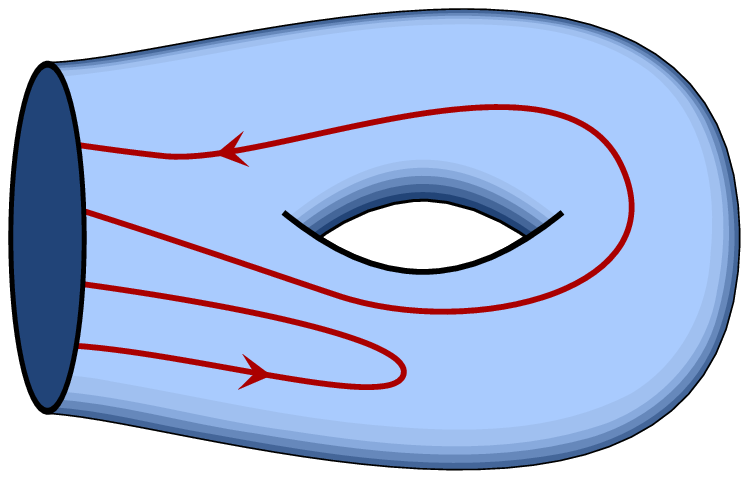}}}
  \end{picture}}
$$

\caption{When a sum over intermediate states is inserted on a circle
that intersects defect lines, the intermediate states lie in a
twisted space of states.}\label{fig:factor-ws}
\end{figure}

\begin{figure}[bt]

$$
  \raisebox{-30pt}{\begin{picture}(80,60)
  \put(0,0){\scalebox{0.60}{\includegraphics{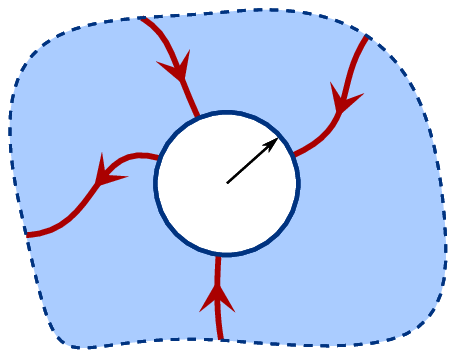}}}
  \put(0,0){
     \setlength{\unitlength}{.60pt}\put(-28,-16){
     \put( 58, 45)   {\scriptsize $|\hspace{-.5pt}\phi\rangle$ }
     \put(100, 63)   {\scriptsize $r$ }
     }\setlength{\unitlength}{1pt}}
  \end{picture}}
  \quad \xrightarrow{~~ r \, \rightarrow \, 0 ~~} \quad
  \raisebox{-30pt}{\begin{picture}(80,60)
  \put(0,0){\scalebox{0.60}{\includegraphics{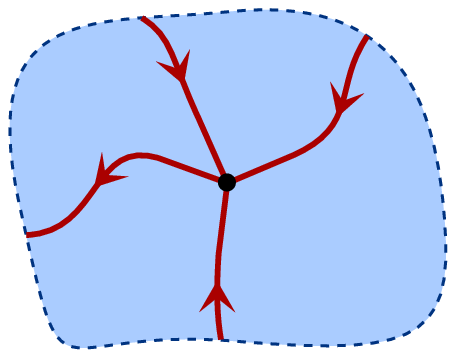}}}
  \put(0,0){
     \setlength{\unitlength}{.60pt}\put(-28,-16){
     \put( 78, 53)   {\scriptsize $\phi$ }
     }\setlength{\unitlength}{1pt}}
  \end{picture}}
$$

\caption{States $\,|\phi\rangle\,$ in a twisted space of states
correspond to fields $\,\phi\,$ at defect junctions via the
state-field correspondence.}\label{fig:state-field}
\end{figure}

There are at least two reasons why one should look at defect
junctions once defect lines are allowed. The first reason is
provided by the quantised sigma model and the factorisation
properties of the path integral, as explained in detail in section
\ref{sec:CFT-sew}. Consider, for example, the quantised sigma model
on a world-sheet such as the one in figure \ref{fig:factor-ws}. By
the factorisation of the path integral we mean that we can cut the
world-sheet along any circle and express the original amplitude as a
sum over intermediate states. If the circle along which we cut
intersects the defect lines $\,D_1,D_2,\ldots,D_n\,$ then the states
we sum over live in a Hilbert space $\,\Hc_{D_1 D_2\ldots D_n}\,$ of
`twisted' field configurations on the circle, cf., again, figure
\ref{fig:factor-ws}. That is, the field on the circle can have
discontinuities where the defect lines $\,D_1,D_2,\ldots,D_n\,$
intersect the circle, and the allowed jumps in the value of the
field are constrained by the defect condition. If the quantised
sigma model is conformal -- for example, if we are considering the
WZW model -- then there is a correspondence between states and
fields. This correspondence works by starting with a boundary circle
labelled by a state $\,|\phi\rangle\,$ and taking the radius of the
circle to zero, using the scale transformations to transport
$\,|\phi\rangle\,$ from one radius to another. What remains when the
radius reaches zero is a field inserted at the centre of the circle.
If several defect lines end on the boundary circle then the
resulting field sits at a junction point of these defect lines. This
is illustrated in figure \ref{fig:state-field} and discussed again
in section \ref{sec:CFT-topdef}.

\begin{figure}[bt]

$$
\Si_L ~=~ \raisebox{-45pt}{\begin{picture}(110,95)
  \put(0,0){\scalebox{0.65}{\includegraphics{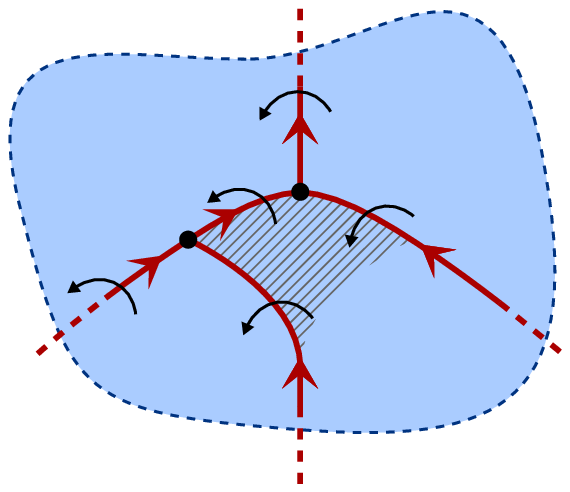}}}
  \put(0,0){
     \setlength{\unitlength}{.65pt}\put(-13,-5){
     \put(107,121)   {\scriptsize $\cdot x y z $ }
     \put( 60, 98)   {\scriptsize $\cdot x y $ }
     \put( 66, 53)   {\scriptsize $\cdot y $ }
     \put( 30, 72)   {\scriptsize $\cdot x $ }
     \put(116, 93)   {\scriptsize $\cdot z $ }
     }\setlength{\unitlength}{1pt}}
\end{picture}}
\,,\qquad\qquad \Si_R ~=~ \raisebox{-45pt}{\begin{picture}(110,95)
  \put(0,0){\scalebox{0.65}{\includegraphics{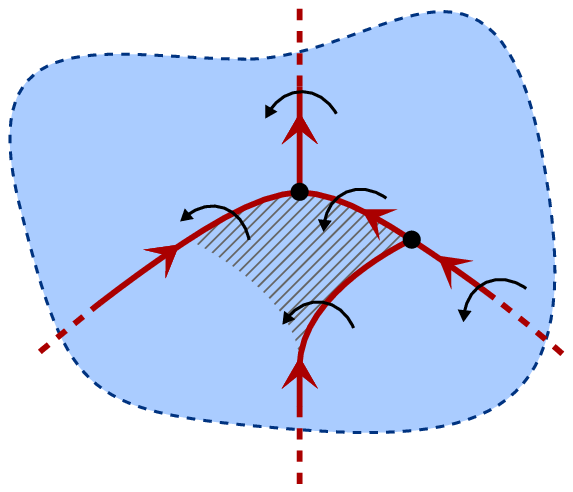}}}
  \put(0,0){
     \setlength{\unitlength}{.65pt}\put(-13,-5){
     \put(107,121)   {\scriptsize $\cdot x y z $ }
     \put( 60, 93)   {\scriptsize $\cdot x $ }
     \put(116, 55)   {\scriptsize $\cdot y $ }
     \put(153, 72)   {\scriptsize $\cdot z $ }
     \put(116, 96)   {\scriptsize $\cdot y z $ }
     }\setlength{\unitlength}{1pt}}
\end{picture}}
$$

\caption{The relevant part of the two world-sheets $\,\Si_L\,$ and
$\,\Si_R\,$ used in the definition of the 3-cocycle on $\,Z(\Gx)$.\
The field jumps by multiplication with the indicated element of
$\,Z(\Gx)\,$ when crossing the defect line. The values of the field
on $\,\Si_L\,$ and $\,\Si_R\,$ differ only in the shaded region.}
\label{fig:defect-for-cocycle}
\end{figure}

The second reason to consider defect junctions is that they allow to
extract interesting data from the classical theory, which one may
next compare to the corresponding quantities in the quantised model.
We illustrate this on the example of jump defects in the WZW model
for a compact simple connected and simply connected Lie group
$\,\Gx\,$ (with Lie algebra $\,\ggt$). Let $\,\Gc\,$ be the gerbe on
$\,\Gx\,$ with the curvature given by the Cartan 3-form
\be
H(g)=\tfrac{1}{3}\,{\rm tr}_\ggt\bigl(g^{-1}\,\sfd g\wedge g^{-1}\,
\sfd g\wedge g^{-1}\,\sfd g\bigr)\,,\qquad\qquad g\in\Gx\,.
\ee
We shall use the gerbe $\,\Gc^{\star\sfk}\,$ for some integer
$\,\sfk\ge 0$,\ which is given by the $\sfk$-fold product of
$\,\Gc\,$ with itself (cf.\ section \ref{sec:gerbes-local-data}) and
thus has curvature $\,\sfk H$.\ The jumps we allow are those by
elements $\,z\,$ of the centre $\,Z(\Gx)\,$ of $\,\Gx$.\ The
corresponding defects have the property that they are {\em
topological} in the sense that the defect line can be moved on the
world-sheet without modifying the value of the action functional.
This is described in more detail in section \ref{sec:conf-top-def}.
There also exist topological defect junctions for these jump
defects, which can similarly be moved on the world-sheet without
affecting the holonomy. Consider the world-sheets $\,\Si_L\,$ and
$\,\Si_R\,$ which contain the respective networks $\,\G_L\,$ and
$\,\G_R\,$ of defect lines. We take $\,\G_L\,$ and $\,\G_R\,$ to
differ only in the subset of the world-sheet shown in figure
\ref{fig:defect-for-cocycle}. In this figure, a defect line is
labelled by the element of $\,Z(\Gx)\,$ by which the field jumps.
Let $\,X_L(\zeta)\,$ be the sigma-model field on $\,\Si_L$,\ and
$\,X_R(\zeta)\,$ the corresponding field on $\,\Si_R$.\ We choose
$\,X_L\,$ and $\,X_R\,$ such that they are equal outside of the
shaded region of the world-sheet shown in figure
\ref{fig:defect-for-cocycle}. In the shaded region, they are related
as $\,X_R(\zeta)=y\cdot X_L(\zeta)$.\ In this way, $\,X_R\,$ is
uniquely determined by $\,X_L$.\ Let us denote by $\,S[(\G,X);\g]\,$
the action functional for a field $\,X(\zeta)\,$ and a defect
network $\,\G\,$ embedded in a world-sheet $\,\Si\,$ with metric
$\,\g$.\ One then finds that
\be
\exp\bigl(-S[(\G_L,X_L);\g]\bigr)=\psi_{\Gc^{\star\sfk}}(x,y,z)
\cdot\exp\bigl(-S[(\G_R,X_R);\g]\bigr)
\labl{eq:intro-hol-compare}
holds for a ${\rm U}(1)$-valued function $\,\psi_{\Gc^{\star\sfk}}
(x,y,z)\,$ which is {\em independent} of the choice of $\,X_L\,$
(this choice then fixes $\,X_R$), and which is invariant under
deformations of the defect lines, provided that we do not move one
vertex past another. We treat this example in detail in section
\ref{sec:holo}, where we also demonstrate that $\,\psi_{\Gc^{\star
\sfk}}\,$ is a 3-cocycle on $\,Z(\Gx)\,$ and defines a class $\,[
\psi_{\Gc^{\star\sfk}}]\,$ in $\,H^3(Z(\Gx),{\rm U}(1))$,\ the third
cohomology group of $\,Z(\Gx)\,$ with values in $\,{\rm U}(1)\,$
(with trivial $Z(\Gx)$-action). The configuration shown in figure
\ref{fig:defect-for-cocycle} was studied in \cite{Jureit:2006yf}
from the point of view of interacting orbifold string theories.
There, figure \ref{fig:defect-for-cocycle} was used to show that the
triviality of $\,[\psi_{\Gc^{\star\sfk}}]\,$ is necessary to have a
consistent interaction of closed strings in the orbifolded theory.

The comparison of \eqref{eq:intro-hol-compare} can also be carried
out in the quantised WZW model for the affine Lie algebra
$\,\ggtk$,\ where the integer $\,\sfk\,$ is the one determining the
gerbe $\,\Gc^{\star\sfk}\,$ used above. This is done in section
\ref{sec:def-of-CFT}, with the following result. The topological
defects of the quantum WZW model for the affine Lie algebra
$\,\ggtk\,$ which commute with the Ka\v c--Moody currents are
labelled by irreducible integrable highest-weight representations
$\,\la\,$ of $\,\ggtk$.\ One can assign a representation $\,\la_z\,$
to each element $\,z\in Z(\Gx)$.\ The representations $\,\la_z\,$
are precisely the simple currents of the WZW model (with the one
exception of $\,\widehat{\gt{e}(8)}_2$,\ which has a simple current
even though $\,Z({\rm E}(8))=\{e\}$,\ see \cite{Fuchs:1990wb}).
Comparing correlators on the world-sheets $\,\Si_L\,$ and
$\,\Si_R\,$ in figure \ref{fig:defect-for-cocycle} gives
\be
\mathrm{Corr}(\G_L,\Si_L)=\psi_{\ggtk}(x,y,z)\cdot
\mathrm{Corr}(\G_R,\Si_R)
\ee
for a ${\rm U}(1)$-valued function $\,\psi_{\ggtk}(x,y,z)\,$ which
is, again, a 3-cocycle on $\,Z(\Gx)\,$ and which defines a
cohomology class $\,[\psi_{\ggtk}]\in H^3(Z(\Gx),{\rm U}(1))$.\ We
compute $\,\psi_{\ggtk}\,$ in section \ref{sec:CFT-cocycle}.

The second main result of this paper is the observation that the
cohomology classes obtained in the classical and quantum
computations coincide,
\be
[\psi_{\Gc^{\star\sfk}}]=[\psi_{\ggtk}]\,.
\labl{eq:3-class-equal}

In the classical theory, the class $\,[\psi_{\Gc^{\star\sfk}}]\,$
determines the obstruction to the existence of a $\Zc$-equivariant
gerbe, for $\,\Zc\subset Z(\Gx)\,$ a subgroup. The condition
$\,[\psi_{\Gc^{\star\sfk}}|_{\Zc}]=1\,$ imposes selection rules on
$\,\sfk$.\ If the condition holds $\Zc$-equivariant gerbes exist and
can be used to define the sigma model on the orbifold $\,\Gx/\Zc\,$
\cite{Felder:1988sd,Gawedzki:2002se,Gawedzki:2003pm}. Similarly, in
the quantum WZW model, $\,[\psi_{\ggtk}]\,$ is the obstruction to
the existence of a simple-current orbifold
\cite{Schellekens:1989am,Schellekens:1990xy,Kreuzer:1994,Fuchs:2004dz}.
Thus, one way to read \eqref{eq:3-class-equal} is that the classical
obstruction to the orbifolding of the sigma model is preserved by
quantisation.

A related way to interpret \eqref{eq:3-class-equal} is as follows:
If a discrete symmetry group $\,S\,$ of a CFT is implemented by
defects then this group automatically comes with the additional
datum of a class $\,[\psi]\in H^3(S,{\rm U}(1))\,$ (this will be
explained in section \ref{sec:CFT-group} below). The same is true
for the classical sigma model. Equation \eqref{eq:3-class-equal}
states that, for the WZW model and for $\,S=Z(\Gx)$,\ the class
$\,[\psi]\,$ in $\,H^3(S,{\rm U}(1))\,$ is not changed when
quantising the model.

That for a given subgroup $\,\Zc\subset Z(\Gx)\,$ the values of
$\,\sfk\,$ for which $\,[\psi_{\Gc^{\star\sfk}}|_\Zc]=1\,$ are
precisely those for which $\,[\psi_{\ggtk}|_\Zc]=1\,$ is already
known from \cite{Gawedzki:2002se,Gawedzki:2003pm}. The novelty in
\eqref{eq:3-class-equal} is the statement that the cohomology
classes coincide for all $\,\sfk$,\ and on all of $\,Z(\Gx)$.\
Defect junctions thus give an explicit way to extract a
non-perturbative CFT datum -- the fusing matrix (6j-symbols)
restricted to the simple-current sector -- from a classical
calculation with gerbes.

\medskip

The paper is organised as follows: We start in section
\ref{sec:holo} by
reviewing the concept of the holonomy for
world-sheets without defects and for those with circular defects.
Then we give our construction of the holonomy in the presence of
defect junctions and compute the 3-cocycle for the jump defects in
the classical WZW model. The formulation of the quantum field theory
in the presence of defect lines and the computation of the 3-cocycle
in the quantum theory are given in section \ref{sec:def-of-CFT}.
Finally, the results of the classical and quantum calculation are
compared in section \ref{sec:comparison}.

\bigskip\noindent
{\bf Acknowledgements:} We would like to thank
  J.~Fuchs,
  K.~Gaw{\c{e}}dzki,
  G.~Sarkissian,
  C.~Schweigert,
  D.~Stevenson, and
  K.~Waldorf
for helpful discussions, and K.~Gaw{\c{e}}dzki for comments on a
draft of this paper. This research was partially supported by the
EPSRC First Grant EP/E005047/1, the PPARC rolling grant PP/C507145/1
and the Marie Curie network `Superstring Theory'
(MRTN-CT-2004-512194).

\newpage

\sect{Holonomy for world-sheets with defect
networks}\label{sec:holo}

In this section, we give a prescription for the holonomy for a
world-sheet with an embedded network of defect lines. We begin by
collecting the necessary ingredients, starting with the definition
of a gerbe in terms of its local data, and proceed to describe and
justify the proposed holonomy formula.

\subsection{Gerbes in terms of local
data}\label{sec:gerbes-local-data}

Let $\,M\,$ be a smooth manifold and let $\,\Oc^M = \{\ \Oc^M_i \ |\
i \In \Ic\ \}\,$ be a good open cover of $\,M$.\ Write the $p$-fold
intersection of open sets as $\,\Oc^M_{i_1 i_2 \dots i_p} =
\bigcap_{k=1}^p\, \Oc^M_{i_k}$.\ The qualifier `good' means that
each non-empty $\,\Oc^M_{i_1 i_2\dots i_p}\,$ is contractible.

For $\,p \ge 0\,$ and $\,r \ge 1$,\ let
$\,\check{C}^{p,r}\bigl(\Oc^M\bigr)\,$ be the set whose elements
$\,\om\,$ are collections of smooth differential $r$-forms \be
  \om = \{\ \om_{i_1\dots i_{p+1}} \in
  \Omega^r\bigl(\Oc^M_{i_1 \dots i_{p+1}}\bigr)\ |\ i_k \in \Ic\
  \text{~s.t.~}\ \Oc^M_{i_1 \dots i_{p+1}} \neq \emptyset\ \}\,,
\ee where $\,\om_{i_1\dots i_{p+1}}\,$ is required to be
antisymmetric in all indices. This is a \v Cech $p$-cochain with
values in the sheaf of differential $r$-forms on $\,M$,\ but we
shall not need this background in the present paper. Note that
$\,\check{C}^{p,r}\bigl(\Oc^M\bigr)\,$ inherits the structure of a
vector space from $\,\Omega^r\bigl(\Oc^M_{i_1\dots i_{p+1}}\bigr)$.\
Below, we shall only use that $\,\Omega^r\bigl(\Oc^M_{i_1 \dots
i_{p+1}}\bigr)\,$ is an abelian group, which will be written
additively.

The sets $\,\check{C}^{p,0}\bigl(\Oc^M\bigr)\,$ are defined slightly
differently. Namely, an element $\,\varphi\,$ of
$\,\check{C}^{p,0}\bigl(\Oc^M\bigr)\,$ is a collection
$\,\varphi_{i_1\dots i_{p+1}}$,\ where each $\,\varphi_{i_1\dots
i_{p+1}}\in {\rm U}(1)_{\Oc^M_{i_1 \dots i_{p+1}}}\,$ is a smooth
${\rm U}(1)$-valued function on $\,\Oc^M_{i_1 \dots i_{p+1}}\,$ that
is antisymmetric in all indices. The set
$\,\check{C}^{p,0}\bigl(\Oc^M\bigr)\,$ inherits the structure of an
abelian group from $\,{\rm U}(1)$,\ which will be written
multiplicatively.

In order to describe a gerbe and its gauge transformations, one uses
the first four components of a chain complex
$\,A^\bullet_M\bigl(\Oc^M\bigr)$,\ given by (we drop $\,\Oc^M\,$
from the notation) \bea
  A^0_M = \check{C}^{0,0}\,, \qquad
  A^1_M = \check{C}^{0,1} \ti \check{C}^{1,0}\,, \qquad
  A^2_M = \check{C}^{0,2} \ti \check{C}^{1,1} \ti \check{C}^{2,0}\,,
  \enl
  A^3_M = \check{C}^{0,3} \ti \check{C}^{1,2} \ti \check{C}^{2,1}
    \ti \check{C}^{3,0}\,.
\eear\ee Thus, for example, an element of $\,A^1_M\,$ is a pair
$\,(\Pi,\chi)\,$ where $\,\Pi\,$ is a collection of smooth 1-forms
$\,\Pi_i\,$ on $\,\Oc^M_i$,\ and $\,\chi\,$ is a collection of
smooth ${\rm U}(1)$-valued functions $\,\chi_{ij}\,$ on the overlap
$\,\Oc^M_{ij} = \Oc^M_i\cap \Oc^M_j\,$ which is antisymmetric in its
\v Cech indices in the sense that $\,\chi_{ij}(x) =
\chi_{ji}(x)^{-1}$.\ We shall also write elements of $\,A^1_M\,$ as
$\,(\Pi_i,\chi_{ij})$,\ and similarly for the other components of
$\,A^\bullet_M$.\ Each $\,A^m_M\,$ forms an abelian group under the
addition of the component $r$-forms and the multiplication of the
${\rm U}(1)$-valued functions. For instance, the definition of the
sum of elements of $\,A^2_M\,$ reads
\be
  (B_i, A_{ij}, g_{ijk}) + (B_i', A_{ij}', g_{ijk}')
  = (B_i+B_i'\,,\, A_{ij}+ A_{ij}'\,,\, g_{ijk} \cdot g_{ijk}')\,.
\ee

The Deligne differential $\,D_{(r)} :  A^r_M \rightarrow
A^{r+1}_M\,$ is given by \be \begin{array}{rl} \displaystyle
  D_{(0)}(f_i) \etb= ( -\sfi\,\sfd \log f_i \,,\, f_j^{-1} \cdot
  f_i)\,,
  \enl
  D_{(1)}(\Pi_i,\chi_{ij}) \etb = (
    \sfd \Pi_i \,,\,
    - \sfi\,\sfd \log \chi_{ij} + \Pi_j - \Pi_i \,,\,
    \chi_{jk}^{-1} \cdot \chi_{ik} \cdot \chi_{ij}^{-1} )\,,
  \enl
  D_{(2)}(B_i, A_{ij}, g_{ijk})
  \etb= (\sfd B_i \,,\,
    \sfd A_{ij} -  B_j + B_i \,,
  \\ \etb \qquad
    - \sfi\,\sfd \log g_{ijk} + A_{jk} - A_{ik}  + A_{ij} \,,\,
    g_{jkl}^{-1} \cdot g_{ikl}^{\phantom{-1}} \cdot
    g_{ijl}^{-1} \cdot g_{ijk}^{\phantom{-1}} ) ~,
\eear\ee where we follow the conventions of \cite{Gawedzki:2007uz}.
One verifies that $\,D_{(r+1)} \cir D_{(r)} = 0$.\ Below, we shall
typically just write $\,D\,$ instead of $\,D_{(r)}$.

Mathematically, the appropriate description of $A^\bullet_M$ is in
terms of the \v Cech--Deligne double complex and the resulting
Deligne hypercohomology\footnote{Strictly speaking, the definition
of Deligne hypercohomology requires truncating the de Rham complex.
In our paper, we could equivalently work with the full \v Cech--de
Rham complex but since we want to adhere to the terminology of
Deligne hypercohomology, we choose to truncate the underlying
differential complex to
$\,\underline{U(1)}_M\xrightarrow{\frac{1}{\sfi}\,\sfd
\log}\underline{\Om}^1(M)\xrightarrow{\sfd}\underline{\Om}^2(M)
\xrightarrow{\sfd}\underline{\Om}^3(M)$,\ where the underlining
indicates that we are dealing with the corresponding sheaf.}. We
refer the reader to \cite{Brylinski:1993ab} for a detailed
exposition.

\medskip

With these ingredients in hand, we can now define the notion of a
gerbe in terms of its local data. A {\em gerbe with connection in
terms of local data}, or a {\em gerbe} for short, on a smooth
manifold $\,M\,$ is a pair $\,\Gc = \bigl(\Oc^M, b\bigr)\,$ where
$\,\Oc^M\,$ is a good open cover of $\,M\,$ and $\,b \in A^2_M\,$ is
such that
\be
Db = (H_i, 0,0, 1)\,.
\labl{eq:def-gerbe}
The objects $\,H_i\,$ are 3-forms on $\,\Oc^M_i\,$ but it is not
hard to see that they are, in fact, restrictions of a globally
defined closed 3-form $\,H \in \Omega^3(M)$,\ called the {\em
curvature} of $\,\Gc$.

\medskip

Given two gerbes: $\,\Gc = \bigl(\Oc^M,b\bigr)\,$ and
$\,\Hc=\bigl(\Oc^M,b'\bigr)\,$ defined with respect to the same open
cover of $\,M$,\ a {\em stable isomorphism} $\,\Phi : \Gc
\rightarrow \Hc\,$ {\em in terms of local data} is an element
$\,\Phi \in A^{1}_M\,$ such that $\,b' = b + D \Phi$.\ Given a third
gerbe $\,\Kc = \bigl(\Oc^M,b''\bigr)\,$ and a stable isomorphism
$\,\Psi :  \Hc \rightarrow \Kc$,\ the {\em composition} $\,\Psi
\circ \Phi\,$ is defined to be the stable isomorphism $\,\Psi + \Phi
\in A^1_M\,$ from $\,\Gc\,$ to $\,\Kc$.\ Indeed, $\,b'' = b' + D
\Psi = b + D \Phi + D \Psi$.\ The notion of a stable isomorphism can
be extended to gerbes over the same base $\,M\,$ but with two
different open covers $\,\Oc^M\,$ and $\,\til\Oc^M\,$ by passing to
a common refinement.

Physically, a stable isomorphism
is a gauge transformation of the
gerbe data. Computations of observable quantities carried out with
respect to distinct but gauge-equivalent (i.e.\ stably isomorphic)
gerbes should give the same result. One distinguishing feature of
gerbes is that the local data of gauge transformations between two
given gerbes can themselves be related by another kind of a gauge
transformation. This is captured by the notion of a 2-morphism,
which is defined as follows: Let $\,\Gc = \bigl(\Oc^M,b\bigr)\,$ and
$\,\Hc = \bigl(\Oc^M,b'\bigr)\,$ be gerbes over a manifold $\,M$,\
and let $\,\Phi :  \Gc \rightarrow \Hc\,$ and $\,\Psi :  \Gc
\rightarrow \Hc\,$ be two stable isomorphisms. A 2-{\em morphism}
$\,\varphi : \Phi \Longrightarrow \Psi\,$ is an element of
$\,A^0_M\,$ such that
\be
  \Phi + D \varphi = \Psi\,.
\ee In terms of local data $\,\Phi = (P_i,K_{ij}),\ \Psi =
(P_i',K_{ij}')\,$ and $\,\varphi = (f_i)$,\ this means that \be
  P_i - \sfi\, \sfd \log f_i = P_i'
  \qquad \text{and} \qquad
  K_{ij}\cdot f_i\cdot f_j^{-1} = K_{ij}'\,.
\ee The word `2-morphism' derives from the 2-categorial
interpretation of gerbes given in terms of local data, see
\cite{Stevenson:2000wj,Waldorf:2007mm}.
\medskip

The presentation of the holonomy formul\ae\ below requires some
additional constructions for gerbes, namely trivial gerbes, the
product of gerbes and pullback gerbes. Let us briefly review the
local formulation of these notions.
\\[.3em]
\nxt A {\em trivial gerbe} for a 2-form $\,\om\,$ on $\,M\,$ is the
gerbe $\,I\bigl(\Oc^M,\om\bigr) = \bigl( \Oc^M, (\om_i,0,1)
\bigr)$,\ where $\,\Oc^M\,$ is a good open cover of $\,M\,$ and
$\,\om_i\,$ is the restriction of $\,\om\,$ to $\,\Oc^M_i$.\ The
curvature 3-form of $\,I\bigl(\Oc^M,\om\bigr)\,$ is exact, hence the
name `trivial'. We shall sometimes abbreviate the notation for the
trivial gerbe as $\,I\bigl(\Oc^M,\om\bigr)\equiv I(\om)$.
\\[.3em]
\nxt Given two gerbes: $\,\Gc = \bigl(\Oc^M,b\bigr)\,$ and
$\,\Hc=\bigl(\Oc^M,b'\bigr)\,$ defined with respect to the same good
open cover of $\,M$,\ we take the {\em product gerbe} to be $\,\Gc
\star \Hc = \bigl(\Oc^M, b+b'\bigr)$.\ This is analogous to the
tensor product of line bundles, hence the product notation. The
curvature of $\,\Gc \star \Hc\,$ is $\,H_\Gc + H_\Hc$,\ where
$\,H_\Gc\,$ and $\,H_\Hc\,$ are the curvatures of $\,\Gc\,$ and
$\,\Hc$,\ respectively.
\\[.3em]
\nxt Let $\,\Oc^M\,$ and $\,\Oc^N\,$ be good open covers of
manifolds $\,M\,$ and $\,N$.\ In order to define pullbacks of
gerbes, stable isomorphisms and 2-morphisms, it is not enough to
specify a smooth map from $\,M\,$ to $\,N\,$ -- we also need to know
how the \v Cech indices are related. To this end, we define a {\em
\v Cech-extended map} $\,\check f: M \rightarrow N\,$ to be a pair
$\,(f,\phi)\,$ where $\,f : M \rightarrow N\,$ is a smooth map and
$\,\phi\,$ is an index map, $\,\phi : \Ic^M \rightarrow \Ic^N$,\
such that $\,f\bigl(\Oc^M_i\bigr) \subset \Oc^N_{\phi(i)}$. Since
$\,\phi\,$ need not exist, not every map $\,f : M\rightarrow N\,$
can be turned into a \v Cech-extended map. Given another \v
Cech-extended map $\,\check g = (g,\gamma) : N \rightarrow K$,\
their composition is defined component-wise as $\,\check g \circ
\check f = (g\circ f,\gamma\circ\phi)$.
\\[.3em]
\nxt Let $\,\Gc = \bigl(\Oc^N,b\bigr)\,$ be a gerbe on $\,N\,$ and
let $\,\check f : M \rightarrow N\,$ be a \v Cech-extended map. The
{\em pullback gerbe} is $\,\check f^* \Gc = \bigl(\Oc^M,b'\bigr)\,$
with $\,b'=(B_i',A_{ij}',g_{ijk}') = (f^* B_{\phi(i)}, f^*
A_{\phi(i)\phi(j)}, g_{\phi(i)\phi(j)\phi(k)} \circ f)\equiv\check
f^*b$.\ The pullback $\,\check f^*C\,$ for $\,C=(c_{i_1}, c_{i_1
i_2},\ldots,c_{i_1 i_2\ldots i_{p+1}})\in A^p_M\,$ is defined in the
same way. To unclutter the notation, we shall frequently use the
shorthand
\be
f^*c_{i_1 i_2 \ldots i_{m}} \equiv
f^*c_{\phi(i_1)\phi(i_2)\ldots\phi(i_{m})}\,.
\labl{eq:index-abbrev}
If $\,\check f_1 = (f,\phi)\,$ and $\,\check f_2 = (f,\til\phi)\,$
are two \v Cech-extended maps that differ only in the choice of the
index map, the resulting pullback gerbes are stably isomorphic.
Indeed, one verifies that $\til b'=b'+Dp\,$ for $\,p=(\Pi_i,
\chi_{ij})=\bigl(f^* A_{\phi(i)\til\phi(i)},
\bigl(g_{\phi(i)\phi(j)\til\phi(j)}\cdot
g_{\phi(i)\til\phi(i)\til\phi(j)}^{-1}\bigr)\circ f\bigr)$.
\medskip

A more geometric description of gerbes is given by bundle gerbes,
see \cite{Murray:1994db,Murray:1999ew} and
\cite{Stevenson:2000wj,Johnson:2003,Waldorf:2007mm,Waldorf:2007phd}.
Their main advantage is that the choice of a good open cover of
$\,M\,$ is replaced by the more flexible concept of a surjective
submersion $\,\pi :  Y \rightarrow M$,\ which proves particularly
convenient in the WZW setting, see
\cite{Gawedzki:2002se,Meinrenken:2002,Gawedzki:2003pm}. It deserves
to be stressed that every bundle gerbe is stably isomorphic to a
bundle gerbe whose surjective submersion comes from a good open
cover of $\,M$,\ see \cite{Murray:1999ew}. The reason for us to use
the local description is that the formalism is easier to set up and
that we find the definition of the holonomy formula in terms of
local data more intuitive.

\subsubsection*{The Lie-group example}

As mentioned in the introduction, the example that we consider in
this paper is the WZW model for a compact simple connected and
simply connected Lie group $\,\Gx\,$ at level $\,\sfk \in
\Zb_{>0}\,$ (geometrically, the value of the level sets the size of
the group manifold). The Lie group comes equipped with the so-called
basic gerbe \cite{Gawedzki:2002se,Meinrenken:2002}, which we denote
by $\,\Gc$.\ It is the unique, up to a stable isomorphism, gerbe
with the curvature given by the Cartan 3-form
\qq
  H(g)=\tfrac{1}{3}\,{\rm tr}_\ggt\bigl(g^{-1}\,\sfd g\wedge g^{-1}\,\sfd
  g \wedge g^{-1}\,\sfd g\bigr)\,,\qquad\qquad g\in\Gx\,,
\qqq
the latter being fixed by the requirement of the vanishing of the
Weyl anomaly in the quantised sigma model. Note that the works
\cite{Gawedzki:2002se,Meinrenken:2002} use the language of bundle
gerbes; here, we have to decide once and for all on a good open
cover $\,\Oc^\Gx\,$ of $\,\Gx\,$ and on local gerbe data
$\,b=(B_i,A_{ij},g_{ijk})\,$ such that $\,\Gc =
\bigl(\Oc^\Gx,b\bigr)$.\ In fact, as we shall later want to consider
translations by elements of the centre $\,Z(\Gx)\,$ of $\,\Gx$,\ we
choose the good open cover $\,\Oc^\Gx\,$ to be invariant under such
translations in the sense that there exists an action $\,Z(\Gx) \ti
\Ic^\Gx \rightarrow \Ic^\Gx\ :\  (z,i) \mapsto z.i\,$ such that
\be
  z\bigl(\Oc^\Gx_i\bigr) = \Oc^\Gx_{z.i}
  \qquad \text{for all}\quad
  z \in Z(\Gx)\,,
\labl{eq:ZG-acts-on-cech}
where $\,z\bigl(\Oc^\Gx_i\bigr) = \{ \ z\cdot g \ |\ g\In\Oc^\Gx_i \
\}$.\ We can use this action to turn $\,z\,$ into a \v Cech-extended
map from $\,\Gx\,$ to itself by setting
\be
\check z\ :\ \Gx \rightarrow \Gx\,, \qquad\qquad \check z = (g
\mapsto z\cdot g\,,\, i\mapsto z.i)\,.
\labl{eq:z-cech-extend}
Given a $Z(\Gx)$-invariant cover of $\,\Gx$,\ we obtain a natural
definition of a left-regular action of $\,Z(\Gx)\,$ on
$\,A^r_\Gx\,,\ r\geq 0\,$ by the \v Cech-extended pullback
\qq\label{eq:ext-pullb-act}
(z,\om)\mapsto\check{\bigl(z^{-1}\bigr)}^*\om=z.\om
\qqq
which we are going to encounter frequently in the sequel. Finally,
for the WZW model at level $\,\sfk$,\ one uses the $\sfk$-fold
product $\,\Gc^{\star\sfk}\,$ of the basic gerbe with itself.

\subsection{Surface holonomy in the absence of
defects}\label{sec:hol-no-defect}

Here, we review the definition of the surface holonomy in the
absence of defects \cite{Alvarez:1984es,Gawedzki:1987ak}. In this
case, the world-sheet is an oriented smooth compact two-manifold
$\,\Si\,$ with empty boundary and the target space is a (not
necessarily connected) smooth manifold $\,M\,$ with gerbe $\,\Gc$.

Write the local data for $\,\Gc = \bigl(\Oc^M,b\bigr)\,$ as
$\,b=(B_i,A_{ij},g_{ijk})$.\ Given a once differentiable
map\footnote{By this we mean a $C^1$-function, i.e.\ a function that
is once continuously differentiable.} $\,X : \Si \rightarrow M$,\
the holonomy $\,\Hol_\Gc(X)\,$ is an element of $\,{\rm U}(1)\,$
defined by the following formula
\be
  \Hol_\Gc(X) ~=~
  \prod_{t \in \triangle(\Si)}\, \Bigg\{ \exp\!\Big(i \int_t \widehat
  B_t \Big)\, \prod_{e \subset t}\, \Bigg[ \exp\!\Big(i \int_e \widehat
  A_{te} \Big)\, \prod_{v \in e}\, \widehat g_{tev}(v) \Bigg]
  \Bigg\}\,,
\labl{eq:hol-no-defect} whose ingredients we proceed to explain:
\begin{itemize}
\item $\triangle(\Si)\,$ is a triangulation of $\,\Si\,$ which is
subordinate to $\,\Oc^M\,$ with respect to $\,X\,$ in the sense that
for each triangle $\,t \in \triangle(\Si)\,$ there is an index
$\,i_t \in \Ic\,$ such that $\,X(t) \subset \Oc^M_{i_t}$.\ This
implies that also the image under $\,X\,$ of an edge $\,e\,$ lies in
one of the open sets of the good cover. For each edge $\,e\,$ and
vertex $\,v$,\ we pick an assignment $\,e\mapsto i_e\,$ and $\,v
\mapsto i_v\,$ such that $\,X(e) \subset \Oc^M_{i_e}\,$ and $\,X(v)
\in \Oc^M_{i_v}$.
\item $\widehat B_t = X^* B_{i_t}\,$ is the pullback of
$\,B_{i_t}\,$ to the triangle $\,t\,$ along $\,X$.\ Here, $\,X\,$ is
understood as a map from $\,t\,$ to $\,\Oc^M_{i_t}$.\ The 2-form
$\,\widehat B_t\,$ is integrated over $\,t\,$ using the orientation
of the world-sheet $\,\Si$.
\item $\widehat A_{te} = X^* A_{i_t i_e}\,$ is the pullback of
$\,A_{i_t i_e}\,$ to the edge $\,e\,$ along $\,X$.\ Here, $\,X\,$ is
understood as a map from $\,e\,$ to $\,\Oc^M_{i_t i_e}$.\ The 1-form
$\,\widehat A_{te}\,$ is integrated over $\,e$,\ where the
orientation of $\,e\,$ is the one induced by the triangle $\,t\,$
via the inward-pointing normal. (For example, the orientation of the
edges of a triangle embedded in $\,\Rb^2\,$ is counter-clockwise.)
\item $\widehat g_{tev} = (X^* g_{i_t i_e i_v})^{\eps_{tev}}$,\
where $\,X\,$ maps $\,v\,$ to $\,\Oc^M_{i_t i_e i_v}\,$ and
$\,\eps_{tev} = \pm 1\,$ is determined as follows: The edge $\,e\,$
inherits an orientation from the triangle $\,t$.\ If the vertex sits
at the end of $\,e\,$ with respect to this orientation we set
$\,\eps_{tev} = 1$.\ Otherwise, $\,\eps_{tev} = -1$.\ (For example,
in the interval $\,[0,1] \subset \Rb\,$ with the standard
orientation, the point $\,0\,$ has $\,\eps=-1\,$ and the point
$\,1\,$ has $\,\eps=1$.)
\end{itemize}
The reason why it is sufficient to require $\,X\,$ to be once
differentiable, rather than smooth, is that the holonomy
$\,\Hol_\Gc(X)\,$ and the kinetic term in the action only depend on
the first derivatives of the map $\,X$.\ This will play a r\^ole
when discussing topological defects in section
\ref{sec:conf-top-def} below.
\medskip

The holonomy remains unchanged if we perform a gauge transformation
on the gerbe data, and the relation $\,D b = (H_i,0,0,1)\,$ ensures
that the holonomy is independent of the choice of triangulation. We
shall return to these points in section \ref{sec:just}.

\subsection{Abelian bi-branes}\label{sec:bibrane-def}

In words, an abelian bi-brane of \cite{Fuchs:2007fw} is a
submanifold $\,Q\,$ of the product $\,M \ti M$,\ together with a
2-form $\,\om\,$ on $\,Q\,$ such that the pullbacks of the gerbe
$\,\Gc\,$ by the canonical projections to the two factors in $\,M
\ti M\,$ differ only by the trivial gerbe $\,I(\om)\,$ when
restricted to $\,Q$.\ Since we are working with local data, some
extra choices are involved, making the actual definition lengthier.
We shall also be slightly more general by not restricting ourselves
to the case that $\,Q\,$ is a submanifold of $\,M \ti M$.

Let $\,\Gc = \bigl(\Oc^M,b\bigr)\,$ be a gerbe on $\,M$.\ An {\em
abelian $\Gc$-bi-brane $\,\Bc\,$ in terms of local data} is a tuple
\be
  \Bc = \bigl(Q,\om,\Oc^Q,\check\iota_1,\check\iota_2,\Phi\bigr)\,,
\labl{eq:bi-brane-tuple} where
\begin{list}{-}{\topsep .4em \leftmargin 2.5em \itemsep 0em}
\item[(B.i)] $Q\,$ is a smooth manifold;
\item[(B.ii)] $\om\,$ is a smooth 2-form on $\,Q$,\ called the
curvature of the $\Gc$-bi-brane;
\item[(B.iii)] $\Oc^Q = \bigl\{\ \Oc^Q_i \ |\ i \In \Ic^Q\ \}\,$
is a good open cover of $\,Q$;
\item[(B.iv)] $\check\iota_1=(\iota_1,\phi_1)\,$ and $\,\check
\iota_2=(\iota_2,\phi_2)\,$ are \v Cech-extended maps from $\,Q\,$
to $\,M$;
\item[(B.v)] $\Phi : \check\iota_1^* \Gc \rightarrow \check\iota_2^* \Gc
\star I\bigl(\Oc^Q,\om\bigr)\,$ is a stable isomorphism.
\end{list}

By (B.v), the two pullbacks of the gerbe $\,\Gc\,$ differ by a
trivial gerbe on $\,Q$.\ For the curvature of $\,\Gc$,\ this implies
$\,\iota_1^* H - \iota_2^* H = \sfd \om$.\ Thus, the difference of
the pullbacks of the curvature has to be exact on $\,Q$.

Let the local data for the gerbe and the stable isomorphism be given
by $\,b = (B_i,A_{ij},g_{ijk})\in A^2_M\,$ and $\,\Phi =
(P_i,K_{ij})\in A^1_Q$,\ respectively. The stable isomorphism now
gives the condition (recall the shorthand notation
\eqref{eq:index-abbrev})
\be
\check\iota_1^*( B_{i} , A_{ij} , g_{ijk} ) + D(P_i , K_{ij}) =
\check\iota_2^*( B_{i} , A_{ij} , g_{ijk} ) + (\om,0,1)\,.
\labl{eq:stable-iso-local}

There are a number of differences with respect to the original
definition in \cite{Fuchs:2007fw,Waldorf:2007phd} which we would
like to point out and justify. Namely, in
\cite{Fuchs:2007fw,Waldorf:2007phd},
\begin{list}{-}{\topsep .4em \leftmargin 1em \itemsep 0em}
\item a bi-brane is defined in terms of more general morphisms
\cite{Waldorf:2007mm} between the pullback gerbes, not just stable
isomorphisms. In a nutshell, the data \eqref{eq:bi-brane-tuple} can
be understood as a gerbe-twisted line bundle over $\,Q$,\ while in
the case of a general bi-brane, one also allows gerbe-twisted vector
bundles of higher rank. In this paper, we shall restrict our
attention to the case of abelian bi-branes. This is done for
simplicity.
\item $\,Q\,$ is taken to be a submanifold of $\,M \ti M$.\ This
is recovered in the present definition as a special case upon
choosing $\,\check\iota_1\,$ and $\,\check\iota_2\,$ to be the
canonical projections onto the two factors. The reason why we use a
more general definition is that it will allow us to treat several
bi-branes using only one manifold $\,Q$,\ even if the worldvolumes
of the individual bi-branes would intersect as submanifolds in $\,M
\ti M$.
\item bi-branes between different target spaces $\,M_1\,$ and
$\,M_2\,$ are allowed, in which case $\,Q\,$ is a submanifold of
$\,M_1 \ti M_2$.\ This situation is covered by our approach because
we can take $\,M\,$ to be the disjoint union $\,M_1 \sqcup M_2\,$
and choose $\,\check\iota_1\,,\,\check\iota_2 : Q \rightarrow M\,$
to be the projections for a submanifold $\,Q \subset M_1 \ti M_2
\subset M \ti M$.
\end{list}

Two simple examples illustrate the data describing a bi-brane: The
trivial $\Gc$-bi-brane and D-branes (the latter were dubbed
$\Gc$-branes in the gerbe-theoretic context of
\cite{Gawedzki:2004tu}). The trivial $\Gc$-bi-brane for the gerbe
$\,\Gc\,$ over the target space $\,M\,$ is given by \be
  \Bc_\text{triv} = \bigl( M, 0, \Oc^M, \check\id, \check\id,
  \id_\Gc \bigr)\,.
\labl{eq:triv-bibrane} This corresponds to the diagonal embedding of
$\,M\,$ into $\,M \ti M\,$ with the 2-form $\,\om\,$ set to zero.
For a trivial $\Gc$-bi-brane, the holonomy for the world-sheet with
an embedded defect network given below reduces to the form
\eqref{eq:hol-no-defect}.

In order to describe a D-brane in a target space $\,N$,\ one takes
the manifold $\,M\,$ to be the disjoint union of $\,N\,$ and a
single point, $\,M = N \sqcup \{ \bullet \}$.\ Let $\,D\,$ be a
submanifold of $\,N\,$ with a 2-form field $\,\om$.\ Consider the
bi-brane
\be
  \Bc_\text{bnd} = ( D, \om, \Oc^D,
  \check\iota,\check\bullet,\Phi )\,,
\labl{eq:D-bibrane} where $\,\check\iota = (\iota,\phi_1)\,$ is just
the embedding of $\,D\,$ into $\,N \subset M\,$ and $\,\check\bullet
= (\bullet,\phi_2)\,$ is the constant map $\,D \rightarrow \{
\bullet \}$.\ For simplicity, we take the open cover $\,\Oc^D\,$ of
$\,D\,$ to be the intersection of the open sets in $\,\Oc^N\,$ with
$\,D\,$ and assume that the open sets in $\,\Oc^N\,$ are small
enough for the resulting cover $\,\Oc^D\,$ to be good. Thus,
$\,\phi_1 = \id\,$ in this case. The cover of $\,\{\bullet\}\,$
consists of one element, which we also label $\,\bullet$,\ and
$\,\phi_2\,$ is the constant map $\,\Ic^N \rightarrow \{ \bullet
\}$.\ Condition (B.v) on $\,\Phi\,$ and $\,\om\,$ now reads $\,\Phi
: \Gc|_{D} \rightarrow I(\om)$,\ which is precisely the local data
for a D-brane as defined in \cite{Gawedzki:2002se,Gawedzki:2004tu}.

\subsubsection*{The Lie-group example (cont'd)}

Continuing the Lie group example from section
\ref{sec:gerbes-local-data}, we now want to define a
$\Gc^{\star\sfk}$-bi-brane $\,\Bc_{Z(\Gx)}\,$ for jump defects. The
underlying manifold is
\be
  Q_{Z(\Gx)} = \Gx \ti Z(\Gx)\,,
\labl{eq:lie-ex-B}
and the 2-form $\,\om\,$ on $\,Q_{Z(\Gx)}\,$ is zero. The open cover
$\,\Oc^{Q_{Z(\Gx)}}\,$ is indexed by pairs from $\,\Ic^{Q_{Z(\Gx)}}
= \Ic^\Gx \ti Z(\Gx)$,\ and the corresponding open sets are
$\,\Oc^{Q_{Z(\Gx)}}_{i,x} = \Oc^\Gx_i \ti \{ x \}$.\ The two \v
Cech-extended maps $\,\check\iota_1 = (\iota_1,\phi_1)\,$ and
$\,\check\iota_2 = (\iota_2,\phi_2)\,$ from $\,Q_{Z(\Gx)}\,$ to
$\,\Gx\,$ are given by the formul\ae
\be
  \iota_1(g,x) = g\,,\qquad\phi_1(i,x) = i\,;\qquad\qquad
  \iota_2(g,x) = x^{-1}\cdot g\,,\qquad \phi_2(i,x)=x^{-1}.i\,.
\ee
The pullback of $\,\Gc\,$ along the translation by $\,x\in Z(\Gx)\,$
yields a gerbe stably isomorphic to $\,\Gc$.\ This follows since the
left-regular action of $\,Z(\Gx)\,$ induced by the pullback as in
\eqref{eq:ext-pullb-act} commutes with the Deligne differential
$\,D$,\ and translations by elements of $\,Z(\Gx)\,$ preserve the
Cartan 3-form. This conclusion also holds for all powers of
$\,\Gc$,\ and so one can find a set of 1-morphisms
\qq
\Ac_{Z(\Gx)}=\bigl\{\ \Ac_x\ :\ \Gc^{\star\sfk} \rightarrow
  x.\Gc^{\star\sfk} \ \vert \ x\in Z(\Gx)\ \bigr\}\,,
\qqq
cf.\ \eqref{eq:ext-pullb-act}, constructed explicitly in
\cite[sect.\,3]{Gawedzki:2003pm} in the framework of bundle gerbes
(the present notation conforms with the unified treatment laid out
in \cite[sect.\,1 \& 3]{Gawedzki:2008}). The local data of $\,\Phi :
\check\iota_1^*\Gc^{\star\sfk}\rightarrow\check\iota_2^*\Gc^{\star
\sfk}\,$ on $\,\Oc^{Q_{Z(\Gx)}}_{i,x}\,$ is given by the local data
of $\,\Ac_x\,$ on $\,\Oc^\Gx_i$,\ where we use $\,\Oc^\Gx_i\cong
\Oc^{Q_{Z(\Gx)}}_{i,x}$.\
\medskip

The existence of the stable isomorphisms $\,\Ac_{Z(\Gx)}\,$ is
ensured by the triviality of the cohomology group $\,H^2(\Gx,{\rm
U}(1))$.\ To see how this comes about, let us have a brief look at a
general target space $\,M\,$ with metric $\,G\,$ and 3-form field
$\,H$.\ Let $\,S\,$ be a finite subgroup of the isometry group of
$\,M\,$ which also preserves $\,H$.\ Pick a good open cover
$\,\Oc^M\,$ of $\,M\,$ which is invariant under $\,S\,$ in the sense
that, for all $\,x\in S$,\ we have $\,x(\Oc^M_i) = \Oc^M_{x.i}\,$
for some index $\,x.i$.\ This turns $\,x\,$ into a \v Cech-extended
map as in \eqref{eq:z-cech-extend}. Let $\,\Gc = (\Oc^M,b)\,$ be a
gerbe on $\,M\,$ with curvature $\,H$.\ Clearly, the stable
isomorphisms $\,\Ac_x : \Gc \rightarrow x.\Gc\,$ exist if and only
if the orbit $\,\{\ x.\Gc \ |\ x \in S \ \}\,$ lies entirely within
one stable-isomorphism class. In terms of local data, $\,\Ac_x\,$
obey $\,(\d_S b)_x=D\Ac_x$,\ where $\,(\d_S b)_x=x.b-b\,$ (cf.\
appendix \ref{app:cohom} for a basic reminder on the finite-group
cohomology). We shall call the collection $\,\Ac_S = \{\ \Ac_x \ |\
x \in S \ \}\,$ such that $\,(\d_S b)_x=D\Ac_x\,$ an {\em
element-wise presentation of} $\,S\,$ {\em on} $\,b$.\ The
obstructions to the existence of an element-wise presentation are
contained in $\,\mathbb{H}^2(M)=\Ker D_{(2)} / \Im D_{(1)}$,\ the
set of stable-isomorphism classes of gerbes over $\,M\,$ with
curvature $\,H=0$.\ Indeed, while $\,D(\d_Sb)_x=(0,0,0,1)\,$ holds
true in consequence of $\,x^*H=H$,\ that $\,(\d_Sb)_x\,$ is a
$D$-coboundary is {\em guaranteed} only if $\,\mathbb{H}^2(M)\,$
consists of just one element. The latter cohomology group satisfies
$\,\mathbb{H}^2(M) \cong H^2(M,U(1))$,\ see \cite{Gajer:1996} and
also \cite[sect.\ 2.2]{Johnson:2003}, and so it trivialises for
$\,M=\Gx\,$ a compact simple connected and simply connected Lie
group.

\subsection{Holonomy for world-sheets with circular defect
lines}\label{sec:circle-hol}

We now turn to the holonomy formula for world-sheets with circular
defect lines, but without defect junctions. This is the situation
treated in \cite{Fuchs:2007fw,Waldorf:2007phd}. As in section
\ref{sec:hol-no-defect}, we are given a manifold $\,M\,$ with gerbe
$\,\Gc$,\ and, in addition, we now have a $\Gc$-bi-brane $\,\Bc\,$
with local data as in \eqref{eq:bi-brane-tuple}.

While the formulation of the holonomy itself does not require any
further structure, the conditions to be imposed on the sigma-model
fields at the defect do. The extra structure is a metric $\,G\,$ on
the target space $\,M\,$ and a metric $\,\g\,$ on the world-sheet
$\,\Si$,\ both of which are part of the sigma-model data, entering
explicitly the kinetic term in the action functional
\eqref{eq:sigma-action}.

By a {\em circle-field configuration} on the world-sheet $\,\Si\,$
we mean a pair $\,(\La,X)$,\ where $\,\La\,$ is an oriented
one-dimensional submanifold of $\,\Si\,$ with empty boundary, i.e.\
a collection of oriented circles in $\,\Si$,\ and
\be
  X\ :\ \Si \rightarrow M \sqcup Q
\ee is a map from the world-sheet into the disjoint union of the
target space $\,M\,$ and the $\Gc$-bi-brane world-volume $\,Q$,\
with the following properties:
\begin{list}{-}{\topsep .4em \leftmargin 2.5em \itemsep 0em}
\item[(L1)] $X\,$ maps $\,\Si-\La\,$ to $\,M\,$ and is once differentiable
on $\,\Si-\La$.\ Furthermore, it maps $\,\La\,$ to $\,Q\,$ and is
once differentiable on $\,\La$.
\item[(L2)] Let $\,p \in \La\,$ and let $\,U\,$ be a small neighbourhood of
$\,p$.\ As $\,\Si\,$ and $\,\La\,$ are oriented, $\,\La\,$ splits
$\,U\,$ into two open sets $\,U_1\,$ and $\,U_2$.\ For example, if
$\,U\,$ is the open unit disc in $\,\Rb^2\,$ and $\,\La\,$ is the
real line, both with the standard orientation, then $\,U_1\,$ is the
upper open half-disc and $\,U_2\,$ the lower open half-disc. We
demand that, for $\,\a=1,2$,\ the restriction $\,X\vert_{U_\a}\,$
has a differentiable extension $\,X_{|\a} : \ovl U_\a \rightarrow
M\,$ to the closure $\,\ovl U_\a\,$ of $\,U_\a$,\ such that
$\,X_{|\a}(p) = \iota_\a( X(p) )$.
\item[(L3)] With $\,p\,,\,U_\a\,$ and $\,X_{|\a}\,$ as in (L2), let
$\,\widehat t \in T_p \Si\,$ be the unit vector tangent to $\,\La\,$
in the direction given by the orientation of $\,\La$,\ and let
$\,\widehat n_\a \in T_p \Si\,,\ \a = 1,2\,$ be the unit vectors
normal to $\,\La\,$ and pointing, each, to the side of $\,U_\a\,$
(so that, in particular, $\,\widehat n_1 = - \widehat n_2$). Then,
for all $\,v \in T_{X(p)} Q$,\ we require that the constraint
\be
G_{X_{|1}(p)}(\iota_{1*}v,X_{|1*}\widehat n_1)+
G_{X_{|2}(p)}(\iota_{2*}v,X_{|2*}\widehat n_2) - \tfrac{\sfi}{2} \,
\om_{X(p)}(v,X_*\widehat t)= 0
\labl{eq:defect-cross} be satisfied by the tangent (pushforward)
maps $\,X_{|\a*} :  T \ovl U_\a \rightarrow TM$.\ A derivation of
the above defect constraint is presented in appendix
\ref{app:def-glue}.
\end{list}

Conditions (L2) and (L3) merit some comment. If $\,Q\,$ is a
submanifold of $\,M \ti M\,$ then (L2) just means that the maps to
the left and to the right of $\,\La\,$ have, each, a differentiable
extension to $\,\La$,\ and that $\,\La\,$ gets mapped to $\,Q
\subset M \ti M\,$ under these two extensions. Condition (L2) is a
straightforward generalisation of this requirement to the case when
$\,Q\,$ is not necessarily a submanifold. Condition (L3) is new; it
forms part of the dynamical data of the sigma model, and is
therefore not needed in \cite{Fuchs:2007fw}; it will play an
important r\^ole in the discussion of topological defect lines in
section \ref{sec:conf-top-def} below. It constrains the variation of
the derivatives of the embedding map $\,X\,$ across the defect in a
manner dictated by the principle of least action applied to the
sigma-model action functional.
\medskip

For instance, if $\,\Bc = \Bc_\text{triv}\,$ is the trivial defect
\eqref{eq:triv-bibrane} then condition (L2) enforces the equality
$\,X_{|1}(p) = X(p) = X_{|2}(p)\,$ for any $\,p \in \La$.\ Thus,
$\,X\,$ is a continuous map $\,\Si \rightarrow M$,\ which -- so far
-- is only required to be {\em differentiable} on $\,\Si - \La\,$
and along $\,\La$.\ Condition \eqref{eq:defect-cross} now reads
$\,G_{X(p)}(v,X_{|1*}\widehat n_1) + G_{X(p)}(v,X_{|2*}\widehat n_2)
= 0\,$ for all $\,v \in T_{X(p)}M$.\ If we remember that
$\,X_{|1}\,$ is the differentiable extension of $\,X\,$ to the left
of $\,\La$,\ that $\,X_{|2}\,$ is the differentiable extension of
$\,X\,$ to the right of $\,\La$,\ that $\,\widehat n_1 = - \widehat
n_2$,\ and that $\,G\,$ is a non-degenerate pairing, we see that
\eqref{eq:defect-cross} forces the normal derivative to be
continuous across $\,\La$.

As another example, consider the case in which $\,\Bc =
\Bc_\text{bnd}\,$ describes a D-brane as in \eqref{eq:D-bibrane}.
Condition (L2) forces the neighbourhood of $\,\La\,$ in $\,\Si\,$ to
its right to be mapped to $\,\{ \bullet \} \subset M$,\ while the
neighbourhood of $\,\La\,$ in $\,\Si\,$ to its left is mapped to
$\,N\,$ such that the extension reads $\,X_{|1} = X\,$ on $\,\La$.\
Since $\,X_{|2}\,$ is constant, $\,X_{|2*}=0$,\ and so
\eqref{eq:defect-cross} becomes $\,2\,G_{X(p)}(v,X_*\widehat n_1)-
\sfi\,\om_{X(p)}(v,X_*\widehat t)=0\,$ for all $\,v\in T_{X(p)}D$.\
When written in terms of local coordinates $\,X^\mu\,$ for $\,M$,\
with $\,v = \delta X^\mu \,\partial_\mu\,$ tangent to $\,D$,\ this
yields
\be
\delta X^\mu \, \bigl( G_{\mu\nu}(X) \, \partial_n X^\nu - \sfi \,
\om_{\mu\nu}(X)\,\partial_t X^\nu \bigr)\big\vert_\La = 0\,,
\ee
where $\,\p_t\,$ is the tangent derivative at the boundary $\,\La\,$
of $\,\Si-X^{-1}(\{\bullet\})$,\
$\,\p_n=\tfrac{\g(\p_t,\p_a)}{\sqrt{\det\g}}\,\ep^{ab}\,\p_a\,$ is
the (inward-)normal one, and $\,\om=\om_{\mu\nu}\,\sfd
X^\mu\wedge\sfd X^\nu\,$ with $\,\om_{\mu\nu}\,$ antisymmetric in
its indices. The above are just the standard mixed
Dirichlet--Neumann boundary conditions for a world-sheet with
boundary $\,\La$,\ where the boundary gets mapped to the D-brane
world-volume $\,D\,$ endowed with the global twisted-gauge invariant
2-form $\,\om=F_i+B_i\,$ ($F_i\,$ being the `curvature' of the
gerbe-twisted gauge field),
see \cite[eqn.\,(3.3)]{Callan:1986bc}.
\medskip

One way to think of a configuration $\,(\La,X)\,$ is that it
describes a string moving in a possibly disconnected target space,
where the string is allowed to `tunnel' from one component into
another by passing through $\,Q$.

From a category-theory perspective, $\,M\,$ could be viewed as the
set of objects and $\,Q\,$ as the set of arrows, with $\,\iota_1\,$
and $\,\iota_2\,$ designating the source and the target of a given
arrow, respectively. Unfortunately, we only know how to formulate
composition in very special cases, so the analogy stops here.
\medskip

Note that, in our formulation, the entire $\,\La\,$ gets mapped to
the {\em same} $\Gc$-bi-brane manifold $\,Q$.\ In this sense, every
defect circle, i.e.\ every connected component of $\,\La$,\ carries
the {\em same} defect condition. This may seem a restriction, but it
is really just a convenient way to absorb the possibility of having
different $\Gc$-bi-branes for different defect circles into the map
$\,X$.

More specifically, note, first of all, that $\,\La\,$ divides the
world-sheet $\,\Si\,$ into connected components, which we shall call
{\em patches}. The situation in which different patches get mapped
to different target spaces $\,M_1,M_2, \dots\,$ is accommodated by
taking $\,M = M_1 \sqcup M_2 \sqcup \dots$.\ Since $\,X\,$ maps
$\,\Si-\La\,$ to $\,M$,\ each patch will sit entirely in one of the
components $\,M_k\,$ by the continuity of $\,X$.\ Next, suppose that
we have several $\Gc$-bi-branes $\,\Bc_1, \Bc_2, \dots\,$ on $\,M$,\
and that we want to label each of the defect circles by one of the
$\,\Bc_k$.\ This is accounted for by setting $\,\Bc = \Bc_1 \sqcup
\Bc_2 \sqcup \dots$,\ and choosing the map $\,X|_\La\,$ accordingly.
Since $\,X|_\La\,$ is continuous, the image of each defect circle
has to lie in one of the components $\,\Bc_k$.

\medskip

The holonomy for a circle-field configuration $\,(\La,X)\,$ is a
modification of \eqref{eq:hol-no-defect} which includes an
additional term associated to $\,\La$, \be
  \Hol(\La,X) = \Hol_\Gc(X) \cdot \Hol_\Bc(X|_\La)\,,
\labl{eq:hol-defect-circle-1}
where
\be
  \Hol_\Bc(X|_\La) =
  \prod_{e \in \triangle(\La)}\, \Bigg[ \exp\Big(i \int_e \widehat
  P_e \Big)\, \prod_{v \in e}\, \widehat K_{ev}(v) \Bigg]\,.
\labl{eq:hol-defect-circle-2} Above, $\,\Hol_\Gc(X)\,$ is
given by the same expression\footnote{We emphasise that expression
\eqref{eq:hol-no-defect} loses its fundamental property -- namely,
the invariance under changes of the world-sheet triangulation and
gauge transformations -- in the presence of a defect network, and
hence it now defines a collection of transport operators for the
transgression bundle of \cite{Gawedzki:1987ak} instead of surface
holonomy. It is with this understanding that we choose to denote it
by the same symbol $\,\Hol_\Gc(X)\,$ for the sake of brevity.
Analogous remarks apply to the defect-vertex corrections to the
holonomy.} \eqref{eq:hol-no-defect}, together with a prescription as
to how to deal with the jumps of $\,X\,$ across $\,\La$,\ which we
give shortly. The novel term $\,\Hol_\Bc(X|_\La)\,$ can be
understood as the holonomy of a gerbe-twisted line bundle over
$\,\La$,\ see \cite{Fuchs:2007fw,Waldorf:2007phd}, and also
\cite{Carey:2002,Gawedzki:2002se,Gawedzki:2004tu} for the
corresponding observation for boundary circles instead of defect
circles. In more detail, $\,\Hol_\Gc(X)\,$ and
$\,\Hol_\Bc(X|_\La)\,$ are defined as follows:
\begin{itemize}
\item Let $\,\triangle(\Si)\,$ be a triangulation subordinate to
$\,(\La,X)\,$ in the following sense. For each triangle $\,t \in
\triangle(\Si)$,\ there must exist an index $\,i_t \in \Ic^M\,$ such
that the interior of $\,t\,$ gets mapped to $\,\Oc^M_{i_t}$.\ We
require that $\,\La\,$ be covered by edges of $\,\triangle(\Si)$,\
and we denote by $\,\triangle(\La)\,$ the resulting 1-dimensional
triangulation of $\,\La$.\ For each edge $\,e \in \triangle(\La)$,\
there must be an index $\,i_e \in \Ic^Q\,$ such that $\,X\,$ maps
$\,e\,$ to $\,\Oc^Q_{i_e}$.\ It is understood that the assignments
$\,t \mapsto i_t\,$ and $\,e \mapsto i_e\,$ are made once and for
all.
\item For each edge $\,e\,$ and each vertex $\,v\,$ of
$\,\triangle(\Si)\,$ which do not lie on $\,\La$,\ we fix indices
$\,i_e \in \Ic^M\,$ and $\,i_v \in \Ic^M\,$ such that $\,X\,$ maps
$\,e\,$ to $\,\Oc^M_{i_e}\,$ and $\,v\,$ to $\,\Oc^M_{i_v}$.\ For
vertices $\,v\,$ that lie on $\,\La$,\ we pick an assignment $\,v
\mapsto i_v \in \Ic^Q$.\ As in section \ref{sec:hol-no-defect},
these maps are guaranteed to exist by the continuity properties of
$\,X$.
\item In the expression \eqref{eq:hol-no-defect} for
$\,\Hol_\Gc(X)$,\ we still have $\,\widehat B_t = X^* B_{i_t}\,,\
\widehat A_{te} = X^* A_{i_t i_e}\,$ and $\,\widehat g_{tev} =
\bigl(X^* g_{i_t i_e i_v}\bigr)^{\eps_{tev}}$.\ If one of the edges
of $\,t\,$ lies in $\,\La\,$ then the pullbacks use the
differentiable extension of $\,X\,$ from the interior of $\,t\,$ to
all of $\,t\,$ (which exists by condition (L2) on $\,X$). If $\,e
\subset \La\,$ then $\,i_e \in \Ic^Q$,\ and it is understood that
$\,A_{i_t i_e}\,$ and $\,g_{i_t i_e i_v}\,$ stand for $\,A_{i_t
\phi_1(i_e)}\,$ and $\,g_{i_t \phi_1(i_e) \phi_1(i_v)}$,\
respectively, if the orientation of $\,e$,\ as induced from $\,t$,\
agrees with that of $\,\La$,\ or for $\,A_{i_t \phi_2(i_e)}\,$ and
$\,g_{i_t \phi_2(i_e) \phi_2(i_v)}\,$ otherwise, cf.\ figure
\ref{fig:triangle-attach}.
\item $\widehat P_e = X^* P_{i_e}$,\ where $\,X\,$ is understood as
a map from $\,e \subset \La\,$ to $\,\Oc^Q_{i_e}$.\ The resulting
1-form on $\,e\,$ is integrated using the orientation of $\,\La$.
\item $\widehat K_{ev} = \bigl(X^*K_{i_e i_v}\bigr)^{-\eps_{ev}}\,$
(note the minus sign), where $\,X\,$ is understood as a map from
$\,v \in \La\,$ to $\,\Oc^Q_{i_e i_v}$.\ The edge $\,e\,$ inherits
an orientation from $\,\La$.\ The sign convention reads:
$\,\eps_{ev}=+1\,$ if, with respect to this orientation, $\,v\,$
sits at the end of $\,e$,\ and $\,\eps_{ev}=-1\,$ otherwise.
\end{itemize}
Again, the expression for the holonomy $\,\Hol(\La,X)\,$ is
basically dictated by requiring invariance with respect to the
choice of the triangulation, and with respect to gauge
transformations of the gerbe. This is discussed at length in section
\ref{sec:just}.

\begin{figure}[bt]

$$
  \raisebox{-50pt}{\begin{picture}(185,110)
  \put(0,11){\scalebox{0.60}{\includegraphics{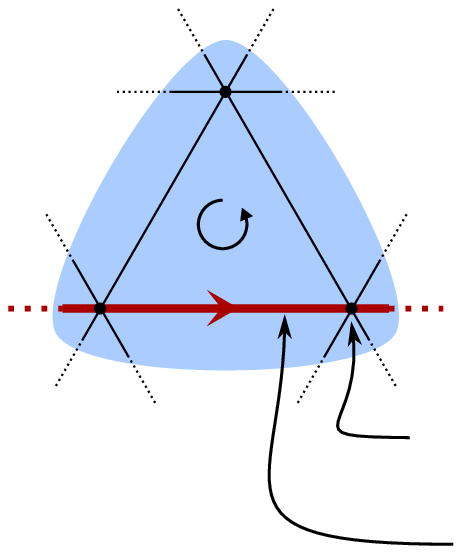}}}
  \put(0,11){
     \setlength{\unitlength}{.60pt}\put(-17,-35){
     \put( 64,120)   {\scriptsize $t$ }
     \put( 62, 94)   {\scriptsize $e$ }
     \put(128,108)   {\scriptsize $\La$ }
     \put(133, 63)   { $\displaystyle
       \big( g_{i_t \phi_1(i_e) \phi_1(i_v)} \circ X_{|1}
       \big)^{\eps_{tev}}$ }
     \put(142, 30)   { $\displaystyle X_{|1}^* A_{i_t \phi_1(i_e)}$ }
     }\setlength{\unitlength}{1pt}}
  \end{picture}}
  \hspace*{1em} \text{vs.} \hspace*{2em}
  \raisebox{-50pt}{\begin{picture}(185,110)
  \put(0,0){\scalebox{0.60}{\includegraphics{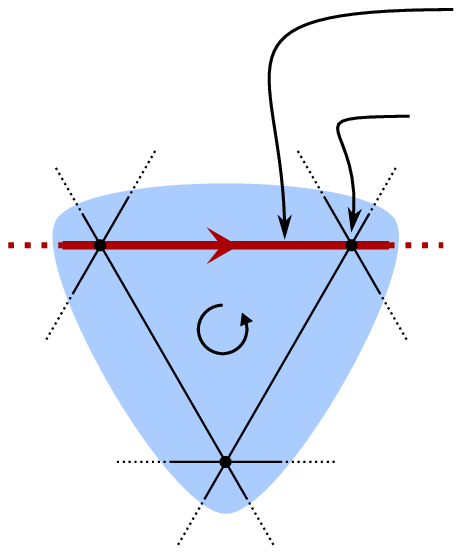}}}
  \put(0,0){
     \setlength{\unitlength}{.60pt}\put(-17,-17){
     \put( 64, 84)   {\scriptsize $t$ }
     \put( 62,109)   {\scriptsize $e$ }
     \put(128,108)   {\scriptsize $\La$ }
     \put(132,134)   { $\displaystyle
       \big( g_{i_t \phi_2(i_e) \phi_2(i_v)} \circ X_{|2}
       \big)^{\eps_{tev}}$ }
     \put(144,166)   { $\displaystyle X_{|2}^* A_{i_t \phi_2(i_e)}$ }
     }\setlength{\unitlength}{1pt}}
  \end{picture}}
$$

\caption{When a triangle $\,t\,$ shares an edge $\,e\,$ with a
defect line, the orientation of the defect either agrees with that
of $\,\partial t\,$ or not. This decides which pullback map to apply
to the connection 1-forms $\,A_{ij}\,$ and the transition functions
$\,g_{ijk}\,$ on $\,M$.} \label{fig:triangle-attach}
\end{figure}
\medskip

For the trivial $\Gc$-bi-brane $\,\Bc = \Bc_\text{triv}$,\ it is
easy to check that $\,\Hol_\Bc(X|_\La) = 1\,$ so that the holonomy
in the presence of defects \eqref{eq:hol-defect-circle-1} reduces to
the holonomy in the absence of defects \eqref{eq:hol-no-defect}.

In the case when the bi-brane describes a D-brane $\,\Bc =
\Bc_\text{bnd}$,\ one can verify that \eqref{eq:hol-defect-circle-1}
reproduces the holonomy for world-sheets with boundary given in
\cite{Carey:2002,Gawedzki:2002se,Gawedzki:2004tu}. The world-sheet
with boundary is obtained as follows: Given a circle-field
configuration $\,(\La,X)\,$ on $\,\Si\,$ for the target $\,M = N
\sqcup \{ \bullet \}\,$ with bi-brane $\,\Bc_\text{bnd}$,\ some
parts of the world-sheet will be mapped to $\,\{ \bullet \}$.\ On
these components of $\,\Si$,\ there are no degrees of freedom, and
so both the kinetic and the topological term of the action vanish.
Hence, we may as well remove these parts of the world-sheet, which
yields a new world-sheet $\,\Si'\,$ with $\,\La\,$ as its boundary.

\subsubsection*{The Lie-group example (cont'd)}

Let us inspect what a circle-field configuration $\,(\La,X)\,$ for
the $\Gc^{\star\sfk}$-bi-brane $\,\Bc_{Z(\Gx)}\,$ with world-volume
\eqref{eq:lie-ex-B} looks like. Condition (L2) means that whenever
$\,p \in \La\,$ gets mapped to $\,X(p) = (g,z) \in G \ti Z(\Gx)\,$
then $\,X_{|1}(p) = \iota_1(g,z) = g\,$ and $\,X_{|2}(p) =
\iota_2(g,z) = z^{-1}\cdot g$.\ Thus,
\be
  \lim_{\zeta \searrow p} X(\zeta) ~=~
  z \cdot \lim_{\zeta \nearrow p} X(\zeta)\,,
\ee
where by $\,\zeta \searrow p\,$ and $\,\zeta \nearrow p\,$ we mean
that $\,\zeta\in\Si\,$ approaches $\,p\,$ in the neighbourhood
$\,U_1\,$ and $\,U_2$,\ respectively, so that, e.g., $\,\lim_{\zeta
\searrow p} X(\zeta) = X_{|1}(p)$.\ Condition (L3) now reads
$\,G_g(v,X_{|1*}\widehat n_1) + G_{z^{-1}\cdot g}(z^{-1}_*
v,X_{|2*}\widehat n_2) = 0\,$ for all $\,v \in T_g\Gx$.\ Since the
(Cartan--Killing) metric on $\,\Gx\,$ is $\Gx$-invariant, this
implies $\,X_{|1*} \widehat n_1 + (z\cdot X_{|2})_*\widehat n_2 =
0$.\ Together with $\,X_{|1*} \widehat t = (z\cdot X_{|2})_*\widehat
t$,\ which follows from the identity $\,X_{|1}(p) = z\cdot
X_{|2}(p)\,$ valid for all $\,p \in \La$,\ this yields an equality
of the tangent maps
\be\label{eq:tang-ext-glue-Lie}
X_{|1*}  = (z\cdot X_{|2})_*\ :\ T_p\Si \longrightarrow T_g\Gx\,.
\ee
In other words, also the first derivative of the field $\,X\,$ has
to jump in a controlled manner across $\,\La$.\ Altogether, we see
that this $\Gc^{\star\sfk}$-bi-brane describes jump defects with the
value of the jump dictated by the value of the field on the defect
circle.

\subsection{Abelian inter-bi-branes}\label{sec:inter-bi-branes}

The introduction of a target-space structure for defect junctions on
the world-sheet calls for the notion of a 2-morphism, as introduced
in section \ref{sec:gerbes-local-data}, as well as for that of the
dual of a stable isomorphism and the dual of a $\Gc$-bi-brane, which
we now define.

Suppose that we are given a target space $\,M\,$ with gerbes
$\,\Gc\,$ and $\,\Hc$,\ and a stable isomorphism $\,\Phi =
(\Pi_i,\chi_{ij}) : \Gc \rightarrow \Hc$.\ We define the {\em dual
stable isomorphism} $\,\Phi^\vee : \Hc \rightarrow \Gc\,$ by the
local data $\,(-\Pi_i, \chi_{ij}^{-1})$.\ We also introduce
2-morphisms (the {\em death 2-morphisms})
\be
d_\Phi\ :\ \Phi^\vee \circ \Phi \Longrightarrow \id_\Gc\,,
\labl{eq:death-def}
with local data $\,d_\Phi = (1)$.\ One may wonder why ever we should
give a special name to a 2-morphism with trivial data. The reason is
that we have made a specific choice for the dual morphisms here;
other choices, differing by 2-isomorphisms (gauge transformations),
are possible which would lead to death 2-morphisms with non-trivial
data. Furthermore, a generic $\,d_\Phi\,$ cannot be avoided in the
framework of bundle gerbes \cite{Waldorf:2007phd}.

Let $\,\Bc\,$ be a $\Gc$-bi-brane of the form
\eqref{eq:bi-brane-tuple}. The $\Gc$-bi-brane {\em dual} to
$\,\Bc\,$ is defined as
\be
\Bc^\vee = \bigl(Q, -\om, \Oc^Q, \check\iota_2, \check\iota_1,
\Phi^\vee\bigr)\,.
\ee
Below, we shall often use the convenient notation $\,\Bc^+ \equiv
\Bc\,$ and $\,\Bc^- \equiv \Bc^\vee$,\ and we shall refer to the
data in $\,\Bc^\pm\,$ by a superscript $(~)^\pm$.\ One can check
that the holonomy \eqref{eq:hol-defect-circle-1} does not change if
we simultaneously reverse the orientation of $\,\La\,$ and replace
$\,\Bc\,$ by $\,\Bc^\vee$.
\medskip

A {\em $(\Gc,\Bc)$-inter-bi-brane} $\,\Jc\,$ for a gerbe $\,\Gc\,$
over a manifold $\,M\,$ and a $\Gc$-bi-brane $\Bc$ is an infinite
tuple
\be
\Jc = (T_n,\Oc^{T_n},\varphi_n,\t_n\ |\ n\in\Zb_{> 0})\,,
\labl{eq:inter-bi-data} where, for every $\,n\in\Zb_{> 0}$,
\begin{list}{-}{\topsep .4em \leftmargin 2.5em \itemsep 0em}
\item[(I.i)] $T_n\,$ is a smooth manifold;
\item[(I.ii)] $\Oc^{T_n} = \{\ \Oc^{T_n}_i\ |\ i \In \Ic^{T_n}\ \}\,$ is a
good open cover of $\,T_n$;
\item[(I.iii)] $\varphi_n\,$ is a 2-morphism;
\item[(I.iv)]
  $\t_n = \big(\,
  \eps_n^{k,k+1}\,,\, \check\pi_n^{k,k+1} \ \big|\ k=1,2,\ldots,n\, \big)\,$ are collections of maps.
\end{list}
The detailed description of the infinite sequence $\,(\varphi_n,
\t_n\ |\ n\in\Zb_{> 0})$,\ and of the conditions which the data
have to obey is somewhat lengthy but straightforward. First of all,
each $\,T_n\,$ carries the data needed to formulate the holonomy for
an $n$-fold junction of defect lines on the world-sheet. We allow
the possibility that $\,T_n\,$ is the empty set. For example, if one
wants to consider world-sheets with three-valent defect vertices
exclusively one could choose $\,T_n = \emptyset\,$ for $\,n\neq 3$.\
In what follows, we shall frequently refer to the manifold $\,T\,$
given by the disjoint union
\be T = \bigsqcup_{n=1}^\infty\,T_n\,,
\ee
which we call the world-volume of the $(\Gc,\Bc)$-inter-bi-brane.

The maps $\,\eps_n^{k,k+1}\,$ in the definition of $\,\t_n\,$ are
continuous functions
\be
\eps_n^{k,k+1}\ :\ T_n\rightarrow\{+1,-1\}\,,\qquad\qquad\eps_n^{n,
n+1}\equiv\eps_n^{n,1}\,.
\labl{eq:inter-bi-brane-epsdef}
They give a decomposition of $\,T_n\,$ into up to 2${}^n$
disconnected pieces (not all combinations of the $n$ signs
$\,\eps_n^{1,2},\eps_n^{2,3},\ldots,\eps_n^{n,n+1}\,$ need occur).
This decomposition of $\,T_n\,$ will be needed to accommodate the
different orientations with which $n$ edges can end on an $n$-valent
defect vertex in the world-sheet. The objects
$\,\check\pi_n^{k,k+1}\,$ are \v Cech-extended maps
\be
\check\pi_n^{k,k+1}=\bigl(\pi_n^{k,k+1},\psi_n^{k,k+1}\bigr)\ :\ T_n
\rightarrow Q\,,\qquad\qquad\check\pi_n^{n,n+1}\equiv\check\pi_n^{n,
1}\,,
\ee
composed of smooth manifold maps $\,\pi_n^{k,k+1}:  T_n\rightarrow
Q\,$ and the attendant index maps $\,\psi_n^{k,k+1}: \Ic^{T_n}
\rightarrow\Ic^Q$,\ subject to the condition
\be
\check\iota_2^{\eps_n^{k-1,k}}\circ\check\pi_n^{k-1,k}=\check
\iota_1^{\eps_n^{k,k+1}}\circ\check\pi_n^{k,k+1}\qquad\text{for}
\quad k=1,2,\ldots,n\,,
\labl{eq:inter-bi-brane-projB}
in which the identifications $\,\eps_n^{0,1}\equiv\eps_n^{n,1}\,$
and $\,\check\pi_n^{0,1}\equiv\check\pi_n^{n,1}\,$ are implicit.
Here, the manifold map from $\,T_n\,$ to $\,M\,$ appearing on the
left-hand side is given by $\,p\mapsto\iota_2^{\eps_n^{k-1,k}(p)}
\circ\pi_n^{k-1,k}(p)$,\ and similarly for the right-hand side.
Recall that the notation $\,\iota_1^\pm\,$ and $\,\iota_2^\pm\,$
refers to the maps from the definition of the $\Gc$-bi-brane
$\,\Bc^+=\Bc\,$ and its dual $\,\Bc^-=\Bc^\vee$.\ Put together with
(B.iv), condition \eqref{eq:inter-bi-brane-projB} enables us to
induce another family of \v Cech-extended maps
\be
\check\pi_n^k=\check\iota_1^{\eps_n^{k,k+1}}\circ\check\pi_n^{k,
k+1}=(\pi_n^k,\psi_n^k)\ :\ T_n \rightarrow M
\labl{eq:V-to-M-extend}
from $\,\check\pi_n^{k,k+1}$.\ Just as for $\Gc$-bi-branes, we shall
not -- for the sake of transparency -- spell out $\,\psi_n^k\,$ and
$\,\psi_n^{k,k+1}\,$ explicitly in formul\ae ~involving pullbacks
from $\,M\,$ or $\,Q\,$ to $\,T_n$.

Using $\,\check\pi_n^k\,$ and $\,\check\pi_n^{k,k+1}$,\ we can pull
back data to $\,T_n\,$ from $\,M\,$ and $\,Q$,\ respectively. Thus,
in particular, we obtain a family of 2-forms on $\,T_n$,
\be\label{eq:omn-pull}
\om^{k,k+1}_n=\bigl(\pi_n^{k,k+1}\bigr)^*\om^{\eps_n^{k,k+1}}\,.
\ee
We demand the sum of all these 2-forms to vanish for each $\,n\in
\Zb_{> 0}$,
\be
\sum_{k=1}^n\,\om^{k,k+1}_n=0\,.
\labl{eq:inter-alpha-cond}

In the light of the invariance arguments to be presented in section
\ref{sec:just}, we could -- more generally -- have postulated the
existence of 1-forms $\,\th_n\,$ on $\,T_n\,$ such that the above
sum is equal to $\,\sfd\th_n\,$ instead of being zero. However, the
analysis of the defect conditions for the fields of the underlying
sigma model, derived in appendix \ref{app:def-glue} through the
application of the variational principle, shows that only those
parts of $\,T_n\,$ can be probed by the sigma-model field in which
$\,\th_n\,$ vanishes. Therefore, we may as well set $\,\th_n=0\,$
for all $\,n\in\Zb_{> 0}\,$ from the start.
\smallskip

In addition to the above, we also obtain, on each $\,T_n$,\ a family
of gerbes and stable isomorphisms
\be
\Gc^k_n=\bigl(\check\pi_n^k\bigr)^*\Gc\,,\qquad\qquad\Phi^{k,k+1}_n
=\bigl(\check\pi_n^{k,k+1}\bigr)^*\Phi^{\eps_n^{k,k+ 1}}\,,
\labl{eq:pullback-G-to-V}
where $\,\Phi^{k,k+1}_n\,$ is readily verified to be a stable
isomorphism $\,\Gc^k_n\rightarrow\Gc^{k+1}_n\star I\bigl(\om^{k,k+
1}_n\bigr)$.\ Here, we have used the identification $\,\Gc_n^{n+1}
\equiv\Gc_n^1\,$ and abbreviated $\,I\bigl(\om^{k,k+1}_n\bigr)
\equiv I\bigl(\Oc^{T_n},\om^{k,k+1}_n\bigr)$,\ and we shall adhere
to these conventions below. Define the stable isomorphisms $\,\Xi_n:
\Gc^1_n\rightarrow\Gc^1_n\,$ on each component $\,T_n\,$ by the
cyclic composition of the stable isomorphisms $\,\Phi^{k,k+1}_n$
\bea
\Xi_n\ :\ \Gc^1_n\xrightarrow{\Phi^{1,2}_n}\Gc^2_n\star I\bigl(
\om^{1,2}_n\bigr)\xrightarrow{\Phi^{2,3}_n\star\id_{I(\om^{1,2}_n
)}}\Gc^3_n\star I\bigl(\om^{2,3}_n\bigr)\star I\bigl(\om^{1,2}_n
\bigr)\longrightarrow\cdots\enl\hspace*{2.5em}
\label{eq:cyclic-sequence}
\xrightarrow{\Phi^{n,1}_n\star\id_{I(\om^{n-1,n}_n)}\star
\ldots\star\id_{I(\om^{1,2}_n)}}\Gc^1_n\star
I\bigl(\om^{n,1}_n\bigr)\star I\bigl(\om^{n-1,n}_n\bigr)\star\cdots
\star I\bigl(\om^{1,2}_n\bigr)\equiv\Gc^1_n\,.
\eear\ee
The last identity follows from the composition rule $\,I(\om)\star
I(\om')=I(\om+\om')$,\ cf.\ section \ref{sec:gerbes-local-data}, and
the condition \eqref{eq:inter-alpha-cond}. We may finally introduce
the 2-morphism $\,\varphi_n\,$ as
\be
\varphi_n\ :\ \Xi_n\Longrightarrow\id_{\Gc^1_n}\,.
\labl{eq:inter-bi-2morph}
This completes the description of the data of an abelian
$(\Gc,\Bc)$-inter-bi-brane and of the conditions it has to satisfy.
\medskip

Upon rewriting \eqref{eq:inter-bi-2morph} in terms of the relevant
local data $\,\Phi^{k,k+1}_n=(P^{k,k+1}_{n,i}, K^{k,k+1}_{n,ij})\in
A^1_{T_n}\,$ and $\,\varphi_n=(f_{n,i})\in A^0_{T_n}\,$ on
$\,\Oc^{T_n}$,\ we obtain the relation
\be
\sum_{k=1}^n\,\bigl(P^{k,k+1}_{n,i},K^{k,k+1}_{n,ij}\bigr)+\bigl(-
\sfi\,\sfd\log f_{n,i},f_{n,i}\,f_{n,j}^{-1}\bigr)=(0,1)\,.
\labl{eq:inter-bi-2morph-loc}
In the sequel, we shall often employ the composite 2-morphism
$\,\varphi=(f_i)\in A^0_T\,$ on the total world-volume $\,T\,$ glued
from the 2-morphisms $\,\varphi_n\,$ as per
\qq\label{eq:2-morph-glued}
\varphi|_{T_n}=\varphi_n\,.
\qqq

The simplest example of an inter-bi-brane is the trivial
$(\Gc,\Bc)$-inter-bi-brane $\,\Jc_\text{triv}\,$ which is defined
for the trivial $\Gc$-bi-brane $\,\Bc=\Bc_\text{triv}$.\ In this
case, one takes $\,T_n\,$ in the form of 2${}^n$ copies of $\,M$,\
one for each possible set of values of the maps $\,\eps_n^{k,k+1}$.\
The projections $\,\check\pi_n^{k,k+1}\,$ all coincide with the
identity map on $\,M$.\ One then finds that $\,\Xi_n\,$ is the
identity stable isomorphism from $\,\Gc\,$ to itself, and one
chooses for $\,\varphi_n\,$ the identity 2-morphism with local data
$\,\varphi_n=(1)$.

\subsubsection*{The Lie-group example (cont'd)}

We shall now tailor down the exposition of the various bi-brane and
inter-bi-brane structures associated with $Z(\Gx)$-jump defects of
the WZW model for $\,\Gx\,$ to the main task at hand which consists
in obtaining the 3-cocycle of \eqref{eq:intro-hol-compare}. The
minimal set of inter-bi-brane data is provided by the two sets
\qq
  T_3=\Gx\ti Z(\Gx)\ti Z(\Gx)\ti\{\pm 1\}^3\,,\qquad\qquad T_4=\Gx
  \ti Z(\Gx)\ti Z(\Gx)\ti Z(\Gx)\ti\{\pm 1\}^4\,,
\qqq
describing three- and four-valent vertices, respectively. As the
notation clearly suggests, the sign maps
\eqref{eq:inter-bi-brane-epsdef} are fixed in the form
\be\begin{array}{l}
\eps_3^{k,k+1}(g,x,y,\eps_{1,2},\eps_{2,3},\eps_{3,1})=\eps_{k,k+1}\,,
\quad\quad\eps_4^{k,k+1}(g,x,y,z,\eps_{1,2},\eps_{2,3},\eps_{3,4},
\eps_{4,1})=\eps_{k,k+1}\,.
\end{array}
\ee
Since we shall only need for our purposes the distinguished
connected components $\,T_{2+1}\equiv T_{3,++-}\,$ and
$\,T_{3+1}\equiv T_{4,+++-}\,$ of $\,T_3\,$ and $\,T_4$,\ we fix the
signs as
\qq
\eps^{1,2}_3=+1=\eps^{2,3}_3\,,\qquad\eps^{3,1}_3=-1\,,\qquad\qquad
\eps^{1,2}_4=+1=\eps^{2,3}_4=\eps^{3,4}_4\,,\qquad\eps^{4,1}_4=-1
\qqq
for the reminder of the discussion. Below, we detail the remaining
elements of the description solely for $\,T_{2+1}$,\ postponing the
construction of $\,T_{3+1}\,$ to section \ref{sec:cluster}.

The good open cover of $\,T_{2+1}\,$ is obtained in the same way as
for the $\Gc^{\star\sfk}$-bi-brane $\,\Bc_{Z(\Gx)}\,$ of
\eqref{eq:lie-ex-B}, that is we choose the open sets $\,\Oc^{T_{2+
1}}_{i,x,y}=\Oc^\Gx_i\ti\{(x,y)\}\,$ with $\,i\in\Ic^\Gx\,$ and
$\,x,y\in Z(\Gx)\,$ (and the redundant signs dropped from the
notation, which is also what we do below). The \v Cech-extended maps
$\,\check\pi^{k,k+1}_{2+1}=\bigl(\pi^{k,k+1}_3,\psi^{k,k+1}_3\bigr)
\big\vert_{T_{2+1}}\,$ for the edges then evaluate on points
$\,(g,x,y)\in\Oc^{T_{2+1}}_{i,x,y}\,$ as
\be
\begin{array}{ll}
\pi_{2+1}^{1,2}(g,x,y)=(g,x)\,,\qquad\quad&\psi^{1,2}_{2+1}(i,x,y)=
(i,x)\,,\cr\cr \pi_{2+1}^{2,3}(g,x,y)=(x^{-1}\cdot
g,y)\,,\qquad\quad&\psi^{2,3}_{2+1}(i,x,y) =(x^{-1}.i,y)\,,\cr\cr
\pi_{2+1}^{3,1}(g,x,y)=(g,x\cdot y)\,,\qquad\quad&\psi^{3,1}_{2+1}
(i,x,y)=(i,x\cdot y)\,.
\end{array}\label{eq:lie-ex-inter-bi-proj}
\ee
These manifestly obey condition \eqref{eq:inter-bi-brane-projB}, for
example $\,\iota_2(\pi_{2+1}^{1,2}(g,x,y)) = x^{-1}\cdot g =
\iota_1(\pi_{2+1}^{2,3}(g,x,y))$.\ The corresponding \v
Cech-extended maps \eqref{eq:V-to-M-extend} for the patches are
\qq\label{eq:Lie-ex-proj-to-G}
\pi_{2+1}^1(g,x,y)=g\,,\qquad\quad\pi_{2+1}^2(g,x,y)=x^{-1}\cdot g
\,,\qquad\quad\pi_{2+1}^3(g,x,y)=(x\cdot y)^{-1}\cdot g\,,
\qqq
and similarly for $\,\psi^k_{2+1}$.

At this stage, we still have to fix the 2-morphisms $\,\varphi_n\,$
from the collection \eqref{eq:inter-bi-data}. We shall only describe
those supported by the subspace $\,T_{2+1}^{x,y}=\Gx\ti\{(x,y)\}
\subset T_{2+1}$,\ which we identify with $\,\Gx$.\ Using
\eqref{eq:Lie-ex-proj-to-G} we get the three pullback gerbes on
$\,T_{2+1}^{x,y}\,$ defined in \eqref{eq:pullback-G-to-V},
\qq
\Gc^1_{2+1}&=&\bigl(\check\pi^1_{2+1}\bigr)^*\Gc^{\star\sfk}=
\Gc^{\star\sfk}\,,\cr\cr
\Gc^2_{2+1}&=&\bigl(\check\pi^2_{2+1}\bigr)^*\Gc^{\star\sfk}=x.
\Gc^{\star\sfk}\,,\\\cr
\Gc^3_{2+1}&=&\bigl(\check\pi^3_{2+1}\bigr)^*\Gc^{\star\sfk}=(x
\cdot y).\Gc^{\star\sfk}\,,\nonumber
\qqq
as well as the pullback 1-morphisms
\qq
\begin{array}{lrl}
\Phi_{2+1}^{1,2}=\bigl(\check\pi^{1,2}_{2+1}\bigr)^*\Phi=& \Ac_x
&:\ \Gc^1_{2+1}\longrightarrow\Gc^2_{2+1}\,,\cr\cr
\Phi_{2+1}^{2,3}=\bigl(\check\pi^{2,3}_{2+1}\bigr)^*\Phi=& x.\Ac_y
&:\ \Gc^2_{2+1} \longrightarrow\Gc^3_{2+1}\,,\cr\cr
\Phi_{2+1}^{3,1}=\bigl(\check\pi^{3,1}_{2+1}\bigr)^*\Phi^\vee=&
\Ac_{x \cdot y}^\vee &:\ \Gc^3_{2+1}\longrightarrow\Gc^1_{2+1}\,.
\end{array}
\qqq
We next fix 2-morphisms
\qq
\til\varphi_{x,y}\ :\ (x.\Ac_y)\circ\Ac_x\Longrightarrow\Ac_{x\cdot
y}\,,
\qqq
where both sides are stable isomorphisms $\,\Gc^{\star\sfk}
\rightarrow(x\cdot y).\Gc^{\star\sfk}$.\ These 2-morphisms can be
read off from \cite[sect.\,3]{Gawedzki:2003pm} upon consulting
\cite[sect.\,1 \& 3]{Gawedzki:2008} whose conventions have been
adopted in our discussion. Finally, we define the 2-morphism
$\,\varphi_{2+
1}:\Phi^{3,1}_{2+1}\circ\Phi^{2,3}_{2+1}\circ\Phi^{1,2}_{2+1}
\Rightarrow\id_{\Gc^1_{2+1}}\,$ on $\,T_{2+1}^{x,y}\,$ as
\qq
\varphi_{2+1}\ :\ \Ac_{x\cdot y}^\vee\circ(x.\Ac_y)\circ\Ac_x
\xLongrightarrow{\id_{\Ac_{x\cdot y}^\vee}\circ\til\varphi_{x,y}}
\Ac_{x\cdot y}^\vee\circ\Ac_{x\cdot y}\xLongrightarrow{d_{\Ac_{x
\cdot y}}}\id_{\Gc^{\star\sfk}}\,.
\qqq
The composition of 2-morphisms represented by the superposition of
the corresponding double arrows is called `vertical' in the
2-categorial language and denoted with the symbol $\,\bullet$,\
e.g., $\,\varphi_{2+1}=d_{\Ac_{x\cdot
y}}\bullet\bigl(\id_{\Ac_{x\cdot y}^\vee}\circ\til\varphi_{x,y}
\bigr)$.\ We shall use the composition symbol in the reminder of the
paper in order to shorten some formul\ae.
\smallskip

The existence of the 2-morphisms $\,\til\varphi_{x,y}\,$ follows
form the triviality of the cohomology group $\,H^1(\Gx,{\rm
U}(1))$.\ In order to see this, let us look at a general symmetry
group $\,S\,$ again, as we did at the end of section
\ref{sec:gerbes-local-data}. Suppose that an element-wise
presentation $\,\Ac_S\,$ of $\,S\,$ on $\,b\,$ exists. In terms of
local data, the 2-morphisms $\,\til\varphi_{x,y} :
(x.\Ac_y)\circ\Ac_x\Longrightarrow\Ac_{x\cdot y}\,$ have to solve
$\,-D\til\varphi_{x,y} = (\d_S \Ac)_{x,y}\,$ for all $\,x,y\in S$,\
where $\,(\d_S \Ac)_{x,y}=x.\Ac_y-\Ac_{x\cdot y}+\Ac_x$.\ We shall
collect the 2-morphisms into a set $\,\til\varphi_S = \{\
\til\varphi_{x,y} \ |\ x,y \in S \ \}\,$ and call the pair
$\,(\Ac_S,\til\varphi_S)\,$ a {\em homomorphic presentation of}
$\,S\,$ {\em on} $\,b$.\ Assuming the existence of the element-wise
presentation $\,\Ac_S$,\ the obstruction to the existence of a
homomorphic presentation is contained in $\,\mathbb{H}^1(M) = {\rm
ker} D_{(1)} / {\rm im} D_{(0)}$,\ the set of isomorphism classes of
flat line bundles over $\,M$.\ Indeed, the equality
$\,D\d_S\Ac\equiv\d_SD \Ac=\d_S^2b=(0,0,1)\,$ always holds due to
$\,\d_S^2=0$,\ but the existence of $\,\tilde\varphi_S\,$ requires
$\,(\d_S \Ac)_{x,y}\,$ to lie in the image of $\,D_{(0)}\,$ for all
$\,x,y \in S$.\ The cohomology group $\,\mathbb{H}^1(M)\,$ satisfies
$\,\mathbb{H}^1(M) \cong H^1(M,U(1))$,\ see
\cite{Gajer:1996,Johnson:2003}, and so it trivialises for
$\,M=\Gx\,$ a compact simple connected and simply connected Lie
group.

\subsection{Holonomy for world-sheets with defect
networks}\label{sec:holo-net}

After all the preparations, we can, at last, describe our
construction of the holonomy for world-sheets with defect networks.

A {\em defect network} $\,\G\,$ on a world-sheet $\,\Si\,$ is an
oriented graph embedded in $\,\Si$,\ together with an ordering of
the edges around each vertex. By this we mean that the edges of
$\,\G\,$ are oriented submanifolds of $\,\Si$,\ and that, for each
vertex of $\,\G$,\ the edges emanating from this vertex have been
labelled in the counter-clockwise order as
$\,e_{1,2},e_{2,3},\dots,e_{n,1}$.\ (Since the world-sheet is
oriented, this is equivalent to marking one of the edges attached to
the vertex.) We allow, in particular, circular edges that are not
attached to any vertex. The set of edges in $\,\G\,$ is denoted by
$\,E_\G$,\ and the set of its vertices by $\,V_\G$.

A {\em network-field configuration} on $\,\Si\,$ for the target
space $\,M\,$ with the $\Gc$-bi-brane $\,\Bc\,$ and the
$(\Gc,\Bc)$-inter-bi-brane $\,\Jc\,$ is a pair $\,(\G,X)$,\ where
$\,\G\,$ is a defect network and \be
  X\ :\ \Si \rightarrow M \sqcup Q \sqcup T
\ee is a map from the world-sheet into the disjoint union of the
target space $\,M$,\ the $\Gc$-bi-brane world-volume $\,Q$,\ and the
$(\Gc,\Bc)$-inter-bi-brane world-volume $\,T$,\ with the following
properties:
\begin{list}{-}{\topsep .4em \leftmargin 2.5em \itemsep 0em}
\item[(N1)] $X\,$ restricts to a once differentiable map $\,\Si-\G
\rightarrow M$,\ and to a once differentiable map $\,\G-V_\G
\rightarrow Q$,\ and it maps $\,V_\G\,$ to $\,T$.\ Furthermore, we
have $\,X(v) \in T_{n_v}\,$ for a vertex $\,v \in V_\G\,$ of valence
$\,n_v$.\
\item[(N2)] In a neighbourhood of a point $\,p \in \G-V_\G$,\ the map
$\,X\,$ obeys conditions (L1)--(L3) for a circle-field configuration
from section \ref{sec:circle-hol}.
\item[(N3)] Let $\,v \in V_\G\,$ be an $n_v$-valent vertex and let
$\,e_{k,k+1}\,$ be an edge converging at $\,v$.\ If the edge is
oriented towards $\,v\,$ we demand that $\,\eps_{n_v}^{k,k+1}(X(v))
= +1$,\ and otherwise that $\,\eps_{n_v}^{k,k+1}(X(v)) = -1$.
\item[(N4)]
Let $\,v\,$ and $\,e_{k,k+1}\,$ be as in (N3). The map $\,X\,$ sends
$\,e_{k,k+1}\,$ with its endpoints removed to $\,Q$.\ We demand that
$\,X\,$ have a differentiable extension $\,X_{k,k+1} : e_{k,k+1}
\rightarrow Q$,\ and that $\,X_{k,k+1}(v) =
\pi_{n_v}^{k,k+1}(X(v))\,$ hold.
\end{list}

Condition (N3) ensures that a vertex gets mapped to the correct
component of $\,T_{n_v}\,$ according to the orientation of the edges
converging at $\,v$,\ and condition (N4) restricts the jump of
$\,X\,$ at the vertex itself. There are two implications of (N4)
that we wish to emphasise.

First, let $\,U \subset \Si\,$ be a small neighbourhood of a vertex
$\,v \in V_\G\,$ of valence $\,n_v$.\ The defect network $\,\G\,$
divides $\,U\,$ into $n_v$ open sets $\,U_1,U_2,\ldots,U_{n_v}$,\
labelled counter-clockwise around $\,v\,$ such that $\,U_k\,$ sits
between the edges $\,e_{k-1,k}\,$ and $\,e_{k,k+1}$.\ The map
$\,X\,$ sends $\,U_k\,$ to $\,M$.\ Condition (N2) implies that it
has a differentiable extension to $\,\ovl U_k - \{v\}$,\ and
condition (N4) ensures that, in fact, $\,X\,$ has a differentiable
extension $\,X_k : \ovl U_k \rightarrow M$,\ and that $\,X_k(v) =
\pi^k_{n_v}(X(v))$.

Second, if $\,\Bc = \Bc_\text{triv}\,$ is the trivial $\Gc$-bi-brane
and $\,\Jc = \Jc_\text{triv}\,$ is the trivial
$(\Gc,\Bc)$-inter-bi-brane then -- as we have already seen in
section \ref{sec:circle-hol} -- $\,X\,$ has a differentiable
extension to all of $\,\Si\,$ for $\,\G\,$ composed solely of
circles. By the same argument, one finds that, for a general defect
network, $\,X$ has a differentiable extension to $\,X-V_\G$.\
However, by the previous remark, it has a differentiable extension
to $\,\ovl U_k\,$ for each of the sectors $\,U_k\,$ around a vertex
$\,v$.\ Thus, it is differentiable on all of $\,\Si$.
\medskip

The holonomy for a network-field configuration $\,(\G,X)\,$ is a
modification of \eqref{eq:hol-defect-circle-2} which includes an
additional ${\rm U}(1)$-factor associated to the vertices of $\G$,
\be
  \Hol(\Gamma,X) = \Hol_\Gc(X) \cdot \Hol_\Bc(X|_{E_{\G}})
  \cdot \Hol_\Jc(X|_{V_{\G}})\,,
\labl{eq:hol-defect-network-1}
where
\be
  \Hol_\Jc(X|_{V_{\G}})
  = \prod_{v \in V_{\G}}\, \widehat f_v(v)\,.
\labl{eq:hol-defect-network-2} $\Hol_\Gc(X)$ is given by the
same expression \eqref{eq:hol-no-defect} and $\Hol_\Bc(X|_{E_{\G}})$
by the expression \eqref{eq:hol-defect-circle-2}, together with a
prescription as to how to treat the vertices of $\G$. Here are the
details:
\begin{itemize}
\item The expressions $\,\Hol_\Gc(X)\,$ and
$\,\Hol_\Bc(X|_{E_\G})\,$ are evaluated with respect to a
triangulation $\,\triangle(\Si)\,$ subordinate to $\,(\G,X)$.\ Such
a triangulation is defined in the same way as the triangulation
subordinate to $\,(\La,X)\,$ from section \ref{sec:circle-hol}, with
the additional requirement that $\,V_\G\,$ is a subset of the set of
the vertices of $\,\triangle(\Si)$,\ and that we have chosen, for
each vertex $\,v\in V_\G$,\ an index $\,i_v\in\Ic^{T_{n_v}}\,$ such
that $\,X(v)\in \Oc^{T_{n_v}}_{i_v}$,\ where $\,n_v\,$ is the
valency of $\,v$.
\item $\Hol_\Gc(X)\,$ is computed as described below
\eqref{eq:hol-defect-circle-1}, except when a vertex $\,v\,$ of a
triangle $\,t\,$ lies in $\,V_\G$.\ Suppose that $\,t\,$ lies
between the defect edges $\,e_{k-1,k}\,$ and $\,e_{k,k+1}$.\ Then,
in the ${\rm U}(1)$-factor $\,g_{i_t i_e i_v}$,\ the index $\,i_v\,$
stands for $\,\psi^k_{n_v}(i_v)$.\ If $\,e\,$ is an edge of $\,\G\,$
then $\,i_e\,$ stands for $\,\phi_1(i_e)\,$ or $\,\phi_2(i_e)$,\
depending on the relative orientation of $\,e\,$ and $\,\partial
t$,\ as explained below \eqref{eq:hol-defect-circle-1}.
\item $\Hol_\Bc(X|_{E_\G})\,$ is computed as described below
\eqref{eq:hol-defect-circle-1}, except when a vertex $\,v\,$ of an
edge $\,e\,$ lies in $\,V_\G$.\ Suppose that the edge $\,e\,$ is the
edge $\,e_{k,k+1}\,$ for the vertex $\,v$.\ Then
\be
\widehat K_{ev} = \bigl(X_{k,k+1}^*K_{i_e\psi^{k,k+1}_{n_v}(i_v)}
\bigr)^{-\eps_{ev}}\,,
\ee where the sign $\,\eps_{ev}\,$ is as detailed below
\eqref{eq:hol-defect-circle-1}. The definition of $\,\widehat P_e =
X^* P_{i_e}\,$ is not affected.
\item Finally, $\,\widehat f_v = X^*f_{i_v}$,\ with
$\,f_{i_v}\vert_{X(v)}=f_{n_v,i_v}\,$ at an $n_v$-valent vertex
$\,v$.
\end{itemize}
We shall discuss in the next section how $\,\Hol(\G,X)\,$ is
determined from the requirement of its independence of the diverse
choices made.

\subsection{Holonomy formul\ae ~from invariance analysis -- a
derivation}\label{sec:just}

In the previous sections, we introduced a host of target-space
structures associated with the gerbe, and used them to postulate the
sigma-model action functional in the presence of defect networks
embedded in the world-sheet. At this stage, we could perform an a
posteriori verification of the invariance of the holonomy formul\ae
~thus obtained under allowed changes of the arbitrary choices made:
the choice of representatives of local data of the gerbe, those of
the stable isomorphisms and 2-morphisms, as well as of the \v Cech
cover of the target space and of the world-sheet triangulation
subordinate to it. This was the route taken in
\cite{Alvarez:1984es,Gawedzki:1987ak} for world-sheets without
defects, in \cite{Gawedzki:2002se,Gawedzki:2004tu} for world-sheet
boundaries, and in \cite{Fuchs:2007fw,Waldorf:2007phd} for circular
defects, and it could readily be adapted to the study of defect
junctions. However, this would leave us with the question as to how
canonical our choices for the specific target-space structures --
that of a bi-brane and that of an inter-bi-brane -- are. Therefore,
we choose to take essentially the reverse route in the present
section in which we successively \emph{derive} all components of the
postulated description from some elementary invariance
considerations. In so doing, we reveal certain twisted gauge
symmetries associated intrinsically with $\Gc$-bi-branes and
$(\Gc,\Bc)$-inter-bi-branes.
\medskip

Let us first look for a modification of the bulk holonomy formula
\eqref{eq:hol-no-defect} necessary to accommodate the embedding of a
collection $\,\La\,$ of non-intersecting defect circles in the
world-sheet $\,\Si$.\ To this end, we compare the value
$\,\Hol_\Gc^p(X)\,$ of the bulk holonomy attained on the
gauge-transformed local data
\qq\label{eq:gerbe-gauge-trans}
b^p=b+Dp\,,\qquad\qquad p=(\Pi_i,\chi_{ij})\in A_M^1
\qqq
of the bulk gerbe with that obtained for the original data
$\,b=(B_i,A_{ij},g_{ijk})$,
\qq
\Hol_\Gc^p(X)=\Hol_\Gc(X)\cdot\prod_{e\in\triangle(\La)}\,\biggl[
\exp \biggl(\sfi\,\int_e\,\bigl(\widehat\Pi_{1,e}-\widehat\Pi_{2,e}
\bigr)\biggr)\,\prod_{v\in e}\,\bigl(\widehat\chi_{1,ev}\bigr)^{-1}
\cdot\widehat\chi_{2,ev}(v)\biggr]\,,
\qqq
where, in the conventions of section \ref{sec:circle-hol},
\begin{itemize}
\item the triangulation $\,\triangle(\La)\,$ is
induced by $\,\triangle(\Si)$;
\item $\,\widehat\Pi_{\a,e}=X^*_{|\a}\Pi_{\phi_\a(i_e)}\,,\ \a=1,2$,\
with the extensions $\,X_{|\a}\,$ understood as maps from $\,e
\subset\La\,$ to $\,\Oc^M_{\phi_\a(i_e)}$;
\item $\,\widehat\chi_{\a,ev}=X^*_{|\a}\chi_{\phi_\a(i_e)\phi_\a(i_v)}
\,,\ \a=1,2$,\ with $\,X_{|\a}\,$ understood as maps from $\,v\in
\La\,$ to $\,\Oc^M_{\phi_\a(i_e)\phi_\a(i_v)}$.
\end{itemize}
Thus, the variation is pushed to the defect $\La\,$ -- the
(gerbe-)gauge symmetry remains unaffected by the presence of the
defect away from it, and -- accordingly -- we should seek a
cancellation of the defect variation through the introduction of
degrees of freedom \emph{localised at the defect}, with
transformation properties dictated by the gauge transformations of
the pullback gerbes on both patches welded by a particular defect
circle. The defect being one-dimensional, we are led to take as the
local data for the defect fields a \v Cech--Deligne cochain
$\,\Phi=(P_i,K_{ij})\in A_Q^1\,$ coupled to the defect as in the
expression $\,\Hol_\Bc(X|_\La)\,$ of \eqref{eq:hol-defect-circle-2}
and transforming as
\qq\label{eq:twisted-gauge-field-transfo}
(P_i,K_{ij})\mapsto(P_i,K_{ij})+\check\iota_2^*\bigl(\Pi_i,\chi_{i
j}\bigr)-\check\iota_1^*\bigl(\Pi_i,\chi_{ij}\bigr)-D( W_i)\,.
\qqq
Here, the second and third term on the right-hand side describe a
twist induced by the bulk transformation $\,p$,\ and the last one,
written in terms of a cochain $\,\eta=(W_i)\in A_Q^0$,\ is an
independent gauge transformation of $\,\Phi\,$ allowed due to the
emptiness of the boundary of $\,\La$.\ The overall transformation
displayed is that of a $\Gc$-(bi-)twisted gauge field over $\,Q$.

Having ensured the invariance of the corrected holonomy formula
$\,\Hol(\La,X)\,$ of \eqref{eq:hol-defect-circle-1} under gauge
transformations of the bulk data, we should now demand that it be
invariant under arbitrary changes of the $(\La,X)$-subordinate
triangulation of $\,\Si$,\ which turns out to constrain the defect
data. The defining relation \eqref{eq:def-gerbe} of $\,\Gc\,$
protects the invariance of $\,\Hol(\La,X)\,$ under all changes which
do not affect the edges and the vertices of $\,\triangle(\Si)\,$
lying within the defect $\,\La$,\ and so the remaining freedom of
man\oe uvre consists in shifting the vertices of
$\,\triangle(\La)\,$ (with the bulk edges converging at them moved
accordingly). In what follows, we consider a particularly simple
example of the general move, which suffices for our purposes. Call
$\,e_v^\pm\,$ the edge of $\,\triangle(\La)\,$ for which $\,v\,$ is
an endpoint with $\,\vep_{e_v^\pm v}=\pm 1\,$ and shift each vertex
$\,v\,$ of the original triangulation $\,\triangle(\La)\,$ along the
defect line to a nearby new location $\,v'\in\La\,$ such that the
segment $\,[v,v']$,\ starting at $\,v\,$ and ending at $\,v'$,\ has
the same orientation as the defect line. Assume, furthermore, that
$\,X([v,v'])\subset\Oc^Q_{i_{e_v^-} i_{e_v^+}}$.\ The shifted
vertices define altogether a new triangulation $\,\triangle'(\La)\,$
of the same defect, compatible with the new triangulation
$\,\triangle'(\Si)\,$ by construction. The only (potential) change
in the assignment of \v Cech indices to the elements of the
triangulation comes from $\,v'\mapsto i_{v'}\in\Ic^Q\,$ replacing
the former $\,v\mapsto i_v\in\Ic^Q$.\ Let us denote by
$\,\Hol'(\La,X)$ the holonomy calculated for the new triangulation
$\,\triangle'(\Si)$.\ After a short calculation, one obtains the
relation
\qq
\Hol'(\La,X)=\Hol(\La,X)\cdot\prod_{v\in\triangle(\La)}\,\biggl[
\exp\biggl(\sfi\,\int_{[v,v']}\,\widehat\om^{(1)}_{e_v^-e_v^+}
\biggr)\cdot\widehat\om^{(0)}_{e_v^-e_v^+v'}(v')\cdot\bigl(
\widehat\om^{(0)}_{e_v^-e_v^+v}\bigr)^{-1}(v)\biggr]\,,
\qqq
where
\begin{itemize}
\item the 1-form in $\,\widehat\om^{(1)}_{e_v^-e_v^+}=X^*
\om^{(1)}_{i_{e_v^-}i_{e_v^+}}\,$ pulled back by $\,X$,\ understood
as a map from $\,[v,v']\,$ to $\,\Oc^Q_{i_{e_v^-}i_{e_v^+}}$,\ is
defined as
\qq\label{eq:om1-def}
\om^{(1)}_{ij}=\check\iota_1^*A_{ij}-\check\iota_2^*A_{ij}+P_j-P_i-
\sfi\,\sfd\log K_{ij}\in\Om^1\bigl( \Oc^Q_{ij}\bigr)\,;
\qqq
\item the ${\rm U}(1)$-valued function in $\,\widehat\om^{(0)}_{e_v^-
e_v^+v^{(')}}=X^*\om^{(0)}_{i_{e_v^-}i_{e_v^+}i_{v^{(')}}}\,$ pulled
back by $\,X$,\ understood as a map from $\,v^\pm\,$ to
$\,\Oc^Q_{i_{e_v^-}i_{e_v^+}i_{v^{(')}}}$,\ is defined as
\qq\label{eq:om0-def}
\om^{(0)}_{ijk}=\check\iota_1^*g_{ijk}\cdot\check\iota_2^*g_{ij
k}^{-1}\cdot K_{jk}^{-1}\cdot K_{ik}\cdot K_{ij}^{-1}\in {\rm
U}(1)_{\Oc^Q_{ijk}}\,.
\qqq
\end{itemize}

The requirement that the unphysical change of the triangulation be
unobservable translates into the constraints
\qq
\om^{(1)}_{ij}= 0\,,\qquad\qquad\om^{(0)}_{ijk}= 1\,.
\qqq
The inspection of \eqref{eq:om1-def} and \eqref{eq:om0-def} reveals
that $\,\om^{(1)}_{ij}\,$ and $\,\om^{(0)}_{ijk}\,$ are, in fact,
the lower-degree components of the \v Cech--Deligne 2-cochain
$\,\Om=\bigl(\om^{(2)}_i,\om^{(1)}_{ij}, \om^{(0)}_{ijk}\bigr)\in
A^2_Q\,$ given by the formula
\qq
\Om=\check\iota_1^*b-\check\iota_2^*b+D\Phi\,.
\qqq
We may now use the identity
\qq\label{eq:Om-through-H}
D\bigl(\check\iota_1^*b-\check\iota_2^*b+D\Phi\bigr)=(\iota_1^*H-
\iota_2^*H,0,0,1)\,,
\qqq
following directly from \eqref{eq:def-gerbe}, to rephrase the former
requirement of invariance as
\qq\label{eq:1-morph-reder}
\check\iota_1^*b-\check\iota_2^*b+D\Phi=(\om,0,1)\,,\qquad\qquad
\sfd\om=\iota_1^*H-\iota_2^*H
\qqq
for a globally defined 2-form $\,\om\in\Om^2(Q)\,,\ \om
\vert_{\Oc^Q_i}\equiv\om^{(2)}_i$.

We have thus retrieved the structure of a 1-morphism of section
\ref{sec:bibrane-def}, with the definition of the global 2-form
$\,\om\,$ included, from elementary invariance considerations.
\medskip

Now that we have identified the local degrees of freedom to be
assigned to the defect line, we may incorporate vertices of a
generic defect network $\,\G\,$ in our description. In analogy with
the previous derivation, we take as the starting point the defect
line-corrected holonomy $\,\Hol(\La,X)$,\ calculated for
$\,\La=E_\G$,\ and study its variation under gauge transformations
replacing the gauge fields $\,b\,$ and $\,\Phi\,$ with the new ones
\qq\label{eq:G-twist-gauge-symm}
\left\{\begin{array}{ll} b^p=b+Dp\,,\qquad & p=(\Pi_i,\chi_{ij})\in
A^1_M\cr\cr \Phi^{p,\eta}=\Phi+\check\iota_2^*p -\check\iota_1^*p -
D\eta\,,\qquad & \eta=(W_i)\in A^0_Q \end{array}\right.\,.
\qqq
Once again, the transformed holonomy, $\,\Hol^{p,\eta}(E_\G,X)$,\
differs from the original one by terms evaluated at the newly
introduced junction points exclusively,
\qq
&\Hol^{p,\eta}(E_\G,X)=\Hol(E_\G,X)\cdot\prod_{v\in V_{\G}}\,
\prod_{k=1}^{n_v}\,\widehat W_{n_v,v}^{k,k+1}(v)^{-1}\,,&\cr&&\\
&\widehat W_{n_v,v}^{k,k+1}=\bigl(X_{k,k+1}^*W_{\psi^{k,k+1}_{n_v}
(i_v)}\bigr)^{\vep_{e_{k,k+1}v}}\,,&\nonumber
\qqq
and it is there that we should localise the new degrees of freedom
$\,\varphi_n=(f_{n,i})\in A^0_{T_n}$.\ They are to be coupled to the
defect as in the expression $\,\Hol_\Jc(X|_{V_{\G}})\,$ of
\eqref{eq:hol-defect-network-1} and to undergo twisted gauge
transformations
\qq\label{eq:2-morph-twgauge}
\varphi_n\too\varphi_n+\sum_{k=1}^n\,\eta^{k,k+1}_n\,,
\qqq
with
\qq
\eta^{k,k+1}_n=\bigl(\check\pi_n^{k,k+1}\bigr)^*\eta^{\eps_n^{k,
k+1}}\,.
\qqq
The $\Phi$-twisted scalar fields $\,\varphi_n\,$ enjoy no proper
gauge freedom for purely dimensional reasons. As we shall see in the
next section, the admissible choices of $\,\varphi_n\,$ turn out to
be very restricted.

The vertex-corrected formula for the holonomy is now invariant with
respect to arbitrary gauge transformations of the local data
involved. What remains to be ascertained at this stage is that it
does not alter under arbitrary changes of the world-sheet
triangulation, taken together with the attendant \v Cech labels.
Just as in the case of a circle-field configuration, we readily
convince ourselves that the relevant changes are those which involve
the vertices of the defect network, and even in this latter case the
ambiguity is very restricted -- the sole freedom that we have is in
the choice of the \v Cech labels assigned to the vertices. Under a
change $\,i_v\too i_v'$,\ the holonomy picks up a phase. The
transformed one, $\,\Hol'(\G,X)$,\ reads
\qq
\Hol'(\G,X)=\Hol(\G,X)\cdot\prod_{v\in V_{\G}}\,\widehat
\th^{(0)}_{n_v,vv}(v)^{-1}\,,
\qqq
where the ${\rm U}(1)$-valued function in $\,\widehat\th^{(0
)}_{n_v,vv}=X^*\th^{(0)}_{n_v,i_v i_v'}\,$ pulled back by $\,X^*$,\
understood as a map from the vertex $\,v\,$ of valence $n_v$ to
$\,\Oc^{T_{n_v}}_{i_v i_v'}$,\ is given by
\qq
\th^{(0)}_{n,ij}=f_{n,i}\cdot f_{n,j}^{-1}\cdot\prod_{k=1}^n\,K^{k,
k+1}_{n,ij}\in {\rm U}(1)_{\Oc^{T_n}_{ij}}\,,
\qqq
with $\,K^{k,k+1}_{n,ij}\,$ as defined above
\eqref{eq:inter-bi-2morph-loc}. We are led to impose the constraint
\qq
\th^{(0)}_{n,ij}= 1\,.
\qqq
The definition of the functions $\,\th^{(0)}_{n,ij}\,$ identifies
them, for every $\,n\in\Zb_{> 0}$,\ as the 0-degree component of
the \v Cech--Deligne 1-cochain $\,\Th_n=\bigl(\th^{(1)}_{n,i},
\th^{(0)}_{n,ij}\bigr)\in A^1_{T_n}\,$ defined as
\qq\label{eq:2-morph-D}
\Th_n=\sum_{k=1}^n\,\Phi^{k,k+1}_n+D\varphi_n
\qqq
and, accordingly, satisfying
\qq
D\Th_n=\sum_{k=1}^n\,\bigl(\om^{k,k+1}_n,0,1\bigr)\,,
\qqq
with $\,\om^{k,k+1}_n\,$ as in \eqref{eq:omn-pull}. The last
identity, in conjunction with the requirement of invariance,
produces the result
\qq\label{eq:2-morph-reder}
\sum_{k=1}^n\,\Phi^{k,k+1}_n+D\varphi_n=(\th_n,1)\,,\qquad\qquad
\sfd\th_n=\sum_{k=1}^n\,\om^{k,k+1}_n
\qqq
for globally defined 1-forms $\,\th_n\in\Om^1(T_n)\,,\ \th_n
\vert_{\Oc^{T_n}_i}\equiv\th^{(1)}_{n,i}$.\ The dynamical arguments
of appendix \ref{app:def-glue} ultimately fix the vertex data by
imposing the constraint
\qq
\th_n=0
\qqq
for all $\,n\in\Zb_{> 0}$.

We have thus recovered the structure of a 2-morphism of section
\ref{sec:inter-bi-branes} from elementary invariance considerations.

\subsection{Defect-vertex data via induction}\label{sec:cluster}

The assignment of the holonomy $\,\Hol(\G,X)\,$ to a given
world-sheet with an embedded defect network involves a number of
choices for the coupled target-space backgrounds $\,(b,\Phi,
\varphi)$,\ reflecting the underlying twisted gauge symmetry.
Besides the unphysical choice of the gauge, cf.\
\eqref{eq:G-twist-gauge-symm} and \eqref{eq:2-morph-twgauge}, there
is also the all-relevant choice of the gauge class which forms an
integral part of the definition of the sigma model.

The $\Phi$-twisted scalar field $\,\varphi\,$ of
\eqref{eq:2-morph-glued} has no proper gauge symmetry but the
possible choices of $\,\varphi\,$ for $\,b\,$ and $\,\Phi\,$ fixed
are strongly constrained. To see this, note that any two such
choices $\,\varphi'\,$ and $\,\varphi\,$ must be related via
\qq\label{eq:2-morph-gauge}
\varphi' = \varphi+\g\,,\qquad\qquad\g=(c_i)\in\Ker D_{(0)}\subset
A^0_T
\qqq
by untwisted ${\rm U}(1)$-valued scalars. Thus, the freedom in the
choice of the $\Phi$-twisted scalar field is parameterised
by (locally) constant phases,
\qq
\sfd c_i=0\,,\qquad\qquad\bigl(c_j\cdot c_i^{-1}\bigr)
\vert_{\Oc^T_{ij}}=1\,,
\qqq
readily seen to compose the group
\qq\label{eq:ker-D0}
\Ker D_{(0)}={\rm U}(1)^{\vert\pi_0(T)\vert}\,,
\qqq
for $\,\pi_0(T)\,$ the set of connected components of $\,T$.\

\medskip

The restricted character of the set of admissible $\Phi$-twisted
scalar fields motivates further investigation of special solutions
to the defining equations \eqref{eq:inter-bi-2morph}. In conformal
field theory, we can generate four-valent defect vertices (or
$n$-valent vertices, for that matter) from three-valent ones as
follows: Recall that a defect vertex corresponds to a defect-field
insertion in CFT (cf.\ figure \ref{fig:state-field}). Consider two
three-valent defect fields joined by one common defect line of a
small length $\,\eps$.\ Taking the limit $\,\eps\rightarrow 0\,$ and
possibly compensating for the resulting divergence leads to a
four-valent defect field. It turns out that we can mimic this
procedure in the classical sigma model.
\medskip

Recall from section \ref{sec:inter-bi-branes} that a
$(\Gc,\Bc)$-inter-bi-brane is defined in terms of a tower of
component world-volumes $\,T= \bigsqcup_{n=1}^\infty\,T_n$,\ with a
2-morphism $\,\varphi_n$ on each $\,T_n$.\ Below, we propose a
method to construct the $\,\varphi_n\,$ with $\,n>3\,$ from
$\,(T_3,\Oc^{T_3},\varphi_3,\t_3)\,$ and some extra data. For the
sake of concreteness, we shall restrict our discussion to the
special case of vertices of valence $\,n=4\,$ with three incoming
edges and one outgoing edge.

The point of departure in our construction is the data
$\,(T_{2+1},\Oc^{T_{2+1}}, \varphi_{2+1},\t_{2+1})\,$ for the
three-valent vertex with two incoming edges and one outgoing edge.
It consists of the $(\Gc,\Bc)$-inter-bi-brane world-volume
$\,T_{3,++-}\equiv T_{2+1}$,\ mapped to the $\Gc$-bi-brane
world-volume $\,Q\,$ by each of the three \v Cech-extended maps
\qq
\check\pi_{2+1}^{1,2}\,,\ \check\pi_{2+1}^{2,3}\,,\ \check
\pi_{2+1}^{3,1}\ :\ T_{2+1}\too Q
\qqq
satisfying the constraints
\qq\label{eq:pi3-constr}
\check\iota_1\circ\check\pi_{2+1}^{2,3}=\check\iota_2\circ\check
\pi_{2+1}^{1,2}\,,\qquad\qquad\check\iota_1\circ\check\pi_{2+1}^{3,1}=
\check\iota_1\circ\check\pi_{2+1}^{1,2}\,,\qquad\qquad\check\iota_2
\circ\check\pi_{2+1}^{3,1}=\check\iota_2\circ\check\pi_{2+1}^{2,3}\,,
\qqq
and of a 2-morphism
\qq
\varphi_{2+1}\ :\ \bigl(\Phi_{2+1}^{3,1}\star\id_{I(\om^{1,2}_{2+1}+
\om^{2,3}_{2+1})}\bigr)\circ\bigl(\Phi_{2+1}^{2,3}\star\id_{I(
\om^{1,2}_{2+1})}\bigr)\circ\Phi_{2+1}^{1,2}\Longrightarrow
\id_{\Gc^1_{2+1}}\,,
\qqq
defined for $\,\Gc^1_{2+1}=\bigl(\check\iota_1\circ \check\pi^{1,
2}_{2+1}\bigr)^*\Gc$.\ The latter canonically induces another
2-morphism
\qq
\til\varphi_{2+1}\ :\ \bigl(\Phi_{2+1}^{2,3}\star\id_{I(
\om^{1,2}_{2+1})}\bigr)\circ\Phi_{2+1}^{1,2}\Longrightarrow\Phi_{2+
1}^{1,3}\,,
\qqq
with $\,\Phi_{2+1}^{3,1}=\bigl(\Phi_{2+1}^{1,3}\bigr)^\vee$,\ giving
a decomposition of $\,\varphi_{2+1}\,$ of the form
\qq
\varphi_{2+1}=d_{\Phi_{2+1}^{1,3}}\bullet\bigl(\id_{\Phi_{2+1}^{3,1}
\star\id}\circ\til\varphi_{2+1}\bigr)\,.
\qqq
Next, we assume that we are given a manifold $\,T_{3+1}\equiv
T_{4,+++-}\,$ together with four \v Cech-extended maps
\qq
\check\vv_{I,J,K}=(\vv_{I,J,K},\nu_{I,J,K})\ :\ T_{3+1}\too
T_{2+1}\,,\qquad\qquad 1\leq I<J<K\leq 4
\qqq
subject to the conditions
\be
\begin{array}{ll}\displaystyle
\check\pi^{1,2}_{2+1}\circ\check
\vv_{1,3,4}=\check\pi^{3,1}_{2+1}\circ\check\vv_{1,2,3}\,,\qquad
\etb
\check\pi^{2,3}_{2+1}\circ\check\vv_{1,2,4}=\check\pi^{3,1}_{2+1}
\circ\check\vv_{2,3,4}\,, \enl
\check\pi^{1,2}_{2+1}\circ\check\vv_{1,2,3}=
\check\pi^{1,2}_{2+1}\circ\check\vv_{1,2,4}\,,\qquad \etb
\check\pi^{2,3}_{2+1}\circ\check\vv_{1,2,3}=
\check\pi^{1,2}_{2+1}\circ\check\vv_{2,3,4}\,, \enl
\check\pi^{2,3}_{2+1}\circ\check\vv_{1,3,4}=
\check\pi^{2,3}_{2+1}\circ\check\vv_{2,3,4}\,,\qquad \etb
\check\pi^{3,1}_{2+1}\circ\check\vv_{1,3,4}=\check\pi^{3,1}_{2+1}
\circ\check\vv_{1,2,4}\,.
\end{array}
\labl{eq:pi-cluster-constr}
Their existence is the basis of our construction, and we shall
provide examples of such maps presently. In order to understand the
index structure, one should have a look at figure
\ref{fig:associator} below. For example, the right-hand side of the
last equation in \eqref{eq:pi-cluster-constr} can be understood as
passing from the image of $\,v \in \Si_{L|R}\,$ in $\,T_{3+1}\,$ to
the image of $\,v \in \Si_R\,$ in $\,T_{2+1}\,$ (with adjacent
patches 1,2,4), and subsequently to that of the endpoint of the edge
between patches 1 and 4 (the edge $\,e^{3,1}\,$ with respect to the
ordering for the vertex $\,v \in \Si_R$). For the left-hand side of
that equation, one uses $\,\Si_L\,$ instead.

The maps $\,\check\vv_{I,J,K}\,$ are readily seen to induce the
inter-bi-brane structure $\,T_{3+1}\,$ for the four-valent vertices.
Indeed, first of all, they provide us with the data of
$\,\t_{3+1}\,$ as per
\qq
\check\pi^{1,2}_{3+1}=\check\pi^{1,2}_{2+1}\circ\check\vv_{1,2,3}
\,,\qquad\check\pi^{2,3}_{3+1}=\check\pi^{2,3}_{2+1}\circ\check
\vv_{1,2,3 }\,,\cr\\
\check\pi^{3,4}_{3+1}=\check\pi^{2,3}_{2+1}\circ\check\vv_{1,3,4}\,,
\qquad\check\pi^{4,1}_{3+1}=\check\pi^{3,1}_{2+1}\circ\check\vv_{1,3,
4}\,,\nonumber
\qqq
and hence also with the patch maps $\,\check\pi^k_{3+1}\,,\ k=1,2,3,
4$.\ The latter give us the pullback gerbes $\,\Gc^k_{3+1}=\bigl(
\check\pi^k_{3+1}\bigr)^*\Gc\,$ on $\,T_{3+1}$.\ It is a simple
exercise to verify that the conditions \eqref{eq:pi-cluster-constr}
in conjunction with \eqref{eq:pi3-constr} ensure that the maps
$\,\check\pi^{k,k+1}_{3+1} : T_{3+1}\too Q\,$ satisfy the
constraints \eqref{eq:inter-bi-brane-projB}. We supplement the above
collection with the extra definitions
\qq
\check\pi^{1,3}_{3+1}\equiv\check\pi^{3,1}_{3+1}=\check\pi^{3,
1}_{2+1}\circ\check\vv_{1,2,3}\,,\qquad\qquad\check\pi^{2,4}_{3+1}
\equiv\check\pi^{4,2}_{3+1}=\check\pi^{2,3}_{2+1}\circ\check\vv_{1,
2,4}\,,
\qqq
allowing us to write down all the pullback 1-morphisms
\qq
\Phi_{3+1}^{I,J}=\bigl(\check\pi^{I,J}_{3+1}\bigr)^*\Phi\ &:&\
\Gc^I_{3+1}\rightarrow\Gc^J_{3+1}\,,\qquad\qquad I<J\,,\qquad
(I,J)\neq (1,4)\,,\cr\cr
\Phi_{3+1}^{4,1}=\bigl(\check\pi^{4,1}_{3+1}\bigr)^*\Phi=\bigl(
\Phi_{3+1}^{1,4}\bigr)^\vee\ &:&\ \Gc^4_{3+1}\rightarrow\Gc^1_{3+1}
\,.
\qqq
We can use these to give the two different definitions of the
defect-vertex 2-morphism
\be\begin{array}{l}
\varphi_{3+1}^L = d_{\Phi_{3+1}^{1,4}}\bullet\bigl(\id\circ\til
\varphi^{1,3,4}\bigr)\bullet\bigl(\id\circ\til\varphi^{1,2,3}\bigr)
\,,\enl \varphi_{3+1}^R =
d_{\Phi_{3+1}^{1,4}}\bullet\bigl(\id\circ\til
\varphi^{1,2,4}\bigr)\bullet\bigl(\id\circ(\til\varphi^{2,3,4}\star
\id_{\id_{I(\om^{1,2}_{3+1})}})\circ\id\bigr)\,,
\end{array}
\labl{eq:phiL-phiR-def}
acting as
\be \begin{array}{l}
\bigl(\Phi^{4,1}_{3+1}\star\id_{I(\om^{1,
2}_{3+1}+\om^{2,3}_{3+1}+\om^{3,4}_{3+1})}\bigr)\circ\bigl(\Phi^{3,
4}_{3+1}\star\id_{I(\om^{1,2}_{3+1}+\om^{2,3}_{3+1})}\bigr)\circ
\bigl(\Phi^{2,3}_{3+1}\star\id_{I(\om^{1,2}_{3+1})}\bigr)\circ
\Phi^{1,2}_{3+1}\xLongrightarrow{\varphi_{3+1}^{L,R}}
\id_{\Gc^1_{3+1}}\,.
\end{array}
\ee
They are expressed in terms of the corresponding pullback
2-morphisms
\qq
\til\varphi^{I,J,K}=\check\vv_{I,J,K}^*\til\varphi_{2+1}\ :\ \bigl(
\Phi_{3+1}^{J,K}\star\id_{I(\om^{I,J}_{3+1})}\bigr)\circ\Phi_{3+
1}^{I,J}\Longrightarrow\Phi_{3+1}^{I,K}\,,
\qqq
and the death 2-morphism $\,d_{\Phi_{3+1}^{1,4}}$.\ Clearly, the two
definitions, $\,\varphi_{3+1}^L\,$ and $\,\varphi_{3+1}^R$,\
correspond to the two inequivalent ways of clustering the incoming
defect-lines converging at a given four-valent defect vertex. It is
worth underlining that while each of the two definitions in
\eqref{eq:phiL-phiR-def} requires only two of the four maps
$\,\check\vv_{I,J,K}$,\ the verification of the constraints
\eqref{eq:inter-bi-brane-projB} for the induced maps
$\,\check\pi^{k,k+1}_{3+1}\,$ uses \emph{all four} $\,\check
\vv_{I,J,K}$.
\medskip

A generic example of an induced $(\Gc,\Bc)$-inter-bi-brane structure
can be obtained from the $\Gc$-bi-brane world-volume $\,Q\subset M
\ti M\,$ and the $(\Gc,\Bc)$-inter-bi-brane world-volumes
$\,T_n\subset M\ti M\ti\cdots\ti M\,$ embedded as submanifolds in
the respective multiple direct products of the target space $\,M\,$
with itself, with $\,\pi^{k,k+1}_n : M_{(1)}\ti M_{(2)}\ti\cdots\ti
M_{(n)}\too M_{(k)}\ti M_{(k+1)}\,$ given by the canonical
projections ($M_{(l)}\equiv M\,,\ l= 1,2,\ldots,n$). In this
setting, given the world-volume $\,T_{2+1}\subset M\ti M\ti M\,$ of
the $(\Gc,\Bc)$-inter-bi-brane, we choose for the world-volume
$\,T_{3+1}\subset M\ti M\ti M\ti M\,$ the common intersection of the
preimages $\,\vv_{I,J,K}^{-1}(T_{2+1})\,$ of $\,T_{2+1}\,$ under the
canonical projections $\,\vv_{I,J,K}\equiv\pi_{3+1}^{I,J,K} :
M_{(1)}\ti M_{(2)}\ti M_{(3)}\ti M_{(4)}\too M_{(I)}\ti M_{(J)}\ti
M_{(K)}$.\ The conditions \eqref{eq:pi-cluster-constr} are trivially
satisfied.

\subsubsection*{The Lie-group example (cont'd)}

We now proceed to demonstrate how the data $\,(T_{2+1},
\Oc^{T_{2+1}},\varphi_{2+1},\t_{2+1})\,$ for three-valent vertices
with signature $\,(+1,+1,-1)$,\ introduced in section
\ref{sec:inter-bi-branes}, can be used to induce the data
$\,(T_{3+1},\Oc^{T_{3+1}},\varphi_{3+1},\t_{3+1})\,$ for four-valent
vertices with signature $\,(+1,+1,+1,-1)\,$ in accord with the
general scheme discussed above. We start with the definition of the
\v Cech-extended maps $\,\check\vv_{I,J,K} : T_{3+1}\rightarrow
T_{2+1}$,\ which -- for $\,(g,x,y,z)\in \Oc^{T_{3+1}}_{i,x,y,z}$,\
written in the previously adopted shorthand notation with the
redundant signs dropped -- reads
\qq
\begin{array}{ll}
\vv_{1,2,3}(g,x,y,z)=(g,x,y)\,,\qquad\qquad &\nu_{1,2,3}(i,x,y,z)=(
i,x,y)\,,\cr\cr \vv_{1,3,4}(g,x,y,z)=(g,x\cdot y,z)\,,\qquad\qquad
&\nu_{1,3,4}(i,x,y,z)=(i,x\cdot y,z)\,,\cr\cr
\vv_{2,3,4}(g,x,y,z)=(x^{-1}\cdot g,y,z)\,,\qquad\qquad
&\nu_{2,3,4}(i,x,y,z)=(x^{-1}.i,y,z)\,,\cr\cr \vv_{1,2,4}(g,x,y,z)=
(g,x,y\cdot z)\,,\qquad\qquad&\nu_{1,2,4}(i,x,y,z)=(i,x,y\cdot z)\,.
\end{array}
\qqq
One readily verifies that $\,\check\vv_{I,J,K}\,$ obey condition
\eqref{eq:pi-cluster-constr}, and so they can be used to pull back
the data $\,(T_{2+1},\Oc^{T_{2+1}},\varphi_{2+1},\t_{2+1})\,$ to
$\,T_{3+1}$.\ Thus, we induce the relevant \v Cech-extended maps
$\,\check\pi_{3+1}^{I,J}\,$ in the form
\qq
\begin{array}{ll}
\pi_{3+1}^{1,2}(g,x,y,z)=(g,x)\,,\qquad\qquad&\pi_{3+1}^{2,3}(g,x,y
,z)=(x^{-1}\cdot g,y)\,,\cr\cr \pi_{3+1}^{3,4}(g,x,y,z)=((x\cdot
y)^{-1}\cdot g,z)\,,\qquad\qquad&\pi_{3+1}^{4,1}(g,x,y,z)=(g,x\cdot
y\cdot z)\,,\cr\cr \pi_{3+1}^{3,1}(g,x,y,z)=(g,x\cdot y)\,,\qquad
\qquad&\pi_{3+1}^{4,2}(g,x,y,z)=(x^{-1}\cdot g,y\cdot z)\,,
\end{array}
\qqq
and similarly for $\,\psi^{I,J}_{3+1}$.\ These, in turn, give us the
\v Cech-extended maps $\,\check\pi_{3+1}^k\,$ for the patches
\qq
\begin{array}{ll}
\pi_{3+1}^1(g,x,y,z)=g\,,\qquad\qquad&\pi_{3+1}^2(g,x,y,z)=x^{-1}
\cdot g\,,\cr\cr \pi_{3+1}^3(g,x,y,z)=(x\cdot y)^{-1}\cdot g\,,
\qquad\qquad&\pi_{3+1}^4(g,x,y,z)=(x\cdot y\cdot z)^{-1}\cdot g\,,
\end{array}
\qqq
and similarly for $\,\psi^k_{3+1}$.\ With the help of the induced
maps, we then obtain on $\,T^{x,y,z}_{3+1}=\Gx\ti\{(x,y,z)\}\subset
T_{3+1}\,$ (again, identified with $\,\Gx$) the pullback gerbes
\qq
&\Gc^1_{3+1}=\Gc^{\star\sfk}\,,\qquad\qquad\Gc^2_{3+1}=x.\Gc^{\star
\sfk}\,,&\cr&&\\
&\Gc^3_{3+1}=(x\cdot y).\Gc^{\star\sfk}\,,\qquad\qquad\Gc^4_{3+1}=
(x\cdot y\cdot z).\Gc^{\star\sfk}\,,&\nonumber
\qqq
the pullback 1-morphisms
\qq
&\Phi_{3+1}^{1,2}=\Ac_x\,,\qquad\qquad\Phi_{3+1}^{2,3}=x.\Ac_y\,,
\qquad\qquad\Phi_{3+1}^{3,4}=(x\cdot y).\Ac_z\,,\qquad\qquad
\Phi_{3+1}^{4,1}=\Ac_{x\cdot y\cdot z}^\vee\,,\cr\cr
&\Phi_{3+1}^{1,3}=\Ac_{x\cdot y}\,,\qquad\qquad\Phi_{3+1}^{2,4}
=x.\Ac_{y\cdot z}\,,
\qqq
and the pullback 2-morphisms
\be
\til\varphi^{1,2,3}=\til\varphi_{x,y}\,,\qquad\quad\til\varphi^{1,
3,4}=\til\varphi_{x\cdot y,z}\,,\qquad\quad\til\varphi^{2,3,4}=x.
\til\varphi_{y,z}\,,\qquad\quad\til\varphi^{1,2,4}=\til\varphi_{x,
y\cdot z}\,,
\ee
where we have used the action $\,x.\til\varphi_{y,z}\equiv
\check{\bigl(x^{-1}\bigr)}^*\til\varphi_{y, z}\,$ of $\,Z(\Gx)\,$
given in \eqref{eq:ext-pullb-act}. Putting all the pieces together,
we arrive at the two definitions of the 2-morphism on $\,T_{3+1}$
\qq
\varphi^L_{3+1}\big\vert_{T_{3+1}^{x,y,z}}&=&d_{\Ac_{x\cdot y\cdot
z}}\bullet\bigl(\id\circ\til\varphi_{x\cdot y,z}\bigr)\bullet\bigl(
\id\circ\til\varphi_{x,y}\bigr)\,,\cr&&\\
\varphi^R_{3+1}\big\vert_{T_{3+1}^{x,y,z}}&=&d_{\Ac_{x\cdot y\cdot
z}} \bullet\bigl(\id\circ\til\varphi_{x,y\cdot z}\bigr)\bullet
\bigl(\id\circ x.\til\varphi_{y,z}\circ\id\bigr)\,,\nonumber
\qqq
acting as
\qq
\varphi^{L,R}_{3+1}\big\vert_{T_{3+1}^{x,y,z}}\ :\ \Ac_{x\cdot
y\cdot z}^\vee\circ(x\cdot y).\Ac_z\circ x.\Ac_y\circ\Ac_x
\Longrightarrow\id_{\Gc^{\star\sfk}}
\qqq
and differing at most by a constant on each connected component
$\,T_{3+1}^{x,y,z}\,$ of the world-volume $\,T_{3+1}\,$ (recall that
the Lie group $\,\Gx\,$ was assumed connected).

We may now compare the two induced 2-morphisms $\,\varphi^L_{3+1}\,$
and $\,\varphi^R_{3+1}\,$ on each $\,T_{3+1}^{x,y,z}$,\ identified
with $\,\Gx\,$ itself, by applying \eqref{eq:2-morph-gauge} and
\eqref{eq:ker-D0} to the setting under consideration. Let $\,g\in
\Oc^\Gx_i\,$ be an arbitrary point in $\,T_{3+1}^{x,y,z}$,\ and let
$\,\bigl(f_i^{L,R}\bigr)\in A^0_{T_{3+1}^{x,y,z}}\,$ be the local
data of $\,\varphi^{L,R}_{3+1}\big\vert_{T_{3+1}^{x,y,z}}$.\ We have
the identity
\qq\label{eq:phiR-vs-phiL-WZW}
f^R_i(g)=\psi(x,y,z)\cdot f^L_i(g)
\qqq
for the ${\rm U}(1)$-valued constant
\qq\label{eq:gerbe-assoc-3-cocycle-def}
\psi(x,y,z)=\bigl[\bigl(x^{-1}\bigr)^*f_{x^{-1}.i}(y,z)\cdot\bigl(
f_i(x\cdot y,z)\bigr)^{-1}\cdot f_i(x,y\cdot z)\cdot\bigl(f_i(x,y)
\bigr)^{-1}\bigr](g)
\qqq
written in terms of the local data of the 2-morphism
$\,\til\varphi_{x,y}=(f_i(x,y))\in A^0_{T^{x,y,z}_{3+1}}$.\ By
virtue of \eqref{eq:2-morph-gauge}, the expression $\,\psi(x,y,z)\,$
depends neither on the specific point $\,g\in\Gx$,\ nor on the
attendant \v Cech index $\,i\in\Ic^\Gx$.\ This permitted us to drop
both $\,g\,$ and $\,i\,$ when writing $\,\psi(x,y,z)\,$ in
\eqref{eq:phiR-vs-phiL-WZW} and
\eqref{eq:gerbe-assoc-3-cocycle-def}. We emphasise that only the
particular combination $\,\psi(x,y,z)\,$ of the locally smooth
functions $\,f_i(x,y)\,$ is constant on $\,\Gx\,$ -- in general,
none of the component terms has this property.

Note that $\,\psi(x,y,z)\,$ rewrites as
\qq\label{eq:psi-as-delS}
\psi(x,y,z)=\bigl[\bigl(\d_{Z(\Gx)}f_i\bigr)(x,y,z)\bigr](g)\,,
\qqq
where we consider the local data of the 2-morphisms
$\,\til\varphi_{x,y}\,$ as elements of the (left) $Z(\Gx)$-module
$\,\unl{{\rm U}(1)}_{T_{3+1}}\,$ of (the sheaf of) locally smooth
${\rm U}(1)$-valued functions on $\,T_{3+1}$.\ The centre
$\,Z(\Gx)\,$ acts on $\,\unl{{\rm U}(1)}_{T_{3+1}}\,$ by the \v
Cech-extended pullbacks
\qq
(x.f)_i(y,z)=\bigl(x^{-1}\bigr)^*f_{x^{-1}.i}(y,z)\,.
\qqq
Despite the form of \eqref{eq:psi-as-delS}, the object
$\,(\psi(x,y,z)\,\vert\,x,y,z\in Z(\Gx))$,\ regarded as a 3-cochain
on $\,Z(\Gx)\,$ with values in the trivial $Z(\Gx)$-module $\,{\rm
U}(1)$,\ is \emph{not} a 3-coboundary -- it is not in the image of
$\,\d_{Z(\Gx)}\ :\ C^2(Z(\Gx),{\rm U}(1)) \rightarrow
C^3(Z(\Gx),{\rm U}(1))$.\ Being an element of the kernel of the
Deligne differential $\,D_{(0)}\,$ on the connected Lie group
$\,\Gx$,\ it is, on the other hand, $\d_{Z(\Gx)}$-closed,
\qq
\bigl(\d_{Z(\Gx)}\psi\bigr)(x,y,z,w)&=&\frac{\psi(y,z,w)\cdot\psi(x
,y\cdot z,w)\cdot\psi(x,y,z)}{\psi(x\cdot y,z,w)\cdot\psi(x,y,z\cdot
w)}\cr\cr
&\equiv&\frac{\Big(\check{\bigl(x^{-1}\bigr)}^*\psi(y,z,w)\Big)\cdot
\psi(x,y\cdot z,w)\cdot\psi(x,y,z)}{\psi(x\cdot y,z,w)\cdot\psi(x,y,
z\cdot w)}\\\cr
&=&\bigl[\bigl(\d_{Z(\Gx)}^2f_i\bigr)(x,y,z)\bigr](g)=1\,.
\nonumber
\qqq
Above, the passage to the second line exploits the stated
independence of $\,\psi(x,y,z)\,$ of the choice of the argument and
of the \v Cech index of the constituent functions $\,f_i(x,y)\,$ by
simply replacing the original expression with the pullback
\qq
\check{\bigl(x^{-1}\bigr)}^*\psi(y,z,w)&=&\bigl((x\cdot y)^{-1}
\bigr)^*f_{(x\cdot y)^{-1}.i}(z,w)\cdot\bigl(\bigl(x^{-1}\bigr)^*
f_{x^{-1}.i}(y\cdot z,w)\bigr)^{-1}\cr&&\\
&&\cdot\bigl(x^{-1}\bigr)^*f_{x^{-1}.i}(y,z\cdot w)\cdot\bigl(
\bigl(x^{-1}\bigr)^*f_{x^{-1}.i}(y,z)\bigr)^{-1}(g)\,.\nonumber
\qqq
Thus, $\,(\psi(x,y,z)\,\vert\,x,y,z \in Z(\Gx))\,$ is a ${\rm
U}(1)$-valued 3-cocycle on $\,Z(\Gx)$.\ As shall become clear in the
next section, it is the very associator 3-cocycle that we have been
after all along.

\subsection{Conformal and topological defects}\label{sec:conf-top-def}

Having specified a sigma-model description
\eqref{eq:hol-defect-network-1} of the coupling of target-space
structures $\,\Gc,\Phi\,$ and $\,\varphi\,$ to a world-sheet
$\,\Si\,$ with an embedded defect network $\,\G$,\ we are now ready
to discuss the local symmetries of the thus established
two-dimensional field theory. They descend from the local-symmetry
group of the sigma model without defects, which is the semidirect
product $\,\textrm{Diff}(\Si)\ltimes\textrm{Weyl}(\g)\,$ of the
group $\,\textrm{Diff}(\Si)\,$ of orientation-preserving
diffeomorphisms $\,\si\mapsto f(\si)\,$ of the world-sheet and the
group $\,\textrm{Weyl}(\g)\,$ of Weyl rescalings $\,\g(\si)\mapsto
\exp(2w(\si))\cdot\g(\si)\,\,$ of the world-sheet metric tensor
$\,\g$.\ Weyl rescalings remain a symmetry in the presence of
defects as the holonomy formula does not involve the world-sheet
metric at all. As a consequence, the energy-momentum tensor
\qq
T^{ab}=-\frac{1}{\sqrt{\det\g}}\,\frac{\d S}{\d\g_{ab}}
\qqq
is traceless. Let $\,f : \Si\rightarrow\Si\,$ be an
orientation-preserving diffeomorphism. Given a network-field
configuration $\,(\G,X)$,\ we obtain a new network-field
configuration $\,(f(\G),X\circ f^{-1})$.\ Clearly, for
$\,S[(\G,X);\gamma]= S_{\rm kin}[X,\g]+\log\Hol(\G,X)$,\ we find
\qq
S[(\G,X);\g] = S[(f(\G),X\circ f^{-1});(f^{-1})^*\g]\,.
\qqq
In this sense, the sigma model for the world-sheet with the defect
network possesses diffeomorphism invariance. In particular, we may
fix a metric $\,\g_0\,$ on $\,\Si\,$ and take $\,f_{\rm c} : \Si
\rightarrow\Si\,$ to be a conformal transformation. Due to the
diffeomorphism invariance, and owing to the Weyl symmetry, the
action obeys
\qq
S[(\G,X);\g_0]=S[(f_{\rm c}(\G),X\circ f_{\rm c}^{-1}); \g_0]\,.
\qqq
If $\,f_{\rm c}\,$ maps the defect network $\,\G\,$ to itself, it is
a symmetry of the model. The defects we describe are therefore
classically conformally invariant.

It is convenient to pass to local complex coordinates $\,z=\si^1+
\sfi\,\si^2\,$ close to a defect line, such that the defect line
coincides with the real axis and such that we can choose a gauge in
which $\,\g_0\,$ is the unital metric $\,\delta_{ab}\,\sfd\si^a\,
\oti\sfd\si^b$.\ We shall use the complex derivatives
$\,\p=\frac{1}{2} \,(\p_1-\sfi\,\p_2)$ and $\,\ovl\p
=\frac{1}{2}\,(\p_1+\sfi\,\p_2)$. \ The holomorphic and
anti-holomorphic components of the energy-momentum tensor are then
given by
\be
T=G_X(\p X,\p X)\,,\qquad\qquad\ovl T=G_X(\ovl\p X,\ovl\p X)\,.
\labl{eq:T-hol-antihol}
Inserting the choice $\,v=X_*\widehat t\,$ in the defect condition
\eqref{eq:defect-cross} yields $\,G_{X_{|1}(p)}( \p_1 X_{|1},\p_2
X_{|1})-G_{X_{|2}(p)}(\p_1 X_{|2},\p_2 X_{|2})=0$,\ or,
equivalently,
\be
T_1(p)-\ovl T_1(p)=T_2(p)-\ovl T_2(p)\,,
\labl{eq:con-def-cond}
where $\,p\,$ is a point on the real axis and $\,T_\a$,\ for
$\,\a=1,2$,\ stands for \eqref{eq:T-hol-antihol} with $\,X\,$
replaced by the extension $\,X_{|\a}$.\ Thus, the classical
energy-momentum tensor indeed obeys the defining equation of a
conformal defect as given in \cite{Oshikawa:1996dj}.
\medskip

Ultimately, we are interested in topological defects, i.e.\ defects
which one can move freely on the world-sheet. For simplicity, we
restrict the following discussion to circle-field configurations.
Let $\,(\La,X)\,$ be a circle-field configuration. If we deform the
embedded defect circles from $\,\La\,$ to $\,\La_\eps\,$ then we
need to extend the map $\,X\,$ to the domain swept during the
deformation in order to obtain a new circle-field configuration. We
shall now describe how this can be achieved.

Let $U$ be a tubular neighbourhood of $\La$. An {\em extension of}
$X$ {\em on} $U$ is a map $\widehat X : U \rightarrow Q$ with the
following properties. The defect circles $\La$ split $U$ into $U_1$
and $U_2$. We demand
\be
\widehat X=X ~\text{on}~ \La\,,\qquad\qquad \iota_1\circ\widehat X=X
~\text{on}~ U_1\,,\qquad\qquad\iota_2\circ\widehat X=X ~\text{on}~
U_2\,,
\labl{eq:extend-X-1}
as well as, for all $\,p\in U\,$ and $\,v\in T_{\widehat X(p)}Q$,
\be
\D G_{\widehat X(p)}\bigl(v,\widehat X_*\widehat u_2\bigr)=
\tfrac{\sfi}{2}\,\omega_{\widehat X(p)}\bigl(v,\widehat X_*\widehat
u_1\bigr)\,,\qquad\qquad\D G=\iota_1^*G-\iota_2^*G\,,
\labl{eq:pretty-defect-cond}
where $\,\bigl(\widehat u_1,\widehat u_2\bigr)\,$ form a
right-handed orthonormal basis of $\,T_p\Si$.

Consider a deformation $\,\La_\eps\,$ of a segment of the defect
circles $\,\La\,$ as depicted below,
\be
  \raisebox{-15pt}{\begin{picture}(110,40)
  \put(0,0){\scalebox{0.55}{\includegraphics{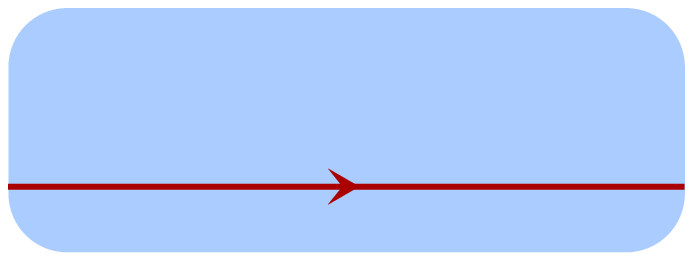}}}
  \put(0,0){
     \setlength{\unitlength}{.55pt}\put(-26,-29){
     \put(87, 54)   {\scriptsize $ \La $}
     }\setlength{\unitlength}{1pt}}
  \end{picture}}
  \quad \longrightarrow \quad
  \raisebox{-15pt}{\begin{picture}(110,40)
  \put(0,0){\scalebox{0.55}{\includegraphics{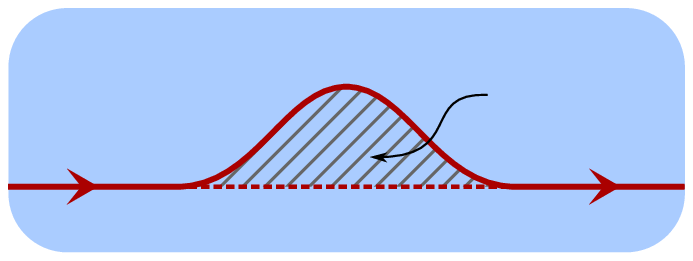}}}
  \put(0,0){
     \setlength{\unitlength}{.55pt}\put(-26,-29){
     \put( 78, 64)   {\scriptsize $ \La_{\eps} $}
     \put(168, 69)   {\scriptsize $ A_\eps $}
     }\setlength{\unitlength}{1pt}}
  \end{picture}}\quad.
\labl{eq:defect-bump}
If we are given a circle-field configuration $\,(\La,X)\,$ and an
extension $\,\widehat X\,$ of $\,X\,$ on a neighbourhood $\,U\,$ of
$\,\La\,$ then we can define a new circle-field configuration
$\,(\La_\eps,X_\eps)\,$ by setting $\,X_\eps=\widehat X\,$ on
$\,\La_\eps$,\ and $\,X_\eps=\iota_2\circ\widehat X\,$ in the shaded
region $\,A_\eps$.\ Outside of $\,\La\,$ and $\,A_\eps$,\ we choose
$\,X_\eps=X$.\ The conditions \eqref{eq:extend-X-1} and
\eqref{eq:pretty-defect-cond} guarantee that $\,(\La_\eps,X_\eps)\,$
is, again, a valid network-field configuration. In particular, it
obeys the defect condition \eqref{eq:defect-cross}, which can be
seen by rewriting \eqref{eq:pretty-defect-cond} in the form
\be
G_{\iota_1\circ\widehat X(p)}\bigl(\iota_{1*}v,\bigl(\iota_1\circ
\widehat X\bigr)_*\widehat u_2\bigr)-G_{\iota_2\circ\widehat X(p)}
\bigl(\iota_{2*}v,\bigl(\iota_2\circ\widehat X\bigr)_*\widehat u_2
\bigr)-\tfrac{\sfi}{2}\,\omega_{\widehat X(p)}\bigl(v,\widehat X_*
\widehat u_1\bigr)=0\,.
\labl{eq:extend-X-2}
In this way, an extension of $\,X\,$ on $\,U\,$ enables us to deform
defect lines. We shall now address the questions of the uniqueness
of an extension and of the behaviour of the sigma-model action under
the replacement of $\,(\La,X)\,$ by $\,(\La_\eps,X_\eps)$.

Suppose that $\,(\iota_1,\iota_2) : Q\rightarrow M\ti M\,$ is an
immersion (i.e.\ the tangent map is everywhere injective). This is,
in particular, the case if $\,Q\,$ is a submanifold of $\,M\ti M$.\
Then, if an extension of $\,X\,$ on $\,U\,$ exists it is {\em
unique}. To see this, use the local coordinates introduced above,
such that defect $\,\La\,$ lies on the real line and such that their
orientations agree. On the real line itself, $\,\widehat X\,$ is
fixed by $\,X$.\ Set $\,\widehat e_a= \p/\p\si^a\,,\ a=1,2\,$ and
consider \eqref{eq:extend-X-2} for $\,\widehat u_a=\widehat e_a$.\
For a point $\,p=(\si^1,\si^2)\,$ with $\,\si^2>0\,$ (say), we have
$\,\iota_1\circ\widehat X=X$,\ and so $\,(\iota_1\circ\widehat
X)_*\widehat e_2\,$ is fixed. The metric $\,G_{\iota_2\circ\widehat
X(p)}\,$ is still non-degenerate when restricted to the image of
$\,\iota_{2*}$,\ hence condition \eqref{eq:extend-X-2} determines
$\,(\iota_2\circ\widehat X)_*\widehat e_2\,$ uniquely in terms of
$\,(\iota_1\circ\widehat X)_*\widehat e_2\,$ and $\widehat
X_*\widehat e_1$.\ Since $\,(\iota_1,\iota_2)\,$ is an immersion,
this -- in turn -- determines $\,\widehat X_*e_2$.\ If it exists the
solution to the resulting Cauchy problem is unique.

We do not have much to say regarding the existence of an extension
$\,\widehat X$.\ We merely point out that an extension typically
does not exist in the special case of D-branes, as condition
\eqref{eq:pretty-defect-cond} would imply that the classical
energy-momentum tensor vanishes identically on the boundary (this
follows from \eqref{eq:con-def-cond} and \eqref{eq:T-plus-T} below),
while for the jump defects treated in the Lie group example, we
shall see below that an extensions always exists.
\medskip

Next, we compute the difference between the values of the action for
the original circle-field configuration $\,(\La,X)\,$ and its
deformation $\,(\La_\eps,X_\eps)\,$ illustrated in
\eqref{eq:defect-bump}. By a straightforward specialisation of the
calculation from appendix \ref{app:assoc-move}, the change in the
holonomy term of the action is given by the integral of $\,\widehat
X^*\om\,$ over the shaded region $\,A_\eps$.\ Together with a
computation of the change in the kinetic term, this leads to
\qq
&&S[(\G_\eps,X_\eps);\gamma_0]-S[(\G,X);\gamma_0]\cr&&
\label{eq:shift-action-diff}\\
&&=\int_{A_\eps}\,\sfd\si_1\wedge\sfd\si_2\,\Big(\sum_{a=1,2}\,\big[
G_{X_2}(\p_a X_2, \p_a X_2)-G_{X_1}(\p_a X_1,\p_a X_1)\big]-\sfi\,
\om_{\widehat X}(\p_1\widehat X,\p_2\widehat X)\Big)\,,\nonumber
\qqq
where we have abbreviated $\,X_\a=\iota_\a\circ\widehat X\,,\ \a=1,
2$.\ Let $\,D_1\,$ be the left-hand side of condition
\eqref{eq:extend-X-2} for $\,\widehat u_1=\widehat e_1\,,\ \widehat
u_2=\widehat e_2\,,\ v=\widehat X_*\widehat e_2$,\ and let $\,D_2\,$
be the same expression for the choice $\,\widehat u_1=\widehat
e_2\,,\ \widehat u_2=-\widehat e_1\,,\ v=\widehat X_*\widehat e_1$.\
Then, $\,D_2-D_1\,$ is equal to the integrand in
\eqref{eq:shift-action-diff}, and hence the difference between the
values of the action vanishes. Thus, given a circle-field
configuration for which an extension exists, we can shift the
position of the defect line without modifying the value of the
action. This is the hallmark of a topological defect. Indeed,
computing $\,D_1+D_2\,$ results in the identity
\qq\label{eq:T-plus-T}
T_1(p)+\ovl T_1(p)=T_2(p)+\ovl T_2(p)
\qqq
at a point $\,p\in\La$.\ Together with \eqref{eq:con-def-cond}, this
implies that both $\,T\,$ and $\,\ovl T\,$ are continuous across the
defect line, which is the defining property of a topological defect
as given in \cite{Petkova:2000ip}.
\medskip

If the defect under consideration is topological, the symmetry of
the sigma model on a world-sheet with defect circles $\,\La\,$ is
enhanced to include conformal transformations which do not obey
$\,f(\La)=\La$.\ Indeed, if $\,\La'=f_\eps(\La)\,$ and $\,X'=X\circ
f_\eps^{-1}\,$ for an infinitesimal conformal transformation
$\,f_\eps\,$ then -- as we saw at the beginning of the section --
the action for $(\La,X)$ is the same as that for $(\La',X')$, and we
know from the preceeding discussion that we can move the defect
$\,\La'\,$ back to its original position $\,\La$.\ In this manner,
we have produced a new field configuration $\,(\La,X')\,$ with the
same value of the action, where outside of a small neighbourhood of
$\,\La$,\ $\,X'\,$ is related to $\,X\,$ via $\,X'=X\circ
f_\eps^{-1}$.
\medskip

\begin{figure}[bt]

\qq\nonumber
\Si_L \hspace{3.5cm} & \Si_{L|R} & \hspace{3.7cm} \Si_R\cr
   \raisebox{-40pt}{\begin{picture}(85,80)
   \put(0,0){\scalebox{0.55}{\includegraphics{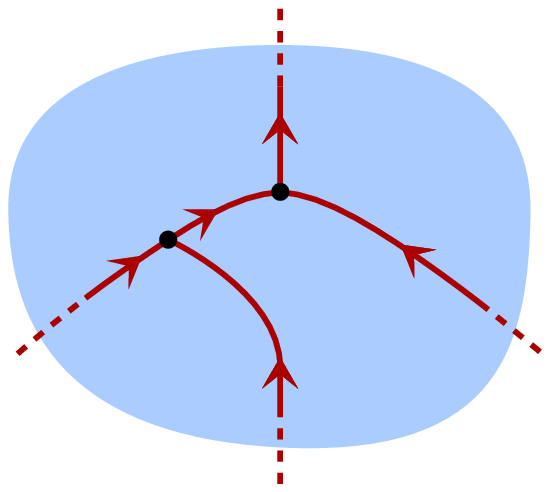}}}
   \put(0,0){
      \setlength{\unitlength}{.55pt}\put(-18,-4){
      \put(102, 95)   {\scriptsize $ v $}
      \put( 70, 90)   {\scriptsize $ \eps^L $}
      \put( 45, 92)   {\scriptsize $ 1 $}
      \put( 64, 42)   {\scriptsize $ 2 $}
      \put(120, 45)   {\scriptsize $ 3 $}
      \put(133, 97)   {\scriptsize $ 4 $}
      }\setlength{\unitlength}{1pt}}
   \end{picture}}
   ~~ \xrightarrow{~~\eps^L\rightarrow 0~~} ~~
   &\raisebox{-40pt}{\begin{picture}(85,80)
   \put(0,0){\scalebox{0.55}{\includegraphics{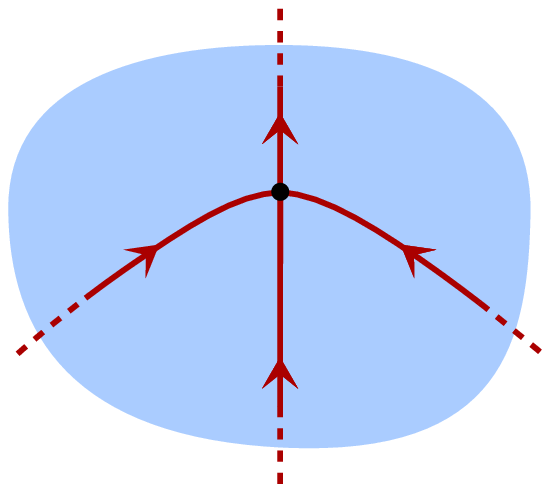}}}
   \put(0,0){
      \setlength{\unitlength}{.55pt}\put(-18,-4){
      \put(102, 95)   {\scriptsize $ v $}
      \put( 45, 92)   {\scriptsize $ 1 $}
      \put( 64, 42)   {\scriptsize $ 2 $}
      \put(120, 45)   {\scriptsize $ 3 $}
      \put(133, 97)   {\scriptsize $ 4 $}
      }\setlength{\unitlength}{1pt}}
   \end{picture}}&
   ~~ \xleftarrow{~~\eps^R\rightarrow 0~~} ~~
   \raisebox{-40pt}{\begin{picture}(85,80)
   \put(0,0){\scalebox{0.55}{\includegraphics{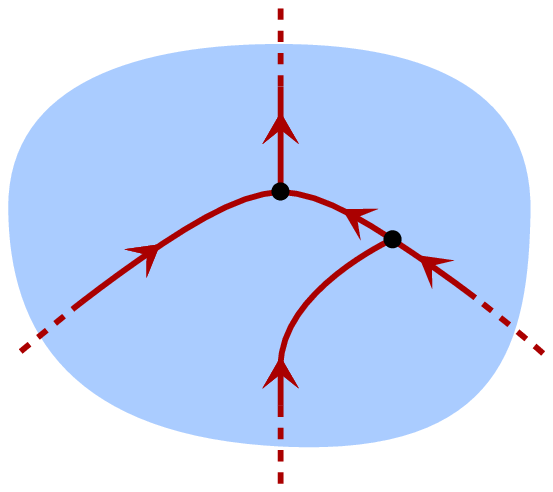}}}
   \put(0,0){
      \setlength{\unitlength}{.55pt}\put(-18,-4){
      \put( 85, 95)   {\scriptsize $ v $}
      \put(112, 90)   {\scriptsize $ \eps^R $}
      \put( 45, 92)   {\scriptsize $ 1 $}
      \put( 64, 42)   {\scriptsize $ 2 $}
      \put(120, 45)   {\scriptsize $ 3 $}
      \put(140, 97)   {\scriptsize $ 4 $}
      }\setlength{\unitlength}{1pt}}
   \end{picture}}
\qqq

\caption{The four-valent defect vertex in $\,\Si_{L|R}\,$ obtained
as a result of collapsing a pair of three-valent vertices in two
inequivalent ways, whereby the two 2-morphisms $\,\til\varphi^L\,$
and $\,\til\varphi^R\,$ are induced at the vertex.}
\label{fig:associator}
\end{figure}

Consider a pair of network-field configurations $\,\bigl(\G_L,X_L
\bigr)\,$ and $\,\bigl(\G_R,X_R\bigr)\,$ with topological defect
conditions at $\,\G_L\,$ and $\,\G_R$,\ differing exclusively within
the region $\,\Si_L\,$ resp. $\,\Si_R\,$ of the world-sheet shown in
the left- resp. rightmost drawing of figure \ref{fig:associator}.
Since the defects are topological, we can take the limits
$\,\eps_L,\eps_R\rightarrow 0\,$ without modifying the value of the
action. Under the assumption of the existence of suitable \v
Cech-extended maps $\,\check\vv^{I,J,K}\ :\ T_{3+1}\too T_{2+1}\,$
with the properties detailed in section \ref{sec:cluster}, we may
readily compare the values attained by the exponentiated sigma-model
action functional $\,\exp \bigl(-S[(\G,X);\g_0]\bigr)\,$ on the two
network-field configurations. After a little thought, one finds that
the value for $\,\bigl(\G_L,X_L\bigr)\,$ is equal to the value that
$\,\exp\bigl(-S[(\G,X);\g_0]\bigr)\,$ takes on the network-field
configuration displayed in the middle drawing of figure
\ref{fig:associator} in which the four-valent defect vertex in
$\,\Si_{L|R}\,$ is understood to carry the pullback data of the
2-morphism $\,\varphi^L_{3+1}\,$ defined in
\eqref{eq:phiL-phiR-def}. By the same token, that for
$\,\bigl(\G_R,X_R\bigr)\,$ is equal to the value that
$\,\exp\bigl(-S[(\G,X);\g_0]\bigr)\,$ takes on the network-field
configuration from the middle drawing but, this time, with the
four-valent defect vertex taken to carry the pullback data of the
2-morphism $\,\varphi^R_{3+1}$.\ Adducing the reasoning of section
\ref{sec:cluster}, we conclude that the two values are related by a
phase as per
\qq\label{eq:expact-LR-phase-rel}
\exp\bigl(-S\bigl[(\G_L,X_L);\g_0\bigr]\bigr)=u(X(v))\cdot\exp
\bigl(-S\bigl[(\G_R,X_R);\g_0\bigr]\bigr)\,,
\qqq
with the function
\qq\label{eq:expact-LR-phase}
u=f^{2,3,4}_{i_v}\cdot\bigl(f^{1,3,4}_{i_v}\bigr)^{-1}\cdot f^{1,2,
4}_{i_v}\cdot\bigl(f^{1,2,3}_{i_v}\bigr)^{-1}\,,
\qqq
expressed in terms of the local data $\,(f^{I,J,K}_i)\in
A^0_{T_{3+1}}\,$ of the induced 2-morphisms
$\,\til\varphi^{I,J,K}$.\ As argued before, $\,u\,$ is constant on
each connected component of $\,T_{3+1}$.

Thus, for classical topological defects with induced data on
$\,T_{3+1}$,\ the operation of pulling one three-valent defect
vertex past another changes the exponentiated action by a phase
determined by the underlying local data
$\,\bigl(T_3,\Oc^{T_3},\varphi_3,\t_3\bigr)$.

\subsubsection*{The Lie-group example (cont'd)}

It is easy to convince oneself that the jump defects introduced
previously satisfy the conditions listed in the preceding section
and hence give us an example of topological defects for the WZW
model. Indeed, this is an immediate consequence of the following
facts: First of all, the extension is fixed as
\qq
(\iota_1,\iota_2)\circ\widehat X|_{U_1}=(X,z^{-1}\cdot X)\,,\qquad
\qquad(\iota_1,\iota_2)\circ\widehat X|_{U_2}=(z\cdot X,X)
\qqq
at the defect line associated with the jump of the embedding field
by $\,z\in Z(\Gx)$.\ Secondly, the curvature $\,\om\,$ of the
$\Gc^{\star\sfk}$-bi-brane $\,\Bc_{Z(\Gx)}\,$ vanishes and the
Cartan--Killing metric on the Lie group is $\Gx$-invariant so that
$\,\D G_{\widehat X(p)}=0\,$ in \eqref{eq:pretty-defect-cond}. Let
us now consider the pair of world-sheets with network-field
configurations $\,(\G_L,X_L)\,$ and $\,(\G_R,X_R)\,$ and jumps
across the defect lines as indicated in figure
\ref{fig:defect-for-cocycle}. Since the data of the
$\bigl(\Gc^{\star\sfk},\Bc_{Z(\Gx)}\bigr)$-inter-bi-brane can be
induced from that for three-valent defect vertices, we derive -- as
a corollary to the general statement
\eqref{eq:expact-LR-phase-rel}-\eqref{eq:expact-LR-phase}, and using
the explicit results for the induced data
$\,\bigl(T_{3+1},\Oc^{T_{3+1}},\varphi^{L,R}_{3+1},\t_{3+1}\bigr)
\,$ from section \ref{sec:cluster} -- the compact relation
\qq
\exp\bigl(-S[(\G_L,X_L);\g_0]\bigr)=\psi_{\Gc^{\star\sfk}}(x,y,
z)\cdot\exp\bigl(-S[(\G_R,X_R);\g_0]\bigr)
\qqq
advertised in the introduction, in which we may now identify the
associator 3-cocycle as the one given by
\eqref{eq:gerbe-assoc-3-cocycle-def}. In the path-integral approach
to the quantisation of the WZW model, an analogous statement could
be inferred for the correlators.

In fact, the 3-cocycle has already appeared in the literature, to
wit, in \cite{Gawedzki:2002se,Gawedzki:2003pm} in order to define
$\Zc$-equivariant gerbes, and in \cite{Jureit:2006yf}, where it was
employed in the path-integral quantisation of the orbifold string
theory. Let us now elaborate on the former point.

Consider a symmetry group $\,S\,$ as at the end of section
\ref{sec:inter-bi-branes} and assume in addition that $\,M\,$ is
connected. We call a homomorphic presentation
$\,(\Ac_S,\tilde\varphi_S)\,$ of $\,S\,$ on $\,b\,$ {\em
associative} if the two 2-morphisms from $\,((x\cdot y).\Ac_z) \circ
(x.\Ac_y) \circ \Ac_x\,$ to $\,\Ac_{x\cdot y\cdot z}\,$ constructed
from $\,\tilde\varphi_S\,$ are equal, or, equivalently, if
$\,\bigl(\d_S\til\varphi\bigr)_{x,y,z}=1\,$ for all $\,x,y,z\in S$.\
Note that, because of
$\,\bigl(\d_S\Ac\bigr)_{x,y}=-D\til\varphi_{x,y}$,\ any two
homomorphic presentations $\,(\Ac_S,\tilde\varphi_S)\,$ and
$\,(\Ac_S,\tilde\varphi'_S)\,$ (with the same underlying
element-wise presentation) are related by a 2-cochain\footnote{
  If we had not assumed $\,M\,$ connected then
  $\,v\,$ would, instead, take values in
  $\,C^2\bigl(S,{\rm U}(1)^{\vert\pi_0(M)\vert}\bigr)$.
} $\,v\in C^2(S,{\rm U}(1))\,$ via $\,\tilde\varphi_{x,y} = v(x,y)
\cdot \tilde\varphi'_{x,y}$.\ Thus, an associative homomorphic
presentation for a given element-wise presentation $\,\Ac_S\,$
exists if and only if
$\,\bigl(\d_S\til\varphi\bigr)_{x,y,z}=\d_Sv(x,y,z)\,$ for some
$\,v\in C^2(S,{\rm U}(1))$,\ where $\,S\,$ acts by the \v
Cech-extended pullback on the local data of $\,\til\varphi_{x,y}$,\
and trivially on $\,v(x,y)$.\ Since
$\,D\d_S\til\varphi=-\d_S^2\Ac=(0,1)$,\ we readily see how the
cohomology class of $\,\psi=\d_S\til\varphi\,$ determines the
obstruction to associativity. Finally, an $S$-equivariant structure
on the gerbe $\,\Gc=(\Oc^M,b)\,$ is an associative homomorphic
presentation of $\,S\,$ on $\,b$.\ It is a prerequisite of
projecting the sigma model on $\,M\,$ to the quotient target space
$\,M/S\,$ (the orbifold) by dividing out the action of the symmetry
group $\,S$,\ see \cite{Gawedzki:2002se,Gawedzki:2003pm}.

From the present point of view, the data needed to define a
classical orbifold consists of a topological bi-brane and an
inter-bi-brane with world-volume $\,T = T_3 \sqcup T_4\,$ which is
associative in the sense that the two limits in figure
\ref{fig:associator} agree. This quite beautifully matches the
construction in \cite{Fuchs:2002cm,Frohlich:2006ch} of a general
rational conformal field theory starting from the Cardy case. There,
one equally fixes a topological defect $\,B\,$ and endows it with an
associative 3-valent vertex. In both cases, the orbifold amplitudes
are obtained by embedding sufficiently fine defect networks into the
world-sheet. In fact, for the CFT one can obtain {\em all} theories
with a given chiral symmetry in this way, including the exceptional
modular invariants \cite{Fuchs:2002cm}.
\medskip

The intermediate steps leading to the explicit form of
$\,\psi_{\Gc^{\star\sfk}}\,$ are rather involved technically (in
particular, the geometric description of the WZW gerbe as a
particular bundle gerbe of \cite{Meinrenken:2002} is used heavily),
which is why we only cite the result that can be read off from
\cite[sect.\,3]{Gawedzki:2003pm} and \cite[sect.\,2]{Gawedzki:2008}.
It is given by
\qq
\psi_{\Gc^{\star\sfk}}(x,y,z)=\exp\bigl(-2\pi\sfi\,\sfk\,\langle
\t_{x^{-1}0},\bx_{y,z}\rangle\bigr)\,,\qquad\qquad x,y,z\in Z(\Gx)
\qqq
for $\,\langle\cdot,\cdot\rangle\,$ the standard scalar product on
the Cartan subalgebra $\,\gt{t}\subset\ggt\,$ (normalised as in
\cite{Gawedzki:2003pm} and employed to identify $\,\gt{t}^*\,$ with
$\,\gt{t}$), $\,\t_{x^{-1}0}\in\gt{t}\,$ the simple coweight of
$\,\ggt\,$ determined, up to an irrelevant element of the coroot
lattice, by the condition\footnote{The condition realises the
isomorphism $\,Z(\Gx)\cong P^\vee(\ggt)/Q^\vee(\ggt)$,\ in which
$\,P^\vee(\ggt)\,$ and $\,Q^\vee(\ggt)\,$ are the coweight lattice
and the coroot lattice of $\,\ggt$,\ respectively.}
\qq
x=\exp\bigl(-2\pi\sfi\,\t_{x^{-1}0}\bigr)\,,
\qqq
and $\,\bx_{y,z}\,$ a particular 2-cocycle on $\,Z(\Gx)\,$ defined
(modulo $\,Q^\vee(\ggt)$) as follows: Let us denote by $\,\a_i\,$
the simple roots of $\,\ggt$,\ by $\,\th\,$ its highest root, and by
$\,\Ac_W(\ggt)\,$ its fundamental Weyl alcove,
\qq
\Ac_W(\ggt)=\bigl\{\ \la\in\gt{t}\quad \big\vert\quad\langle\la,
\th\rangle\leq 1\quad\land\quad\langle\la,\a_i\rangle\geq 0\,,\ i=1,
2,\ldots,{\rm rank}\,\ggt\ \bigr\}\,.
\qqq
The action of the centre $\,Z(\Gx)\,$ on the group $\,\Gx\,$ by
multiplication maps conjugacy classes into conjugacy classes, and
every conjugacy class $\,C\,$ can be represented by a unique element
$\,\t\in\Ac_W(\ggt)\subset\gt{t}\,$ of the fundamental Weyl alcove
of $\,\ggt\,$ such that $\,\exp(2\pi\sfi\,\t)\in C$.\ Accordingly,
the action of $\,Z(\Gx)\,$ induces an affine map $\,\t\mapsto
x.\t\,$ of $\,\Ac_W(\ggt)\,$ to itself, determined by the relation
\qq
x\cdot\exp\bigl(2\pi\sfi\,\t\bigr)=w_x^{-1}\cdot\exp\bigl(2\pi\sfi
\,(x.\t)\bigr)\cdot w_x
\qqq
satisfied by a certain element $\,w_x\,$ of the normaliser
$\,N(T)\,$ of the Cartan subgroup $\,T\subset\Gx$.\ In particular,
$\,\t_{x^{-1}.0}\,$ is the preimage of the weight $\,\t=0\,$ under
this action. The element $\,w_x\,$ is fixed only up to the
multiplication $\,w_x\mapsto t\cdot w_x\,$ by an arbitrary element
$\,t\in T$,\ hence it is only the class $\,[w_x]\in N(T)/T\,$ of
$\,w_x\,$ in the Weyl group $\,N(T)/T\,$ of $\,\Gx\,$ that is
determined uniquely. The assignment $\,x\mapsto [w_x]\,$ is an
injective homomorphism, however, $\,w_x\,$ itself cannot -- in
general -- be chosen to depend multiplicatively on $\,x$,\ that is
we cannot set $\,w_{x\cdot y}\,$ equal to $\,w_x\cdot w_y\,$ for all
$\,x,y\in Z(\Gx)$.\ Nevertheless, the condition $\,w_x\cdot w_y\cdot
w_{x\cdot y}^{-1}\in T\,$ is always satisfied, which leads us to the
definition
\qq
w_x\cdot w_y\cdot w_{x\cdot y}^{-1}=\exp\bigl(2\pi\sfi\,\bx_{x,y}
\bigr)
\qqq
of the 2-cocycle $\,\bx_{x,y}$,\ defined modulo $\,Q^\vee(\ggt)$.\
The action of $\,Z(\Gx)\,$ on $\,\Ac_W(\ggt)\,$ and the elements
$\,\t_{x^{-1}0}, \bx_{x,y}\,$ for all simple Lie groups with a
nontrivial centre were listed in \cite[sect.\ 4]{Gawedzki:2003pm}.
These data were subsequently used to compute the 3-cocycles
$\,\psi_{\Gc^{\star\sfk}}$,\ see also
\cite{Gawedzki:2002se,Gawedzki:2007uz} (we use the conventions of
\cite{Gawedzki:2007uz}, in terms of which
$\,u_{x,y,z}=\psi_{\Gc^{\star\sfk}}(x,y,z)$).

\sect{World-sheets with defect networks in
CFT}\label{sec:def-of-CFT}

In `constructive' conformal field theory, one tries to determine the
correlation functions of the theory from their symmetries and from a
set of consistency relations known as sewing constraints
\cite{Friedan:1986ua,Vafa:1987ea,Sonoda:1988fq}. For oriented closed
conformal field theories, this approach was given a mathematical
framework in \cite{Segal:2002}. In this section, we describe its
straightforward generalisation to surfaces with defect lines and
outline the simplifications that occur for topological defects. We
shall show that if a discrete symmetry group of a CFT is implemented
by defects it gets equipped with an associator 3-cocycle.

\subsection{Sewing constraints for world-sheets with defects}
\label{sec:CFT-sew}

From \cite{Segal:2002}, we know that a convenient way to encode the
sewing constraints is to use the language of functors. We shall
describe a symmetric monoidal category $\,\WD\,$ of `world-sheets
with defect lines' and define a two-dimensional euclidean quantum
field theory in the presence of defect lines as a symmetric monoidal
functor from $\,\WD\,$ to $\,\Tvec$,\ the symmetric monoidal
category of locally convex topological vector spaces (see, e.g., the
foreword to \cite{Segal:2002}, and section 2 in \cite{Stolz}).
\medskip

\begin{figure}[bt]
$$
  \raisebox{-55pt}{\begin{picture}(160,115)
  \put(0,0){\scalebox{0.75}{\includegraphics{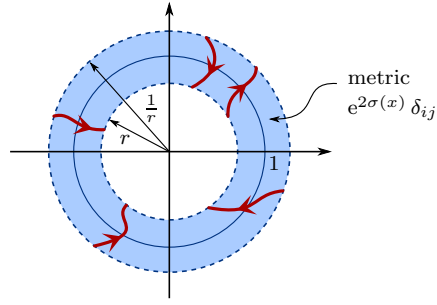}}}
  \put(0,0){
     \setlength{\unitlength}{.75pt}\put(-14,-16){
     \put(145,82)   {\scriptsize$ 1 $}
     \put(70,95)   {\scriptsize$ r $}
     \put(82,110)   {\scriptsize$ \frac1r $}
     \put(185,125)   {\scriptsize\text{metric}}
     \put(185,110)   {\scriptsize$ {\rm e}^{2 \sigma(x)} \, \delta_{ij} $}
     }\setlength{\unitlength}{1pt}}
  \end{picture}}
$$
\caption{An annulus with arcs $\,O=(r,\sigma,L)$.\ Indicated are the
inner and outer radius $\,r\,$ and $\,1/r$,\ the metric defined in
terms of the function $\,\si$,\ and the oriented submanifold $\,L\,$
which describes the defect lines.} \label{fig:ann-arc}
\end{figure}

An {\em annulus with arcs} $\,O\,$ is a triple $\,(r,\si,L)\,$ with
the following constituents (cf.\ figure \ref{fig:ann-arc}):
\begin{list}{-}{\topsep .4em \leftmargin 2.5em \itemsep 0em}
\item[(A.i)] $0<r<1\,$ is a real number. It defines the annulus $\,A_r
= \{\ z\,{\in}\,\Cb\ \big|\ r\,{<}\,|z|\,{<}\,r^{-1}\ \}$.
\item[(A.ii)] $\sigma : A_r \rightarrow \Rb\,$ is a smooth function. It
defines a metric in conformal gauge on $\,A_r\,$ via $\,g_{ij}(x) =
{\rm e}^{2\si(x)}\,\delta_{ij}$.
\item[(A.iii)] $L\,$ is a smooth oriented one-dimensional submanifold of
$\,A_r$.\ If $\,L\,$ has $n$ connected components then, for each
concentric circle $\,C\subset A_r$,\ the intersection $\,C \cap L\,$
is demanded to consist of $n$ points.
\end{list}
Note that we obtain an ordering of the connected components of
$\,L\,$ upon labelling them by $\,1,2,\ldots,n\,$ in the order in
which they intersect the circle $\,|z|=1\,$ starting from the point
$\,z=1$.

Given an annulus with arcs $\,O$,\ by $\,O^+\,$ we mean the subset
$\,\{\ z \in \Cb \ |\ 1 \le |z| < 1/r \ \}\,$ endowed with the
metric and the one-dimensional submanifold inherited from $\,O$,\
and by $\,O^-\,$ we denote the analogous restriction to $\,\{\ z \in
\Cb \ |\ r < |z| \le 1 \ \}$.\ By $\,O_{(m)}\,$ we mean an ordered
list $\,(O_1,O_2,\ldots,O_m)\,$ of a finite number of annuli with
arcs.

\begin{figure}[bt]
$$
  \raisebox{-75pt}{\begin{picture}(300,130)
  \put(0,0){\scalebox{0.55}{\includegraphics{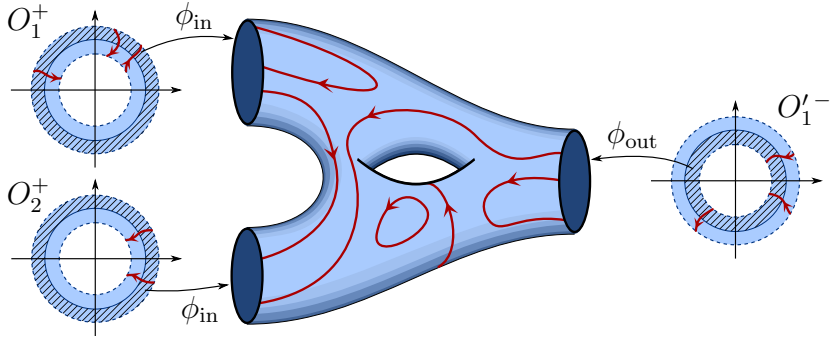}}}
  \put(0,0){
     \setlength{\unitlength}{.55pt}\put(-22,-8){
     \put(136,224)   {$ \phi_\text{in} $}
     \put(140, 22)   {$ \phi_\text{in} $}
     \put(433,142)   {$ \phi_\text{out} $}
     \put(550,160)   {$ O'_{1}{}^- $}
     \put( 20,220)   {$ O_{1}^+ $}
     \put( 20,100)   {$ O_{2}^+ $}
     }\setlength{\unitlength}{1pt}}
  \end{picture}}
$$
\caption{A world-sheet with defect lines from $\,(O_1,O_2)\,$ to
$\,(O'_{1})$.\ The shaded regions of the annuli indicate the subsets
$\,O^+_1\,,\ O^+_2\,$ and $\,O'_{1}{}^-$.\ The maps
$\,\phi_\text{\rm in}\,$ and $\,\phi_\text{\rm out}\,$ are defined
in the shaded regions and map the solid circle $\,|z|=1\,$ to the
boundary of the world-sheet.} \label{fig:wsh-defect}
\end{figure}

A {\em world-sheet with defect lines} $\,\Si\,$ {\em from}
$\,O_{(m)}\,$ {\em to} $\,O_{(n)}'$,\ to be denoted as
$\,O_{(m)}\xrightarrow{\Si}O_{(n)}'\,$ in what follows, is a tuple
$\,(W,L,\phi_{\rm in},\phi_{\rm out})$,\ where (cf.\ figure
\ref{fig:wsh-defect}):
\begin{list}{-}{\topsep .4em \leftmargin 2.5em \itemsep 0em}
\item[(W.i)] $W\,$ is a smooth oriented two-dimensional manifold
with riemannian metric, possibly with a non-empty boundary.
\item[(W.ii)] $L\,$ is a smooth oriented one-dimensional submanifold
of $\,W$.
\item[(W.iii)] $\phi_{\rm in}\,$ is a smooth injective isometry from the disjoint
union $\,O_1^+\sqcup O_2^+\sqcup\cdots\sqcup O_m^+\,$ to $\,W\,$
which preserves the orientation, the boundaries, and the
one-dimensional submanifolds with their orientation.
\item[(W.iv)] $\phi_{\rm out}\,$ is a smooth injective isometry from the disjoint
union $\,{O_1'}^-\sqcup {O_2'}^-\sqcup\cdots\sqcup {O_n'}^-\,$ to
$\,W\,$ with the same properties as in (W.iii).
\end{list}
We refer to the boundary components of $\,W\,$ in the image of
$\,\phi_\text{in}\,$ as {\em in-going}, and to those in the image of
$\,\phi_\text{out}\,$ as {\em out-going}. A {\em defect line} is a
connected component of $\,L$.\ Note that $\,\phi_\text{in}\,$
induces a numbering of the in-going boundary components by assigning
the number $k$ to the component which lies in
$\,\phi_\text{in}(O_k^+)$.\ Similarly, out-going boundary components
are numbered by $\,\phi_\text{out}$.

Given world-sheets $\,O_{(k)}\xrightarrow{\Si_1}O_{(l)}'\,$ and
$\,O_{(l)}'\xrightarrow{\Si_2}O_{(m)}''$,\ we can obtain the glued
world-sheet $\,\Si_2\circ\Si_1\,$ by identifying the boundaries
parameterised by $\,O_{(l)}'$.\ The fact that we work with annuli
and arcs instead of just circles and marked points ensures that the
gluing results again in a smooth manifold with a smooth metric, and
a smooth submanifold.

Two world-sheets with defect lines are {\em equivalent} if there is
a smooth orientation-preserving isometry between them that is
compatible with the parameterisations
$\,\phi_\text{in/out}\,$ and
preserves the one-dimensional submanifolds with their orientation.
\medskip

We can now describe the category $\,\WD$.\ The objects\footnote{We
should really define the objects to be germs of annuli with arcs
because we can always restrict an annulus $\,O = (r,\si,L)\,$ to one
with a smaller radius $\,r'<r$,\ and this should not affect the
amplitude of the QFT. We have avoided this point to make the
exposition less technical.} of $\,\WD\,$ are ordered lists
$\,O_{(m)}$.\ The morphisms from $\,O_{(m)}\,$ to $\,O_{(n)}'\,$ are
equivalence classes $\,[\Si]\,$ of world-sheets from $\,O_{(m)}\,$
to $\,O_{(n)}'$,\ and, if $\,m=n$,\ all $\,\pi\in S_m\,$ (the group
of permutations of $m$ objects) for which $\,O_i = O'_{\pi(i)}\,,\
i=1,2,\ldots,m$.\ The permutations account for the freedom to choose
a different numbering of the boundary components of a world-sheet
$\,\Si$.\ The four possible compositions are defined as follows:
\begin{list}{-}{\topsep .4em \leftmargin 2.5em \itemsep 0em}
\item[] $O_{(k)} \overset{[\Si_1]}{\longrightarrow} O_{(l)}'
\overset{[\Si_2]}{\longrightarrow} O_{(m)}''\,$ is the equivalence
class of the glued world-sheet $\,[\Si_2\circ\Si_1]$;
\item[] $O_{(k)} \overset{\pi}{\longrightarrow} O_{(k)}'
\overset{[\Si_2]}{\longrightarrow} O_{(m)}''\,$ is defined by
precomposing the parameterisation $\,\phi_\text{in}\,$ with $\,\pi$;
\item[] $O_{(k)} \overset{[\Si_1]}{\longrightarrow} O_{(l)}'
\overset{\pi}{\longrightarrow} O_{(l)}''\,$ is defined by
precomposing the parameterisation $\,\phi_\text{out}\,$ with
$\,\pi^{-1}$;
\item[] $O_{(k)} \overset{\pi_1}{\longrightarrow} O_{(k)}'
\overset{\pi_2}{\longrightarrow} O_{(k)}''\,$ is the composition of
permutations $\,\pi_2 \circ \pi_1$.
\end{list}
Since we are using equivalence classes of world-sheets, the
composition is strictly associative. The identity morphism of
$\,O_{(m)}\,$ is the identity permutation. The tensor product is the
concatenation of lists on objects and disjoint union on morphisms.
Both will be written as $\,\sqcup$.\ The symmetry isomorphism
$\,O_{(m)} \sqcup O_{(n)}' \rightarrow O_{(n)}' \sqcup O_{(m)}\,$ is
the obvious permutation $\,\pi\in S_{m+n}$.
\medskip

Having said all this, we define a {\em euclidean quantum field
theory with defect lines} as a symmetric monoidal functor $\,C : \WD
\rightarrow \Tvec\,$ which depends continuously on the world-sheet
metric and on the position of the defect lines.
\medskip

Let us unpack this definition. To each annulus with arcs $\,O$,\ the
functor assigns a space of states $\,C(O)=H(O)$.\ Since $\,C\,$ is
monoidal, we have $\,C(O_{(m)})=H(O_1)\otimes H(O_2)\otimes\cdots
\otimes H(O_m)$.\ The empty list $\,O=()\,$ is the tensor unit of
$\,\WD$,\ and, accordingly, we have $\,C(O)=\Cb$,\ the tensor unit
of $\,\Tvec$.\ Given a morphism $\,[\Si] : O_{(m)} \rightarrow
O_{(n)}'$,\ the functor provides a linear map
\be
C(\Si)\ :\ H(O_1)\otimes H(O_2)\otimes\cdots\otimes H(O_m)
\longrightarrow H(O_1')\otimes H(O_2')\otimes\cdots\otimes H(O_n')
\,,
\ee
the amplitude for the world-sheet $\,\Si$.\ As the morphisms are
equivalence classes of world-sheets, equivalent world-sheets have to
give the same amplitude. That $\,C\,$ is monoidal on morphisms
implies that $\,C(\Si\sqcup\Si')=C(\Si)\otimes C(\Si')$,\ and the
symmetry of $\,C\,$ implies that changing the numbering of the
boundary components of $\,\Si\,$ translates into the corresponding
relabelling of the arguments of the linear map $\,C(\Si)$.\ The most
non-trivial condition is the compatibility with composition, which
amounts to the insertion of a sum over intermediate states in the
path-integral language,
\be
O_{(k)} \overset{[\Si_1]}{\longrightarrow} O_{(l)}'
\overset{[\Si_2]}{\longrightarrow} O_{(m)}'' \qquad \Longrightarrow
\qquad
  C(\Si_2 \circ \Si_1) = C(\Si_2) \circ C(\Si_1)\,.
\labl{eq:C-functorial}

In general, it will be very difficult to construct examples of such
a symmetric monoidal functor $\,C : \WD \rightarrow \Tvec$.\
However, for a special subclass of defects in conformal field
theories, the so-called topological defects, further progress can be
made. This is the topic of the next section.

\subsection{Topological defects in conformal field theory}
\label{sec:CFT-topdef}

Let $\,\Si = (W,L,\phi_\text{in},\phi_\text{out})\,$ and $\,\Si' =
(W,L',\phi_\text{in},\phi_\text{out})\,$ be two world-sheets which
differ only in the choice of defect lines. We say that $\,\Si\,$ and
$\,\Si'\,$ have {\em homotopic defect lines} if $\,L\,$ and $\,L'\,$
are homotopic (as oriented paths) via a homotopy that is constant on
the image of $\,\phi_\text{in}\,$ and on that of
$\,\phi_\text{out}$.\ We call the defects in a 2d-QFT {\em
topological} if $\,C(\Si) = C(\Si')\,$ whenever $\,\Si\,$ and
$\,\Si'\,$ have homotopic defect lines.

Recall that a 2d-QFT is conformal if an amplitude $\,C(\Si)\,$
changes only by an overall factor upon applying a Weyl
transformation $\,\g(x)\mapsto\g'(x)=\Omega(x)\cdot\g(x)\,$ to the
metric (where $\,\Omega \equiv 1\,$ on the image of
$\,\phi_\text{in}\,$ and on that of $\,\phi_\text{out}$). The factor
is computed in terms of the Liouville action and the central charge,
see, e.g., \cite{Gawedzki96lectures} for more details.
\medskip

For a 2d-CFT with topological defects, the functor $\,C\,$
simplifies in two significant ways. First, if $\,O=(r,\si,L)\,$ then
$\,H(O)\,$ does not depend on $\,r\,$ or $\,\si$,\ and it depends on
$\,L\,$ only through the number $n$ of points in the intersection of
$\,L\,$ with the unit circle, and on $n$ signs $\,\eps^{k,k+1}\,,\
k=1,2,\ldots,n$.\ The sign $\,\eps^{k,k+1}\,$ is $+1$ if the $k$-th
connected component of $\,L\,$ is oriented from the outside of the
unit circle to the inside. Otherwise, $\,\eps^{k,k+1}=-1$.\ To
specify $\,C\,$ on objects of $\,\WD$,\ it is thus enough to give
vector spaces
\be
  H_{n,\vec\eps}
  \,,\qquad\qquad
  n \in \Zb_{\ge 0}\,,\qquad
  \vec\eps = \big\{\ \eps^{k,k+1}\,{\in}\,\{\pm 1\}
  \ \big|\ k\,{=}\,1,\dots,n \ \big\}\,.
\ee
Elements of $\,H_{n>0,\vec\eps}\,$ will be called {\em twisted
states}, and those of $\,H_{n=0}\,$ {\em untwisted states}, in
conformity with the physical jargon.

To fix $\,C\,$ on world-sheets, it is enough to give it on a set of
fundamental world-sheets from which all others can be obtained via
gluing. As opposed to the theory without defects, we now need an
infinite set of fundamental world-sheets. One possible choice is
\bea
  D_i = \raisebox{-28pt}{\begin{picture}(56,56)
  \put(0,0){\scalebox{0.55}{\includegraphics{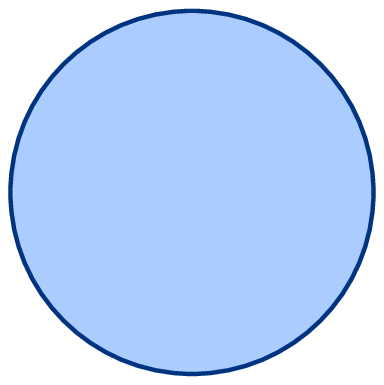}}}
  \put(0,0){
     \setlength{\unitlength}{.55pt}\put(-19,-18){
     \put(112,112)   {\scriptsize in }
     }\setlength{\unitlength}{1pt}}
  \end{picture}}
  \quad , \qquad
  A_{oo} = \raisebox{-28pt}{\begin{picture}(56,56)
  \put(0,0){\scalebox{0.55}{\includegraphics{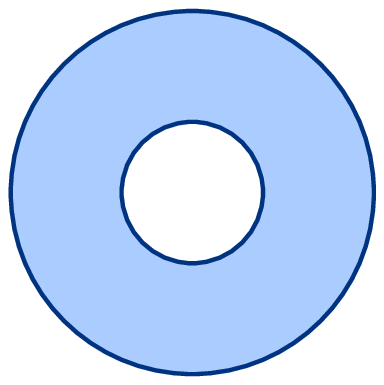}}}
  \put(0,0){
     \setlength{\unitlength}{.55pt}\put(-19,-18){
     \put(112,112)   {\scriptsize out }
     \put( 62, 78)   {\scriptsize out }
     }\setlength{\unitlength}{1pt}}
  \end{picture}}
  \quad ,
\enl
  \hspace{5em}
  A^D_{ii} = \raisebox{-28pt}{\begin{picture}(56,56)
  \put(0,0){\scalebox{0.55}{\includegraphics{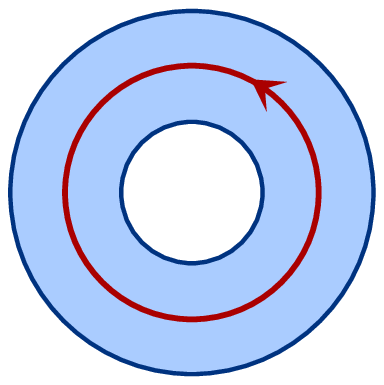}}}
  \put(0,0){
     \setlength{\unitlength}{.55pt}\put(-19,-18){
     \put(112,112)   {\scriptsize in }
     \put( 62, 76)   {\scriptsize in }
     }\setlength{\unitlength}{1pt}}
  \end{picture}}
  \quad , \qquad
  P(n,m,k;L) = \raisebox{-35pt}{\begin{picture}(70,70)
  \put(0,0){\scalebox{0.55}{\includegraphics{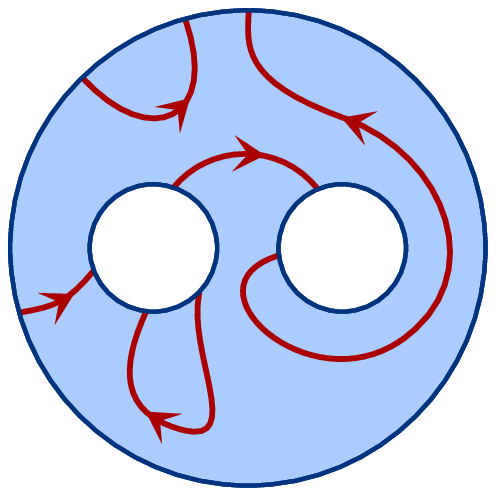}}}
  \put(0,0){
     \setlength{\unitlength}{.55pt}\put(-14,-14){
     \put( 50, 87)   {\scriptsize in }
     \put(103, 87)   {\scriptsize in }
     \put(127,143)   {\scriptsize in }
     }\setlength{\unitlength}{1pt}}
  \end{picture}}\quad.
\eear\labl{eq:def-generate}
In $\,P(n,m,k;L)$,\ the integers $\,n,m,k\in \Zb_{\ge 0}\,$
designate how many defect lines end on each of the three boundary
circles, and $\,L\,$ is the corresponding set of defect lines. The
defect lines are not allowed to contain closed loops (these are
already generated by $\,A^D_{ii}$). Note that, even for fixed
$\,n,m,k$,\ there are an infinite number of possibilities for
$\,L\,$ as a defect line can wind an arbitrary number of times
around one of the holes.
\medskip

In \cite{Sonoda:1988fq,Lewellen:1991tb}, a generators-and-relations
approach to closed and open/closed conformal field theory is given.
In both cases, a finite number of generators and relations are
sufficient. In the presence of defects, already the number of
amplitudes one needs to fix for the fundamental world-sheets
\eqref{eq:def-generate} is infinite, and the list of sewing
constraints that need to be satisfied to allow a consistent
definition of the amplitudes for more complicated world-sheets is
infinite as well. A concrete set of sufficient sewing constraints
has not been worked out to date.

However, there exists an alternative approach to determine the
functor $\,C : \WD \rightarrow \Tvec\,$ for a conformal field theory
with topological defect lines \cite{Fuchs:2002cm,Frohlich:2006ch}.
This approach applies to rational conformal field theories and uses
an associated three-dimensional topological field theory. In the
case of the WZW model, this is just the three-dimensional
Chern--Simons theory \cite{Witten:1988hf,Frohlich:1989gr}. In the
TFT-approach, one makes a proposal for all $\,C(\Si)\,$
simultaneously and then verifies that this, indeed, defines a
symmetric monoidal functor. (Admittedly, a complete proof of this
statement along the lines of \cite{Fjelstad:2005ua} is not yet
available.) The data that determine $\,C\,$ are a rational
vertex-operator algebra $\,\mathcal{V}$,\ a symmetric special
Frobenius algebra $\,A\,$ in the category
$\,\mathrm{Rep}(\mathcal{V})\,$ of representations of $\,\Vc$,\ and
an $A$-$A$-bimodule $\,Q\,$ in $\,\mathrm{Rep}(\mathcal{V})$.\ We
refer to \cite{Frohlich:2006ch} for details; we do not need this
general approach in the present paper. However, let us point out
that in the special case of $\,\mathcal{V} = \Cb$,\ i.e.\ for a
two-dimensional {\em topological} field theory with topological
defect lines, the resulting algebraic structure is very similar to
that of a planar algebra \cite{Jones:1999}.
\medskip

In order to prepare the subsequent discussion of symmetries
implemented by defects, we need to assume some further properties of
$\,C$.\ These are satisfied in the WZW model studied in section
\ref{sec:CFT-cocycle}.

Consider the world-sheet $\,A_{n,\vec\eps}^r\,$ given by an annulus
of outer radius one and inner radius $\,r\,$ with $n$ rays of defect
lines, having orientations given by a list $\,\vec\eps =
(\eps^{1,2},\eps^{2,3},\dots,\eps^{n,1})$,\ e.g.,
\be
A_{5,+--++}^r =  \raisebox{-31pt}{\begin{picture}(100,65)
  \put(20,0){\scalebox{0.55}{\includegraphics{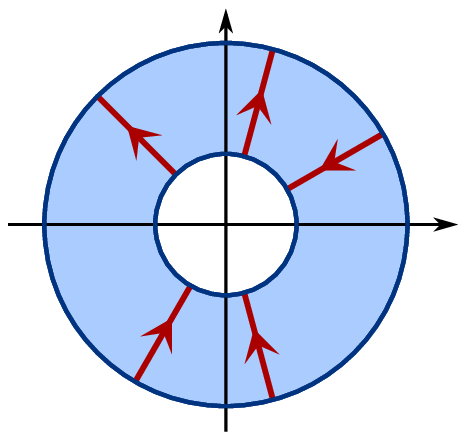}}}
  \put(20,0){
     \setlength{\unitlength}{.55pt}\put(-7,-8){
     \put( 94, 58)   {\scriptsize $r$ }
     \put(126, 56)   {\scriptsize $1$ }
     \put(118,100)   {\scriptsize $\eps^{1,2}=+$ }
     \put( 90,122)   {\scriptsize $\eps^{2,3}=-$ }
     \put(-20,110)   {\scriptsize $\eps^{3,4}=-$ }
     \put(-10, 11)   {\scriptsize $\eps^{4,5}=+$ }
     \put( 85,  6)   {\scriptsize $\eps^{5,1}=+$ }
     }\setlength{\unitlength}{1pt}}
  \end{picture}}\quad.
\ee
We assume that the `propagator'
\be
C(A_{n,\vec\eps}^r)\ :\ H_{n,\vec\eps} \rightarrow H_{n,\vec\eps}
\labl{eq:n-propagator}
is invertible. If we are given an eigenvector $\,\phi\,$ of
$\,C(A_{n,\vec\eps}^r)\,$ such that
\be
C(A_{n,\vec\eps}^r) \, \phi = r^{\Delta_\phi}\, \phi\,,
\ee
with $\,\D_\phi\,$ the conformal weight of $\,\phi$,\ we can define
a {\em field insertion} $\,\phi\,$ to mean
\be
  C\Bigg(~ \raisebox{-23pt}{\begin{picture}(70,52)
  \put(0,0){\scalebox{0.55}{\includegraphics{pic19b.eps}}}
  \put(0,0){
     \setlength{\unitlength}{.55pt}\put(-28,-16){
     \put( 99, 56)   {\scriptsize $\phi$ }
     }\setlength{\unitlength}{1pt}}
  \end{picture}} ~\Bigg)
  ~:=~ r^{-\Delta_\phi} \cdot
  C\Bigg(~ \raisebox{-26pt}{\begin{picture}(70,52)
  \put(0,0){\scalebox{0.55}{\includegraphics{pic19a.eps}}}
  \put(0,0){
     \setlength{\unitlength}{.55pt}\put(-28,-16){
     \put( 75, 65)   {\scriptsize $\phi$ }
     \put(100, 63)   {\scriptsize $r$ }
     }\setlength{\unitlength}{1pt}}
  \end{picture}} ~\Bigg)\,.
\ee
The left-hand side shows a fragment of a world-sheet with the
insertion, and the right-hand side, in which we have drawn a
world-sheet $\,\Si_\phi\,$ with the corresponding hole marked by
$\,\phi$,\ means that the argument of the linear operator
$\,C(\Si_\phi)\,$ corresponding to the (marked) in-going boundary
shown in the figure is set to $\,\phi$.\ The gluing properties of
$\,C\,$ imply that this definition is independent of $\,r$.\ Even if
not made explicit in the notation, a field insertion by definition
carries a local coordinate system since it corresponds to a small
parameterised hole.
\medskip

Denote by $\,T,\ovl T\in H_{0}\,$ the holomorphic and
anti-holomorphic components of the energy-momentum tensor. We demand
that topological defects commute with $\,T\,$ and $\,\ovl T\,$ in
the sense that
\be
  C\Bigg(~ \raisebox{-28pt}{\begin{picture}(56,56)
  \put(0,0){\scalebox{0.55}{\includegraphics{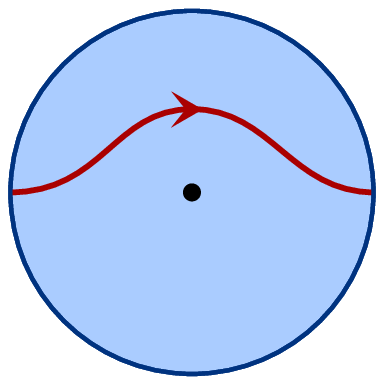}}}
  \put(0,0){
     \setlength{\unitlength}{.55pt}\put(-19,-18){
     \put(112,112)   {\scriptsize out }
     \put( 78, 64)   {\scriptsize $\phi$ }
     }\setlength{\unitlength}{1pt}}
  \end{picture}} ~~~\Bigg)
  ~=~
  C\Bigg(~ \raisebox{-28pt}{\begin{picture}(56,56)
  \put(0,0){\scalebox{0.55}{\includegraphics{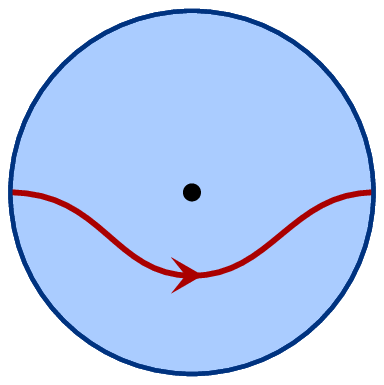}}}
  \put(0,0){
     \setlength{\unitlength}{.55pt}\put(-19,-18){
     \put(112,112)   {\scriptsize out }
     \put( 78, 64)   {\scriptsize $\phi$ }
     }\setlength{\unitlength}{1pt}}
  \end{picture}} ~~~\Bigg)
  \qquad \text{for}~\quad\phi = T~\quad\text{or}\quad~\phi = \ovl
  T\,.
\labl{eq:def-com-T}
This is, in fact, the original definition of topological defects
\cite{Petkova:2000ip} (the name itself was introduced in
\cite{Bachas:2004sy}). For the more general {\em conformal} defects,
treated, e.g., in \cite{Bachas:2001vj,Quella:2006de,Bachas:2007td},
condition \eqref{eq:def-com-T} does not have to hold. The
topological defects in the WZW model we shall be interested in
satisfy property \eqref{eq:def-com-T} also for the Ka\v c--Moody
currents $\,\phi = J^a,\ovl  J^a$.

By virtue of \eqref{eq:def-com-T}, we have an action of
$\,\mathrm{Vir}\oplus\mathrm{Vir}\,$ on each of the state spaces
$\,H_{n,\vec\eps}$.\ We can, in particular, use the operators
$\,L_0\,$ and $\,\ovl  L_0\,$ to make \eqref{eq:n-propagator}
explicit,
\be
C(A^r_{n,\vec\eps}) = r^{L_0+\ovl  L_0}\,.
\ee
We shall be interested in the subspace $\,H_{n,\vec\eps}^{(0)}\,$ of
$\,H_{n,\vec\eps}\,$ consisting of the $\gt{sl}(2,\Cb)$-invariant
states,
\be
  H_{n,\vec\eps}^{(0)}
  = \Big( \bigcap_{m=0,\pm 1}\, \mathrm{ker}L_m\big|_{H_{n,\vec\eps}} \Big)
  \cap \Big( \bigcap_{m=0,\pm 1}\, \mathrm{ker}\ovl  L_m\big|_{H_{n,\vec\eps}} \Big)\,.
\ee
Since an element of $\,H_{n,\vec\eps}^{(0)}\,$ is annihilated by the
generators of translations, $\,L_{-1}\,$ and $\,\ovl  L_{-1}$,\ an
amplitude with an insertion of $\,\phi \in H_{n,\vec\eps}^{(0)}\,$
is independent of the insertion point, e.g., for $\,\phi \in
H_{3,+--}^{(0)}\,$ and $\,\phi' \in H_{4,-+++}^{(0)}$,
\be
  C\Bigg(~ \raisebox{-28pt}{\begin{picture}(56,56)
  \put(0,0){\scalebox{0.55}{\includegraphics{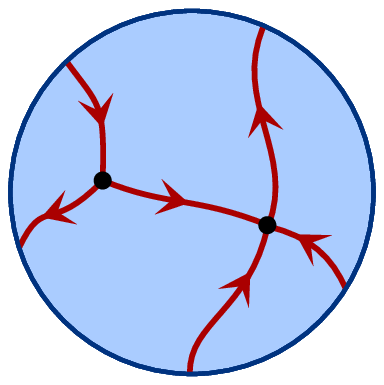}}}
  \put(0,0){
     \setlength{\unitlength}{.55pt}\put(-19,-18){
     \put(112,112)   {\scriptsize out }
     \put( 32, 78)   {\scriptsize $\phi$ }
     \put(102, 68)   {\scriptsize $\phi'$ }
     }\setlength{\unitlength}{1pt}}
  \end{picture}} ~~~\Bigg)
  ~=~
  C\Bigg(~ \raisebox{-28pt}{\begin{picture}(56,56)
  \put(0,0){\scalebox{0.55}{\includegraphics{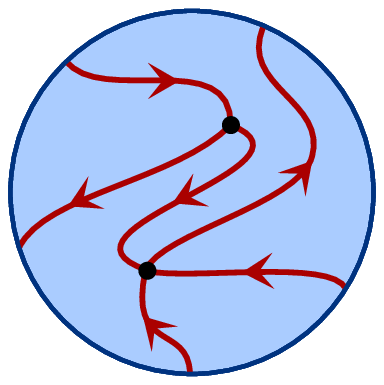}}}
  \put(0,0){
     \setlength{\unitlength}{.55pt}\put(-19,-18){
     \put(112,112)   {\scriptsize out }
     \put( 88, 93)   {\scriptsize $\phi$ }
     \put( 42, 36)   {\scriptsize $\phi'$ }
     }\setlength{\unitlength}{1pt}}
  \end{picture}} ~~~\Bigg)\,.
\ee
Consider the following two world-sheets:
\be
  D^D = \raisebox{-28pt}{\begin{picture}(56,56)
  \put(0,0){\scalebox{0.55}{\includegraphics{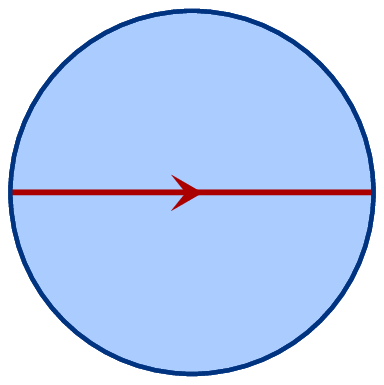}}}
  \put(0,0){
     \setlength{\unitlength}{.55pt}\put(-19,-18){
     \put(112,112)   {\scriptsize out }
     }\setlength{\unitlength}{1pt}}
  \end{picture}}
  \quad,\qquad \qquad
  M^D = \raisebox{-35pt}{\begin{picture}(75,75)
  \put(0,0){\scalebox{0.55}{\includegraphics{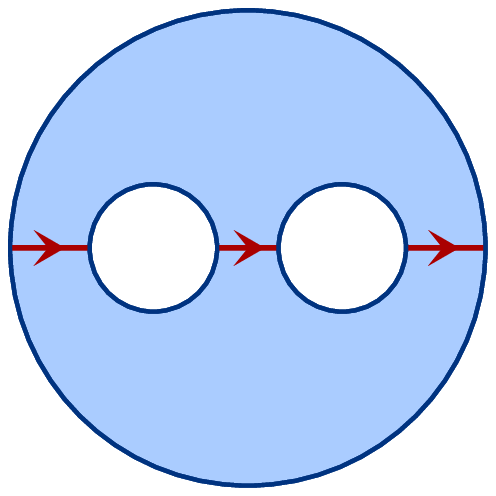}}}
  \put(0,0){
     \setlength{\unitlength}{.55pt}\put(-14,-14){
     \put( 44, 84)   {\scriptsize in,2 }
     \put( 99, 84)   {\scriptsize in,1 }
     \put(127,143)   {\scriptsize out }
     }\setlength{\unitlength}{1pt}}
  \end{picture}}  \quad.
\ee
Let us abbreviate $\,A^D = H_{2,-+}^{(0)}$.\ Define, for $\,a,b \in
A^D$,
\be
  \one^D = C(D^D)\,1
  \,,\qquad \qquad
  m^D(a,b) = C(M^D)(a \otimes b)\,.
\ee
The notation $\,C(D^D)\,1\,$ refers to the fact that $\,[D^D]\,$ is
a morphism from the empty list to $\,O_{(1)}$,\ which the functor
$\,C\,$ takes to a linear map $\,C(D^D) : \Cb \rightarrow
H_{2,-+}$.\ We evaluate the map on $\,1\,$ to get an element of
$\,H_{2,-+}$.

By \eqref{eq:def-com-T}, we have $\,\one^D \in A^D\,$ and also
$\,m^D(a,b) \in A^D$.\ Using the gluing property and the fact that
the elements of $\,A^D\,$ are $\gt{sl}(2,\Cb)$-invariant, one
verifies that $\,\one^D\,$ and $\,m^D\,$ turn $\,A^D\,$ into an
associative unital algebra. That is, for $\,a,b,c \in A^D$,\ we have
\be
  m^D(\one^D,a) = a = m^D(a,\one^D)
  \,,\qquad \qquad
  m^D(a,m^D(b,c)) = a = m^D(m^D(a,b),c)\,.
\ee
The vector $\,\one^D\,$ can be understood as a twisted vacuum state,
or as the identity field on the defect $\,D$.\ We also define the
untwisted vacuum to be simply the correlator of the unit disc
without defect lines, evaluated on $\,1 \in \Cb$,
\be
  \one =
    C\Bigg(~
  \raisebox{-20pt}{\begin{picture}(50,45)
  \put(0,0){\scalebox{0.45}{\includegraphics{pic03a.eps}}}
  \put(0,0){
     \setlength{\unitlength}{.45pt}\put(-20,-19){
     \put(109,120)   {\scriptsize out }
     }\setlength{\unitlength}{1pt}}
  \end{picture}}
  ~\Bigg)\,1 ~ \in ~ H^{(0)}_0\,.
\ee

\subsection{Symmetries implemented by defects}
\label{sec:CFT-group}

Topological defects can implement symmetries of the CFT. This leads
to the notion of `group-like defects'
\cite{Frohlich:2004ef,Frohlich:2006ch}, where one has one such
defect for each element of the symmetry group. In the approach taken
here, we would only have a single type of the defect line, which in
the language of \cite{Frohlich:2006ch} would be a superposition of
all group-like defects.

In the remainder of this section, we explain the notion of a
symmetry that is implemented by defects using the framework
developed in the previous two sections.
\medskip

Let $\,S\,$ be a finite group. We demand that the space $\,A^D =
H^{(0)}_{2,-+}\,$ has a basis $\,\{\ p_g \ |\ g \in S \ \}\,$ such
that
\be\label{eq:group-H02-basis}
  \sum_{g \in S}\, p_g = \one^D
  \qquad \text{and} \qquad
  m^D(p_g,p_h) = \delta_{g,h} \, p_g\,.
\ee
In the approach of \cite{Frohlich:2006ch}, $\,p_g\,$ can be
understood as projectors onto the individual group-like defects.
Consider the annulus $\,A_{n,\vec\eps}^r\,$ with projectors
$\,p_{g_1},p_{g_2},\ldots,p_{g_n}\,$ inserted on the defect lines,
\be
    A_{n,\vec\eps}^r(g_1,\dots,g_n) ~= ~
    \raisebox{-36pt}{\begin{picture}(80,80)
  \put(0,0){\scalebox{0.70}{\includegraphics{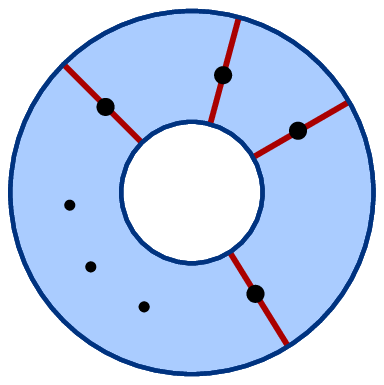}}}
  \put(0,0){
     \setlength{\unitlength}{0.70pt}\put(-17,-15){
     \put(  8, 30)   {\scriptsize out }
     \put( 58, 60)   {\scriptsize in }
     \put(103, 77)   {\scriptsize $p_{g_1}$ }
     \put(118, 97)   {\scriptsize $\eps^{1,2}$ }
     \put( 84,101)   {\scriptsize $p_{g_2}$ }
     \put( 83,124)   {\scriptsize $\eps^{2,3}$ }
     \put( 49,100)   {\scriptsize $p_{g_3}$ }
     \put( 15,111)   {\scriptsize $\eps^{3,4}$ }
     \put( 74, 33)   {\scriptsize $p_{g_n}$ }
     \put( 95, 15)   {\scriptsize $\eps^{n,1}$ }
     }\setlength{\unitlength}{1pt}}
  \end{picture}}\quad.
\ee
We define the linear maps
\be
  P_{n,\vec\eps}(g_1,g_2,\ldots,g_n) =
  C(A_{n,\vec\eps}^r(g_1,g_2,\ldots,g_n)) \, r^{-L_0-\ovl  L_0}\,.
\ee
One verifies, using the gluing properties and
\eqref{eq:group-H02-basis}, that
$\,P_{n,\vec\eps}(g_1,g_2,\ldots,g_n)\,$ are independent of $\,r\,$
and obey
\be
  P_{n,\vec\eps}(g_1,g_2,\ldots,g_n)
  P_{n,\vec\eps}(h_1,h_2,\ldots,h_n)
  = \d_{g_1,h_1}\,\d_{g_2,h_2}\cdots\d_{g_n,h_n}\,
  P_{n,\vec\eps}(g_1,g_2,\ldots,g_n)\,.
\ee
We now impose the condition that a twisted state space contain an
$\gt{sl}(2,\Cb)$-invariant vacuum state only if the overall twist is
trivial, and that the vacuum is unique in this case,
\be
  \dim \Im\big(P_{n,\vec\eps}(g_1,g_2,\ldots,g_n)
  \big|_{H^{(0)}_{n,\vec\eps}}\big) = \begin{cases}
  ~1 & \quad\textrm{if}\quad\prod_{i=1}^n\, g_i^{\eps^{i,i+1}} = e \\
  ~0 & \quad\text{otherwise}
  \end{cases}\,.
\ee
Choose non-zero vectors $\,\varphi_{g,h}\,$ in the image of
$\,P_{3,-++}(g\cdot h,g,h)\,$ applied to $\,H^{(0)}_{3,-++}$.\ Then,
the sum
\be
  \varphi = \sum_{g,h\in S}\,\varphi_{g,h}
\labl{eq:phi-choice}
obeys the condition
\qq
P_{3,-++}(g\cdot h,g,h)\,\varphi = \varphi_{g,h} \neq 0
\qqq
for all $\,g,h\in S$.\ We shall use $\,\varphi\,$ to label all
three-valent junctions with two incoming defect lines and one
outgoing defect line. We demand that there exist a vector $\,\ovl
\varphi \in H^{(0)}_{3,+--}\,$ such that the following two
non-degeneracy conditions for the defect correlators are satisfied
(only the third one involves $\,\ovl \varphi$)
\bea
C\Bigg(
  \raisebox{-25pt}{\begin{picture}(60,55)
  \put(0,0){\scalebox{0.55}{\includegraphics{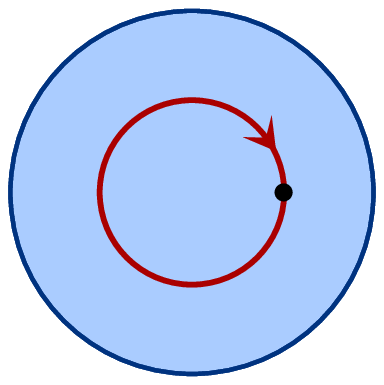}}}
  \put(0,0){
     \setlength{\unitlength}{.55pt}\put(-20,-19){
     \put(106,117)   {\scriptsize out }
     \put(105, 69)   {\scriptsize $p_g$ }
     }\setlength{\unitlength}{1pt}}
  \end{picture}}
  \Bigg) = \chi(g) \, \one
  \,, \qquad\qquad
  C\Bigg(
  \raisebox{-25pt}{\begin{picture}(60,55)
  \put(0,0){\scalebox{0.55}{\includegraphics{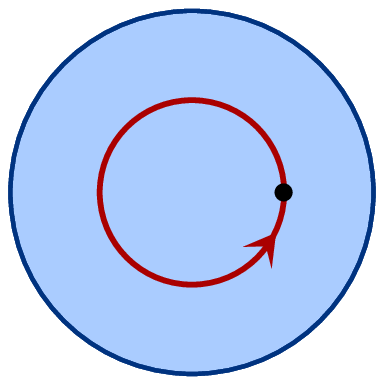}}}
  \put(0,0){
     \setlength{\unitlength}{.55pt}\put(-20,-19){
     \put(106,117)   {\scriptsize out }
     \put(105, 69)   {\scriptsize $p_g$ }
     }\setlength{\unitlength}{1pt}}
  \end{picture}}
  \Bigg) = \chi(g^{-1}) \, \one
  \,,
\\[3em] \displaystyle
  C\Bigg(
  \raisebox{-25pt}{\begin{picture}(60,55)
  \put(0,0){\scalebox{0.55}{\includegraphics{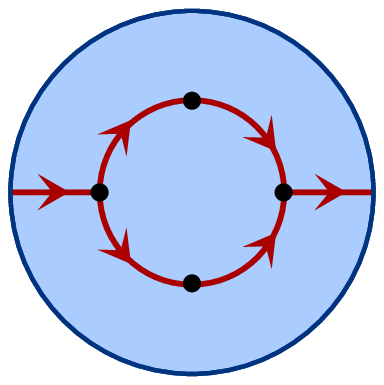}}}
  \put(0,0){
     \setlength{\unitlength}{.55pt}\put(-20,-19){
     \put(106,117)   {\scriptsize out }
     \put( 70, 87)   {\scriptsize $p_g$ }
     \put( 69, 35)   {\scriptsize $p_h$ }
     \put( 53, 69)   {\scriptsize $\ovl \varphi$ }
     \put( 85, 69)   {\scriptsize $\varphi$ }
     }\setlength{\unitlength}{1pt}}
  \end{picture}}
  \Bigg)
  ~=~
  C\Bigg(
  \raisebox{-25pt}{\begin{picture}(60,55)
  \put(0,0){\scalebox{0.55}{\includegraphics{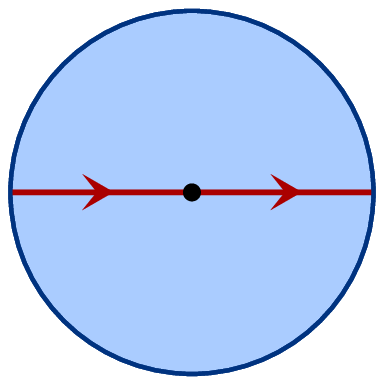}}}
  \put(0,0){
     \setlength{\unitlength}{.55pt}\put(-20,-19){
     \put(106,117)   {\scriptsize out }
\put( 68, 57)   {\scriptsize $p_{g\cdot h}$ }
     }\setlength{\unitlength}{1pt}}
  \end{picture}}
  \Bigg)
\eear
\labl{eq:defect-non-deg}
for some values $\,\chi(g) \in \Cb^\times$.\ This completes the list
of properties that we demand of a symmetry implemented by defects.
\medskip

Let us now look at some consequences of these properties. First, we
shall demonstrate the identity
\be
  C\Bigg(
  \raisebox{-25pt}{\begin{picture}(60,55)
  \put(0,0){\scalebox{0.55}{\includegraphics{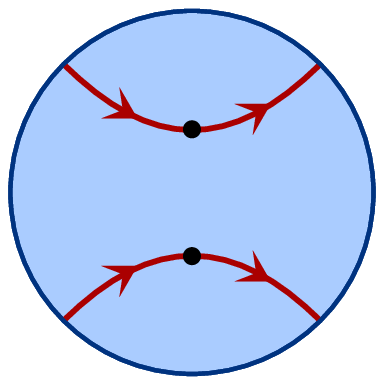}}}
  \put(0,0){
     \setlength{\unitlength}{.55pt}\put(-20,-19){
     \put(106,117)   {\scriptsize out }
     \put( 69, 78)   {\scriptsize $p_g$ }
     \put( 69, 41)   {\scriptsize $p_h$ }
     }\setlength{\unitlength}{1pt}}
  \end{picture}}
  \Bigg)
  ~=~
  C\Bigg(
  \raisebox{-25pt}{\begin{picture}(60,55)
  \put(0,0){\scalebox{0.55}{\includegraphics{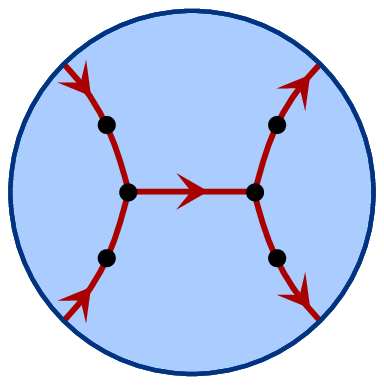}}}
  \put(0,0){
     \setlength{\unitlength}{.55pt}\put(-20,-19){
     \put(106,117)   {\scriptsize out }
     \put( 54, 91)   {\scriptsize $p_g$ }
     \put( 80, 91)   {\scriptsize $p_g$ }
     \put( 54, 47)   {\scriptsize $p_h$ }
     \put( 80, 47)   {\scriptsize $p_h$ }
     \put( 97, 69)   {\scriptsize $\ovl \varphi$ }
     \put( 40, 69)   {\scriptsize $\varphi$ }
     }\setlength{\unitlength}{1pt}}
  \end{picture}}
  \Bigg)\,.
\labl{eq:defect-id-1}
Both sides are in the image of $\,P_{4,-++-}(g,g,h,h)$,\ and the
image is one-dimensional, hence they are proportional. Gluing both
sides into the larger world-sheet
\be
  \raisebox{-40pt}{\begin{picture}(110,80)
  \put(0,0){\scalebox{0.45}{\includegraphics{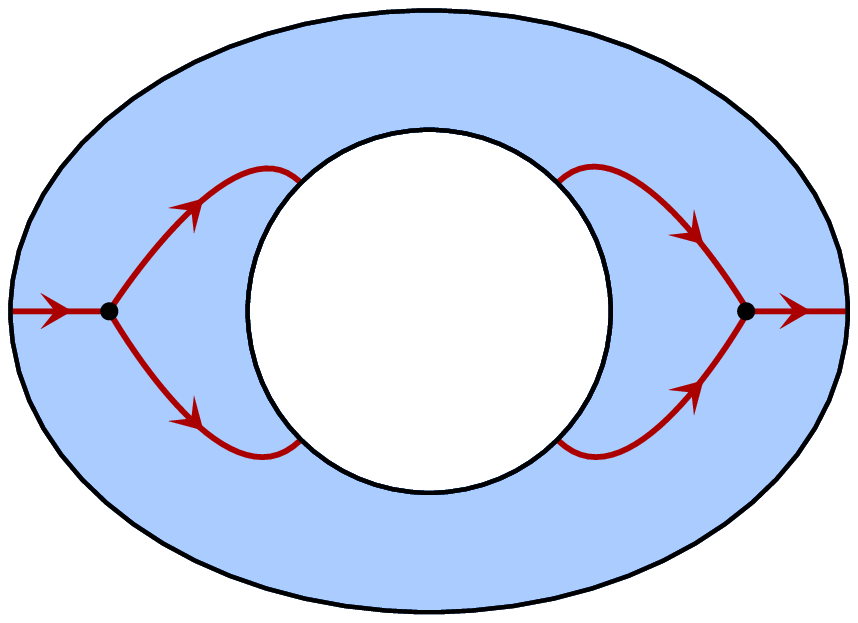}}}
  \put(0,0){
     \setlength{\unitlength}{.45pt}\put(-15,-15){
     \put(216,176)   {\scriptsize out }
     \put(146,136)   {\scriptsize in }
     \put(211, 99)   {\scriptsize $\varphi$ }
     \put( 55, 99)   {\scriptsize $\ovl \varphi$ }
     }\setlength{\unitlength}{1pt}}
  \end{picture}}
\ee
and applying \eqref{eq:group-H02-basis} and
\eqref{eq:defect-non-deg}, one obtains $\,p_{g\cdot h}\,$ in both
cases. The proportionality constant is thus equal to one. This
establishes \eqref{eq:defect-id-1}. Along the same lines, one can
verify the identity
\be
  C\Bigg(\raisebox{-25pt}{\begin{picture}(60,55)
  \put(0,0){\scalebox{0.55}{\includegraphics{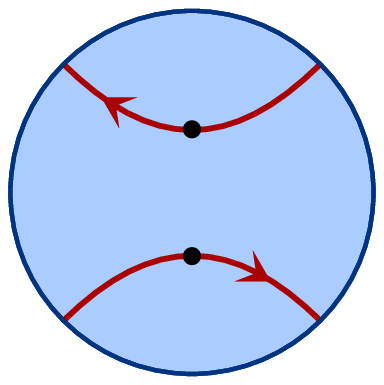}}}
  \put(0,0){
     \setlength{\unitlength}{.55pt}\put(-20,-19){
     \put(106,117)   {\scriptsize out }
     \put( 70, 78)   {\scriptsize $p_g$ }
     \put( 70, 41)   {\scriptsize $p_g$ }
     }\setlength{\unitlength}{1pt}}
  \end{picture}}
  \Bigg)
  ~=~
  \chi(g)
  \cdot
  C\Bigg(
  \raisebox{-25pt}{\begin{picture}(60,55)
  \put(0,0){\scalebox{0.55}{\includegraphics{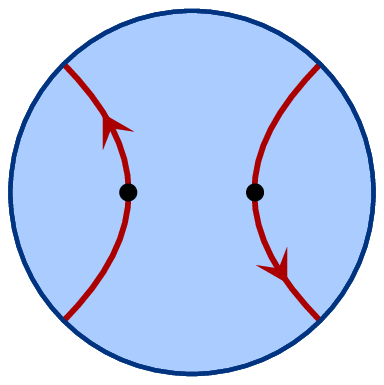}}}
  \put(0,0){
     \setlength{\unitlength}{.55pt}\put(-20,-19){
     \put(106,117)   {\scriptsize out }
     \put( 61, 69)   {\scriptsize $p_g$ }
     \put( 99, 69)   {\scriptsize $p_g$ }
     }\setlength{\unitlength}{1pt}}
  \end{picture}}
  \Bigg)\,.
\labl{eq:defect-id-2}
Finally, consider the world-sheet
\be
A^r(g) ~=~  \raisebox{-28pt}{\begin{picture}(56,56)
  \put(0,0){\scalebox{0.55}{\includegraphics{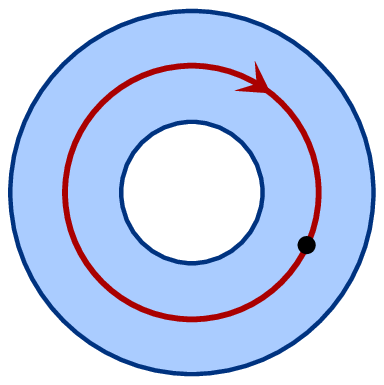}}}
  \put(0,0){
     \setlength{\unitlength}{.55pt}\put(-19,-18){
     \put(112,112)   {\scriptsize out }
     \put( 62, 78)   {\scriptsize in }
     \put( 94, 67)   {\scriptsize $p_{\hspace{-0.9pt}g}$ }
     }\setlength{\unitlength}{1pt}}
  \end{picture}}\quad.
\ee
We define the linear map $\,D_g : H_0 \rightarrow H_0\,$ as
\be
  D_g = C(A^r(g)) \, r^{-L_0-\ovl  L_0}\,.
\ee
This is, again, independent of $\,r$,\ and it follows from the
gluing properties and \eqref{eq:defect-id-1} that
\be
  D_g \, D_h = D_{g\cdot h}\,,
\ee
i.e.\ we obtain a representation of $\,S\,$ on the untwisted state
space $\,H_0$.\ If we apply that identity to $\,\one \in H_0\,$ we
obtain $\,\chi(g)\,\chi(h) = \chi(g\cdot h)$,\ i.e.\ $\,\chi\,$ is a
character of $\,S$.

The operators $\,D_g\,$ implement $\,S\,$ as a symmetry of the CFT
on world-sheets without defect lines. Let $\,O_{(m)}
\xrightarrow{\Si} O'_{(n)}\,$ be a world-sheet without defect lines
(i.e.\ the submanifold $\,L\,$ is empty) but of arbitrary genus.
Then,
\be
  (D_g)^{\otimes n} \circ C(\Si)
  = \chi(g)^{n-m} \,C(\Si) \circ  (D_g)^{\otimes m}\,.
\ee
This follows from repeated application of \eqref{eq:defect-id-2} by
the same arguments as those used in
\cite[sect.\,3.1]{Frohlich:2006ch}.
\medskip

Finally, the associator 3-cocycle on $\,S\,$ is obtained as follows:
The two vectors
\be
  v^L ~=~ C\Bigg(
  \raisebox{-34pt}{\begin{picture}(74,74)
  \put(0,0){\scalebox{0.55}{\includegraphics{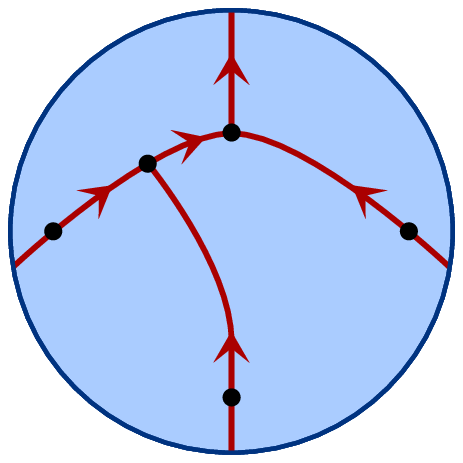}}}
  \put(0,0){
     \setlength{\unitlength}{.55pt}\put(-10,-7){
     \put(116,126)   {\scriptsize out }
     \put( 28, 64)   {\scriptsize $p_g$ }
     \put( 82, 21)   {\scriptsize $p_h$ }
     \put(118, 60)   {\scriptsize $p_k$ }
     \put( 70, 89)   {\scriptsize $\varphi$ }
     \put( 40, 97)   {\scriptsize $\varphi$ }
     }\setlength{\unitlength}{1pt}}
  \end{picture}}
  \Bigg)
  \quad , \qquad\qquad
  v^R ~=~ C\Bigg(
  \raisebox{-34pt}{\begin{picture}(74,74)
  \put(0,0){\scalebox{0.55}{\includegraphics{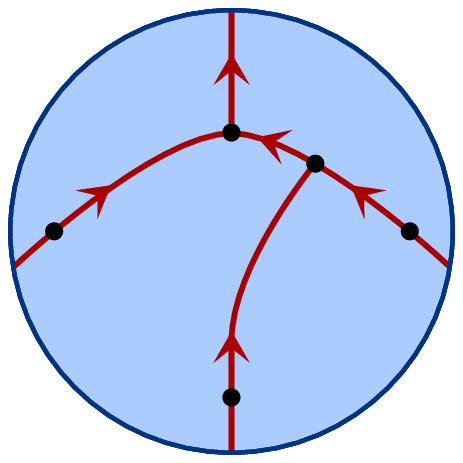}}}
  \put(0,0){
     \setlength{\unitlength}{.55pt}\put(-10,-7){
     \put(116,126)   {\scriptsize out }
     \put( 28, 64)   {\scriptsize $p_g$ }
     \put( 82, 21)   {\scriptsize $p_h$ }
     \put(118, 60)   {\scriptsize $p_k$ }
     \put( 70, 89)   {\scriptsize $\varphi$ }
     \put(104, 97)   {\scriptsize $\varphi$ }
     }\setlength{\unitlength}{1pt}}
  \end{picture}}
  \Bigg)
\labl{eq:def-3cocycl-aux1}
lie in the image of $\,P_{4,-+++}(g\cdot h\cdot k,g,h,k)\,$ and are
therefore linearly dependent. They are also both non-zero. To see
this, embed each of \eqref{eq:def-3cocycl-aux1} into a `mirrored'
picture, e.g.,
\be
  \raisebox{-42pt}{\begin{picture}(90,90)
  \put(0,0){\scalebox{0.55}{\includegraphics{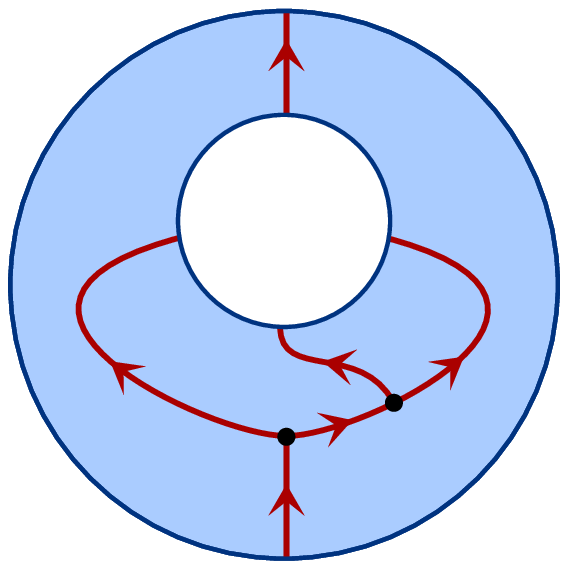}}}
  \put(0,0){
     \setlength{\unitlength}{.55pt}\put(-10,-7){
     \put( 75, 30)   {\scriptsize $ \ovl \varphi $}
     \put(125, 42)   {\scriptsize $ \ovl \varphi $}
     \put( 75, 88)   {\scriptsize in}
     \put(145,150)   {\scriptsize out}
     }\setlength{\unitlength}{1pt}}
  \end{picture}}
\ee
for $\,v^R$,\ and then use \eqref{eq:defect-non-deg} twice. Define a
$\Cb^\times$-valued 3-cochain $\,\psi\,$ on $\,S\,$ via
\be
  v^L = \psi(g,h,k)\, v^R\,.
\labl{eq:psi-CFT-def}
The usual pentagon relation obtained from the two ways of relating
\be
  C\Bigg(
  \raisebox{-34pt}{\begin{picture}(74,74)
  \put(0,0){\scalebox{0.55}{\includegraphics{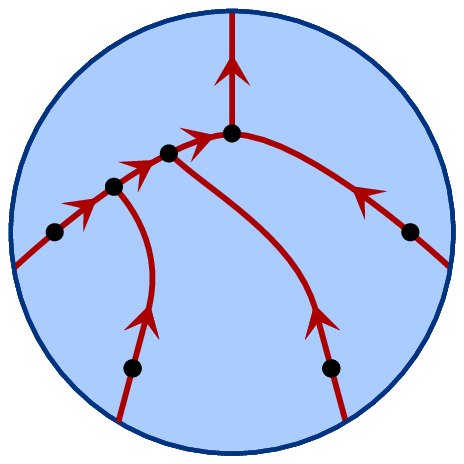}}}
  \put(0,0){
     \setlength{\unitlength}{.55pt}\put(-10,-7){
     \put(116,126)   {\scriptsize out }
     \put( 28, 64)   {\scriptsize $p_g$ }
     \put( 52, 28)   {\scriptsize $p_h$ }
     \put(109, 37)   {\scriptsize $p_k$ }
     \put(118, 60)   {\scriptsize $p_l$ }
     \put( 81,107)   {\scriptsize $\varphi$ }
     \put( 46,100)   {\scriptsize $\varphi$ }
     \put( 29, 90)   {\scriptsize $\varphi$ }
     }\setlength{\unitlength}{1pt}}
  \end{picture}}
  \Bigg)
  \qquad \text{and} \qquad
  C\Bigg(
  \raisebox{-34pt}{\begin{picture}(74,74)
  \put(0,0){\scalebox{0.55}{\includegraphics{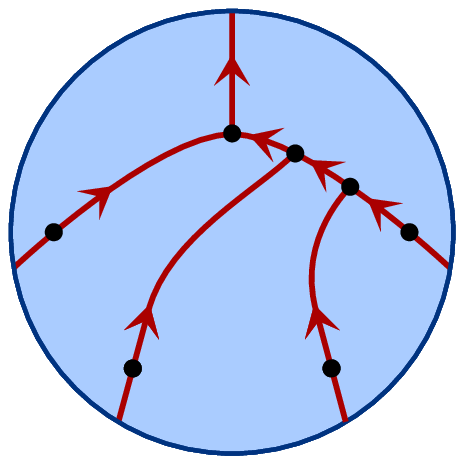}}}
  \put(0,0){
     \setlength{\unitlength}{.55pt}\put(-10,-7){
     \put(116,126)   {\scriptsize out }
     \put( 28, 64)   {\scriptsize $p_g$ }
     \put( 52, 28)   {\scriptsize $p_h$ }
     \put(109, 37)   {\scriptsize $p_k$ }
     \put(118, 60)   {\scriptsize $p_l$ }
     \put( 62,107)   {\scriptsize $\varphi$ }
     \put( 94,103)   {\scriptsize $\varphi$ }
     \put(111, 92)   {\scriptsize $\varphi$ }
     }\setlength{\unitlength}{1pt}}
  \end{picture}}
  \Bigg)
\ee
shows that $\,\delta_S \psi = 1$,\ i.e.\ $\,\psi\,$ is a cocycle.
Furthermore, modifying the choice of vectors $\,\varphi_{g,h}\,$ in
\eqref{eq:phi-choice} amounts to replacing $\,\varphi\,$ by
$\,\varphi' = \sum_{g,h\in S}\, \la(g,h)\, \varphi_{g,h}\,$ for some
2-cochain $\,\la \in C^2(S,\Cb^\times)$.\ The resulting change in
$\,\psi\,$ is $\,\psi = \psi' \cdot \delta_S\la$.\ We can
find\footnote{The argument is as follows (see, e.g., \cite[chap.\
I]{Neukirch:2000}, specifically exercises 4,5 of \S2 and proposition
1.6.1 of \S6): Since $\,\Cb^\times \cong {\rm U}(1) \times
\Rb_{>0}\,$ as multiplicative groups, we have $\,H^n(S,\Cb^\times)
\cong H^n(S,{\rm U}(1)) \times H^n(S,\Rb_{>0})$.\ The isomorphism is
provided by the decomposition $\,\psi = \psi_\theta\, \psi_r$,\
where $\,|\psi_\theta|=1\,$ and $\,\psi_r \in \Rb_{>0}$.\ However,
$\,H^n(S,\Rb_{>0}) = 1\,$ so that $\,\psi_r = \delta_S \chi\,$ for
some $\,\chi$.\ Thus, $\,\psi\,$ is cohomologous to
$\,\psi_\theta$.\ Finally, every class in $\,H^n(S,A)\,$ (for
$\,A\,$ an abelian group) can be represented by a normalised
cochain.} a cocycle $\,\psi'\,$ cohomologous to $\,\psi\,$ which is
a normalised cochain and takes values in $\,{\rm
U}(1)\subset\Cb^\times$.

Altogether, we see that an implementation of the symmetry group
$\,S\,$ by defects provides a cohomology class
\be
  [\psi] \in H^3(S,{\rm U}(1))\,.
\ee

\subsection{3-cocycle from CFT description of the WZW model}
\label{sec:CFT-cocycle}

The charge-conjugation modular invariant CFT constructed from the
affine Lie algebra $\,\widehat\ggt_\sfk\,$ is the WZW model for the
compact simple connected and simply connected Lie group $\,\Gx\,$ of
$\,\ggt\,$ at level $\,\sfk$.\ Let $\,\Oc_{\ggt,\sfk}\,$ be the
category of direct sums of integrable highest-weight representations
of $\,\widehat\ggt_\sfk$.\ It is a semi-simple abelian braided
monoidal category (in fact, it is even modular). The irreducible
representations in $\,\Oc_{\ggt,\sfk}\,$ are labelled by integrable
dominant weights $\,\la \in P^\sfk_+(\ggt)\,$ from the fundamental
affine Weyl alcove
\qq
P^\sfk_+(\ggt)=\bigl\{\ \la\in P(\ggt)\quad\big\vert
\quad\langle\la,\th\rangle
\leq\sfk\quad\land\quad\langle\la,\a_i\rangle\geq 0\,,\
i=1,2,\ldots,{\rm rank}\,\ggt\ \bigr\}\,.
\qqq
We denote the corresponding representation by $\,\widehat V_\la$.

We are interested in the simple-current sector of the model. To each
element in the centre $\,Z(\Gx)\,$ of $\,\Gx$,\ one can assign a
weight $\,\la_z \in P^\sfk_+(\ggt)\,$ such that $\,\widehat
V_{\la_z}\,$ is a simple current, see \cite{Schellekens:1990xy}. The
weights $\,\la_z\,$ for all $\,\widehat\ggt_\sfk\,$ are listed in
section \ref{sec:comparison}. The assignment $\,z \mapsto \la_z\,$
is injective, and it gives all simple currents except for the case
of $\,\widehat{\gt{e}(8)}_2$,\ already discussed in the
introduction. It is also compatible with the group structure in the
sense that for all $\,z,w \in Z(\Gx)$,
\be
  \widehat V_{\la_z} \otimes \widehat V_{\la_w}
  \cong \widehat V_{\la_{z\cdot w}}\,.
\ee

The different possible topological defects in the WZW model for
$\,\widehat \ggt_\sfk\,$ which commute with the Ka\v c--Moody
currents are in a one-to-one correspondence with objects of
$\,\Oc_{\ggt,\sfk}\,$ \cite{Petkova:2000ip,Frohlich:2006ch}. We
choose the object
\be
B = \bigoplus_{z \in Z(\Gx)}\,\widehat V_{\la_z}\,.
\ee

The 3-cocycle associated to the $Z(\Gx)$-symmetry can be computed
within the TFT-approach. There, the CFT correlator is evaluated as
the amplitude of the three-dimensional Chern--Simons theory at level
$\,\sfk\,$ with the gauge group $\,\Gx$,\ where the relevant
three-manifold is a direct product $\,\Si\ti I\,$ of the world-sheet
and an interval, and the defect lines get replaced by a Wilson graph
inside the three-manifold \cite{Frohlich:2006ch}. For
\eqref{eq:psi-CFT-def}, one thus obtains
\be
  C_{\text{CS}}\Bigg(
  \raisebox{-34pt}{\begin{picture}(115,88)
  \put(0,0){\scalebox{0.60}{\includegraphics{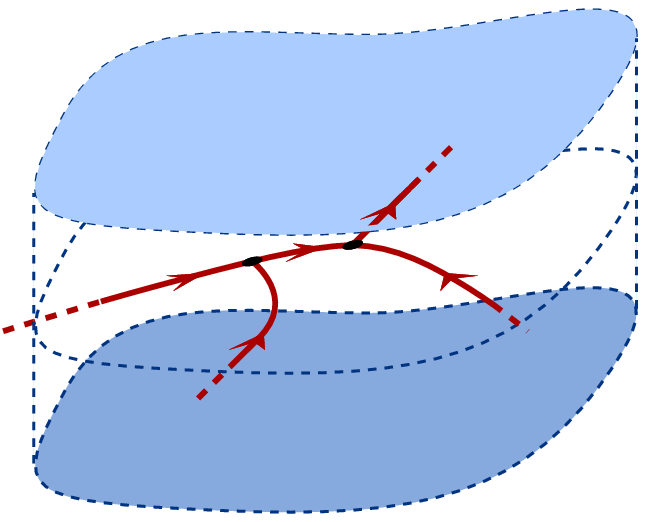}}}
  \put(0,0){
     \setlength{\unitlength}{.60pt}\put(-0,-14){
     \put( 34, 64)   {\scriptsize $\widehat V_{\la_x}$ }
     \put( 76, 50)   {\scriptsize $\widehat V_{\la_y}$ }
     \put(129, 60)   {\scriptsize $\widehat V_{\la_z}$ }
     \put( 68,102)   {\scriptsize $\Phi_{xy,z}$ }
     \put( 40, 91)   {\scriptsize $\Phi_{x,y}$ }
     }\setlength{\unitlength}{1pt}}
  \end{picture}}
  \Bigg)
  ~=~
  \psi_{\ggtk}(x,y,z) \cdot C_{\text{CS}}\Bigg(
  \raisebox{-40pt}{\begin{picture}(115,88)
  \put(0,0){\scalebox{0.60}{\includegraphics{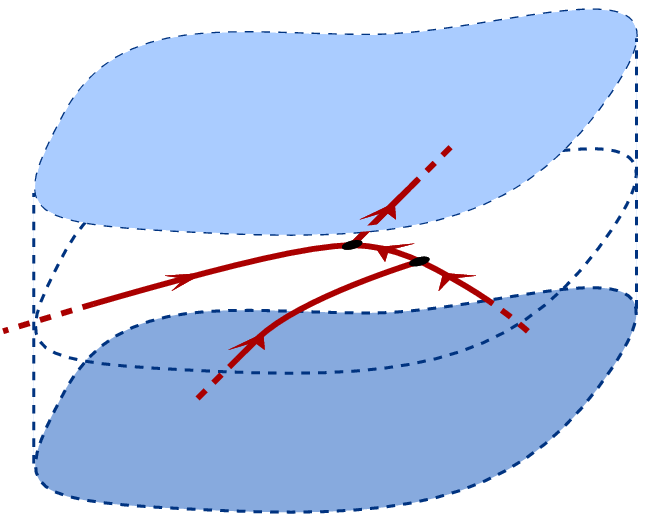}}}
  \put(0,0){
     \setlength{\unitlength}{.60pt}\put(-0,-14){
     \put( 34, 64)   {\scriptsize $\widehat V_{\la_x}$ }
     \put( 76, 50)   {\scriptsize $\widehat V_{\la_y}$ }
     \put(129, 60)   {\scriptsize $\widehat V_{\la_z}$ }
     \put( 68,102)   {\scriptsize $\Phi_{x,yz}$ }
     \put(122, 90)   {\scriptsize $\Phi_{y,z}$ }
     }\setlength{\unitlength}{1pt}}
  \end{picture}}
  \Bigg)\,.
\labl{eq:CS-relation}
The Wilson lines are labelled by objects of $\,\Oc_{\ggt,\sfk}$,\
and the junction points by nonzero morphisms $\,\Phi_{g,h}\in
\Hom(\widehat V_{\la_g} \otimes \widehat V_{\la_h}, \widehat
V_{\la_{g\cdot h}})$.\ The choice of the morphisms $\,\Phi_{g,h}\,$
corresponds to the choice of the states $\,\varphi_{g,h}\,$ in
\eqref{eq:phi-choice}.

By the definition of the Chern--Simons theory, the objects
$\,\psi_{\ggtk}(x,y,z)\,$ in \eqref{eq:CS-relation} are then entries
of the fusing matrix (or 6j-symbols) of the category
$\,\Oc_{\ggt,\sfk}\,$ restricted to the simple-current sector. The
tensor product and the braiding in the simple-current sector of
$\,\Oc_{\ggt,\sfk}\,$ can be described by abelian-group cohomology
\cite[prop.\,3.1]{Joyal:1993} (cf.\ appendix \ref{app:cohom} for a
brief overview of some pertinent facts about abelian-group
cohomology). In fact, once we have chosen the basis $\,\Phi_{g,h}$,\
we obtain an abelian 3-cocycle $\,(\psi,\Om)\,$ on $\,Z(\Gx)\,$ with
values in $\,{\rm U}(1)\,$ (a $Z(\Gx)$-module with the trivial
$Z(\Gx)$-action), see \cite[sect.\,2]{Fuchs:2004dz}. The element
$\,\psi\,$ is an ordinary 3-cocycle on $\,Z(\Gx)\,$ with values in
$\,{\rm U}(1)$,\ and $\,\Om\,$ is a 2-cochain on $\,Z(\Gx)$.\
Together, they satisfy the hexagon condition, cf.\
\eqref{eq:penta-hexa}. Furthermore, the diagonal elements of
$\,\Om\,$ are determined by the conformal weights via (see, e.g.,
\cite[sect.\,2]{Fuchs:2004dz})
\be
  \Om(z,z) = \exp\bigl(2\pi\sfi\,h(\la_z)\bigr)\,,\qquad\qquad
  h(\la_z) = \frac{\langle \la_z,\la_z+2\rho\rangle}{
   2(\sfk+g^\vee)}\,,
\labl{eq:Omega-diag}
where $\,\rho\,$ is the Weyl vector of $\,\ggt\,$ and $\,g^\vee\,$
is its dual Coxeter number. If one chooses a different basis
$\,\Phi_{g,h}\,$ the abelian $3$-cocycle changes by a coboundary.
The basis-independent information describing the tensor product and
the braiding in the simple-current sector is therefore provided by a
class $\,[\psi,\Om] \in H^3_\text{ab}(Z(\Gx),{\rm U}(1))$.

Given an abelian 3-cocycle $\,(\psi,\Om)$,\ we obtain the function
$\,q_{\psi,\Om}(z) = \Om(z,z)\,$ on $\,Z(\Gx)$.\ It is proved in
\cite{Eilenberg:1952,MacLane:1952} that $\,q_{\psi,\Om}\,$ depends
only on the class $\,[\psi,\Om]$,\ and that it determines this class
uniquely, cf.\ \eqref{eq:em-def}. This fact, together with
\eqref{eq:Omega-diag}, makes it feasible to compute a representative
for the 3-cocycle $\,\psi_{\ggtk}\,$ in \eqref{eq:CS-relation}, and
therefore also in \eqref{eq:psi-CFT-def}. We list the results in
table \ref{tab:algebra-data} below.

\section{The classical 3-cocycle vs the quantum
3-cocycle}\label{sec:comparison}

In this final section of our paper, we bring to completion the
discussion of the correspondence between the classical, i.e.\
gerbe-theoretic, and the quantum, i.e.\ conformal-field-theoretic,
description of world-sheets related by an associator move of figure
\ref{fig:defect-for-cocycle} in the setting of the WZW sigma model
on a compact simple connected and simply connected Lie group
$\,\Gx$.\ We do so by demonstrating, through a case-by-case
comparison, that the 3-cocycle component $\,\psi_{\ggtk}\in
Z^3(Z(\Gx),{\rm U}(1))\,$ of a representative $\,(\psi_{\ggtk},
\Omega_{\ggtk})\,$ of the class $\,[\psi_{\ggtk},\Om_{\ggtk}]\in
H^3_{\rm ab}(Z(\Gx),{\rm U}(1))\,$ fixed by \eqref{eq:Omega-diag}
via the Eilenberg--MacLane map (cf.\ appendix \ref{app:cohom}),
coincides with the associator 3-cocycle $\,\psi_{\Gc^{\star\sfk}}\,$
obtained from the analysis of the variation of the action functional
of the WZW model under the associator move of the embedded defect
network.

Here are the details of the comparison. The starting point is the
computation of the diagonal components of the 2-cochain
$\,\Om_{\ggtk}\,$ from \eqref{eq:Omega-diag}, using the data for
$\,\la_x\,$ given in table \ref{tab:algebra-data}, and that for the
metric on $\,P(\ggt)\,$ (the quadratic-form matrix $\,F$) taken,
e.g., from \cite[chap.\ 13]{DiFrancesco:1997nk}. Having found
$\,\Om_{\ggtk}(x,x)\,$ for all $\,x\in Z(\Gx)$,\ we then proceed
according to the type of $\,Z(\Gx)\,$ at hand, to wit:

\medskip

\nxt The acyclic centre $\,Z(\Gx)=\Zb_2\ti\Zb_2\,$ of $\,\Gx= {\rm
Spin}(4s)$.\ In this case, we simply solve the coupled pentagon and
hexagon equations explicitly for $\,\psi_{\widehat{\gt{spin}
(4s)}_\sfk}$,\ employing the definition \eqref{eq:def-cobound-ab} in
the end (that is, dividing out an appropriate trivial 3-cocycle from
the general solution) in order to get the specific representative
from table \ref{tab:algebra-data}. The latter is precisely the
gerbe-theoretic 3-cocycle $\,u_{z_1^{n_1}z_2^{n_2},
z_1^{n_1'}z_2^{n_2'},z_1^{n_1''} z_2^{n_2''}}\,$ for $\,\Gx={\rm
Spin}(4s)\,$ given in \cite[sect.\ 4]{Gawedzki:2003pm}.

\smallskip

\nxt A cyclic centre of an even order, $\,Z(\Gx)=\Zb_{2s}\,$ with
generator $\,z\,$ -- this covers the cases of $\,{\rm SU}(2r+ 1)\,$
(with $\,s=r+1$) and $\,{\rm Spin}(4r+2)\,$ (with $\,s=2$), as well
as $\,{\rm Spin}(2r+1),{\rm Sp}(2r)\,$ and $\,{\rm E}(7)\,$ (all
three with $\,s=1$). We start by considering an auxiliary object,
namely the CFT of the free boson compactified on the circle of a
rational radius squared, $\,R^2=\frac{p}{q}$,\ where $\,p\,$ and
$\,q\,$ are two positive coprime integers (and where we use units in
which the self-dual radius is 1). At these radii, the free-boson CFT
has an enhanced chiral symmetry. The fusion ring of its
representations is given by $\,\Zb_{2pq}\,$ with generator $\,\xi$,\
and the relevant abelian 3-cocycle is
\cite{Brunner:2000wx,Fuchs:2007tx}
\be
\psi_{{\rm FB}(p,q)}\bigl(\xi^n,\xi^{n'},\xi^{n''}\bigr)=(-1)^{n\,
\frac{n'+n''-[n'+n'']_{2pq}}{2pq}}\,,\qquad\Om_{{\rm FB}(p,q)}
\bigl(\xi^n,\xi^{n'}\bigr)=\erm^{\frac{\pi\,\sfi\,n\,n'}{2pq}}\,,
\ee
where $\,0\leq[m]_{2pq}<2pq\,$ is the unique integer such that
$\,[m]_{2pq}=m\mod 2pq\,$. One can now check, for all the
above-mentioned $\,\Gx$,\ that $\,\Om_{\ggtk}\bigl(z^n,z^n\bigr)\,$
obeys, for every $\,n\in\Zb_{2s}$,
\qq
\Om_{\ggtk}\bigl(z^n,z^n\bigr)=\Om_{{\rm FB}(s,1)}\bigl(\xi^n,
\xi^n\bigr)^{P(\ggt,\sfk)}\,,\qquad\qquad P(\ggt,\sfk)\in\Nb\,,
\qqq
for an integer $\,P(\ggt,\sfk)\,$ independent of $\,n$.\ At this
stage, we may adduce the theorem of Eilenberg and MacLane cited in
appendix \ref{app:cohom} to conclude that the entire abelian
3-cocycle of interest can be written as
\qq
(\psi_{\ggtk},\Omega_{\ggtk})=\Big(\bigl(\psi_{{\rm
FB}(s,1)}\bigr)^{P(\ggt, \sfk)},\bigl(\Om_{{\rm
FB}(s,1)}\bigr)^{P(\ggt,\sfk)}\Big)\,.
\qqq
We now readily verify that the 3-cocycle $\,\bigl(\psi_{{\rm
FB}(s,1)}\bigr)^{P(\ggt,\sfk)}\,$ coincides, in each of the cases of
interest, with the corresponding gerbe-theoretic 3-cocycle from
\cite[sect.\ 4]{Gawedzki:2003pm}.

\smallskip

\nxt A cyclic centre of an odd order, $\,Z(\Gx)=\Zb_{2s+1}\,$ with
generator $\,z\,$ -- this accounts for the remaining cases of
$\,{\rm SU}(2r)\,$ (with $\,s=r$) and $\,{\rm E}(6)\,$ (with
$\,s=1$). For each of these groups, we first check that the diagonal
components of $\,\Om_{\ggtk}\,$ obey the identity
\qq
\Om_{\ggtk}\bigl(z^n,z^n\bigr)^{N_n}=1\,,
\qquad\qquad N_n=\frac{{\rm LCM}(2s+1,n)}{n}
\qqq
for $\,{\rm LCM}(2s+1,n)\,$ the least common multiple of $\,2s+1\,$
and $\,n$.\ The number $\,N_n\,$ thus defined is exactly the order
of the element $\,z^n\,$ of the centre, and so we see that
$\,\Om_{\ggtk}\,$ satisfies the assumptions of Lemma 2.17 of
\cite{Fuchs:2004dz}, stated in appendix \ref{app:cohom}.
Consequently, the abelian 3-cocycle $\,(\psi_{\ggtk},
\Omega_{\ggtk})\,$ has a representative with $\,\psi_{\ggtk}=1$,\ in
accord with the gerbe-theoretic result of \cite[sect.\
4]{Gawedzki:2003pm}. The corresponding 2-cochain
$\,\Omega_{\ggtk}\,$ is then fixed by the hexagon equation to be a
bihomomorphism, whence
\qq
\Omega_{\ggtk}\bigl(z^n,z^{n'}\bigr)=\Omega_{\ggtk}
\bigl(z,z\bigr)^{n\,n'}\,,
\qqq
which is the form of the 2-cochain given in table
\ref{tab:algebra-data}.

\medskip

We shall now list the relevant algebraic data and the
representatives of the abelian 3-cocycles obtained in the procedure
detailed above. In so doing, we use the symbol $\,\La^{(\vee)}_i\,$
to denote the $i$-th fundamental (co)weight of $\,\ggt$ (we
follow the labelling conventions of
\cite{DiFrancesco:1997nk}),\ and the
shorthand notation $\,[m]_k\,$ for the unique integer
$\,0\leq[m]_k<k\,$ such that $\,[m]_k=m\mod k\,$ for $\,k\in\Zb_{>
0}$.

\begin{table}[!htp]
\caption{{\bf The comparison data.}}\label{tab:algebra-data}
\vspace{-20pt}
\end{table}
\qq
% &&\rule{37em}{.5pt}\cr\cr
&&\hbox to
4cm{\bf \unl{Algebra}\hfill}{\bf A_r=\gt{su}(r+1)}\cr\cr
&&\hbox to 4cm{centre\hfill}\Zb_{r+1}=\{e,z,z^2,\ldots,z^r\}\,,
\qquad z=\erm^{-2\pi\sfi\,\La_r^\vee}\cr
&&\hbox to 4cm{simple currents\hfill}\la_{z^n}=\sfk\,\La_{r+1-n}\,,
\qquad n\in\ovl{1,r}\,,\qquad\quad h(\la_{z^n})=\frac{\sfk\,n\,
(r+1-n)}{2(r+1)}\cr\cr
&&\hbox to 4cm{abelian\hfill}\psi_{\widehat{\gt{su}(r+1
)}_\sfk}\bigl(z^n,z^{n'},z^{n''}\bigr)=(-1)^{\sfk\,r\,n\,
(n'+n''-[n'+n'']_{r+1})\,/\,(r+1)}\cr&&\hbox{3-cocycle}\cr
&&\hbox to 4cm{\hfill}\Om_{\widehat{\gt{su}(r+1)}_\sfk}\bigl(z^n,
z^{n'}\bigr)=\erm^{\pi\sfi\,\sfk\,r\,n\,n'\,/\,(r+1)}\cr\cr
&&\rule{37em}{.5pt}\cr\cr
&&\hbox to
4cm{\bf \unl{Algebra}\hfill}{\bf B_r = \gt{spin}(2r+1)}\cr\cr
&&\hbox to 4cm{centre\hfill}\Zb_2=\{e,z\}\,,\qquad z=\erm^{-2\pi\sfi
\,\La_1^\vee}\cr
&&\hbox to 4cm{simple current\hfill}\la_z=\sfk\,\La_1\,,\qquad\quad
h(\la_z)=\frac{\sfk}{2}\cr\cr
&&\hbox to 4cm{abelian\hfill}\psi_{\widehat{\gt{spin}(2r+
1)}_\sfk}\bigl(z^n,z^{n'},z^{n''}\bigr)=1\cr&&\hbox{3-cocycle}\cr
&&\hbox to 4cm{\hfill}\Om_{\widehat{\gt{spin}(2r+1)}_\sfk}\bigl(z^n,
z^{n'}\bigr)=(-1)^{\sfk\,n\,n'}\cr\cr
% &&\rule{37em}{.5pt}\cr\cr
% pagebreaks by hand
  \nonumber
  \qqq

  \qq
&&\hbox to 4cm{\bf \unl{Algebra}\hfill}{\bf C_r=\gt{sp}(2r)}\cr\cr
&&\hbox to 4cm{centre\hfill}\Zb_2=\{e,z\}\,,\qquad z=\erm^{-2\pi\sfi
\,\La_r^\vee}\cr\cr
&&\hbox to 4cm{simple current\hfill}\la_z=\sfk\,\La_r\,,\qquad\quad
h(\la_z)=\frac{\sfk\,r}{4}\cr\cr %\cr
&&\hbox to 4cm{abelian\hfill}\psi_{\widehat{\gt{sp}(2r
)}_\sfk}\bigl(z^n,z^{n'},z^{n''}\bigr)=(-1)^{\sfk\,r\,n\,n'\,n''}
\cr&&\hbox{3-cocycle}\cr
&&\hbox to 4cm{\hfill}\Om_{\widehat{\gt{sp}(2r)}_\sfk}\bigl(z^n,
z^{n'}\bigr)=\erm^{\pi\sfi\,\sfk\,r\,n\,n'\,/\,2} \cr \cr
&&\rule{37em}{.5pt} \cr \cr
&&\hbox to 4cm{\bf \unl{Algebra}\hfill}{\bf D_{2s+1}=\gt{spin}(4s+2)}\cr\cr
&&\hbox to 4cm{centre\hfill}\Zb_4=\{e,z,z^2,z^3\}\,,\qquad z=
\erm^{-2\pi\sfi\,\La_{2s+1}^\vee}\cr\cr
&&\hbox to 4cm{simple currents\hfill}\left\{\begin{array}{ll}
\la_z=\sfk\,\La_{2s}\,,\qquad & h(\la_z)=\frac{\sfk\,(2s+1)}{8}\cr
\cr
\la_{z^2}=\sfk\,\La_1\,,\qquad & h(\la_{z^2})=\frac{\sfk}{2}\cr\cr
\la_{z^3}=\sfk\,\La_{2s+1}\,,\qquad & h(\la_{z^3})=\frac{\sfk\,(2s+
1)}{8}
\end{array}\right.\cr\cr\cr
&&\hbox to 4cm{abelian\hfill}\psi_{\widehat{\gt{spin}(4s
+2)}_\sfk}\bigl(z^n,z^{n'},z^{n''}\bigr)=(-1)^{\sfk\,n\,(n'+n''
-[n'+n'']_4)\,/\,4}\cr&&\hbox{3-cocycle}\cr
&&\hbox to 4cm{\hfill}\Om_{\widehat{\gt{spin}(4s+2)}_\sfk}\bigl(z^n,
z^{n'}\bigr)=\erm^{\pi\sfi\,\sfk\,(2s+1)\,n\,n'\,/\,4}\cr %\cr
&&\rule{37em}{.5pt}\cr \cr
&&\hbox to 4cm{\bf \unl{Algebra}\hfill}{\bf D_{2s}=\gt{spin}(4s)}\cr\cr
&&\hbox to
4cm{centre\hfill}\Zb_2\ti\Zb_2=\{e,z_1\}\ti\{e,z_2\}\,,\qquad z_1=
\erm^{-2\pi\sfi\,\La_{2s}^\vee}\,,\quad z_2=\erm^{-2\pi\sfi\,
\La_1^\vee}\cr\cr &&\hbox to 4cm{simple currents\hfill} \left\{
\begin{array}{ll}
\la_{z_1}=\sfk\,\La_{2s}\,,\qquad & h(\la_{z_1})=\frac{\sfk\,s}{4}
\cr\cr
\la_{z_2}=\sfk\,\La_1,,\qquad & h(\la_{z_2})=\frac{\sfk}{2}\cr\cr
\la_{z_1 z_2}=\sfk\,\La_{2s-1}\,,\qquad & h(\la_{z_1 z_2})=
\frac{\sfk\,s}{4}
\end{array}\right.\cr\cr\cr
&&\hbox to 4cm{abelian\hfill}\psi_{\widehat{\gt{spin}(4s
)}_\sfk}\bigl(z_1^{n_1}z_2^{n_2},z_1^{n_1'}z_2^{n_2'},z_1^{n_1''}
z_2^{n_2''}\bigr)=(-1)^{\sfk\,(s\,n_1\,n_1'\,n_1''+n_1\,n_2'\,n_2''
+n_2\,n_1'\,n_1'')}\cr&&\hbox{3-cocycle}\cr
&&\hbox to 4cm{\hfill}\Om_{\widehat{\gt{spin}(4s)}_\sfk}\bigl(
z_1^{n_1}z_2^{n_2},z_1^{n_1'}z_2^{n_2'}\bigr)=\erm^{\pi\sfi\,
\sfk\,(s\,n_1\,n_1'+2n_2\,n_2'+n_1\,n_2'+n_2\, n_1')\,/\,2}\cr\cr
% &&\rule{37em}{.5pt}\cr\cr
% pagebreaks by hand
  \nonumber
  \qqq

  \qq
&&\hbox to 4cm{\bf \unl{Algebra}\hfill}{\bf E_6}\cr\cr &&\hbox to
4cm{centre\hfill}\Zb_3=\{e,z,z^2\}\,,\qquad z=\erm^{-2\pi\sfi\,
\La_5^\vee}\cr\cr
&&\hbox to 4cm{simple currents\hfill}\left\{
\begin{array}{ll}
\la_z=\sfk\,\La_1\,,\qquad & h(\la_z)=\frac{2\sfk}{3}\cr\cr
\la_{z^2}=\sfk\,\La_5\,,\qquad & h(\la_{z^2})=\frac{2\sfk}{3}
\end{array}\right.\cr\cr\cr
&&\hbox to 4cm{abelian\hfill}\psi_{\widehat{\gt{e}(6)}_\sfk}\bigl(
z^n,z^{n'},z^{n''}\bigr)=1\cr&&\hbox{3-cocycle}\cr
&&\hbox to 4cm{\hfill}\Om_{\widehat{\gt{e}(6)}_\sfk}\bigl(z^n,
z^{n'}\bigr)=\erm^{-2\pi\sfi\,\sfk\,n\,n'\,/\,3}\cr\cr
&&\rule{37em}{.5pt}\cr\cr &&\hbox to
4cm{\bf \unl{Algebra}\hfill}{\bf E_7}\cr\cr &&\hbox to
4cm{centre\hfill}\Zb_2=\{e,z\}\,,\qquad z=\erm^{-2\pi\sfi\,
\La_1^\vee}\cr\cr
&&\hbox to 4cm{simple current\hfill}\la_z=\sfk\,\La_6\,,\qquad\quad
h(\la_z)=\frac{3\sfk}{4}\cr\cr\cr
&&\hbox to 4cm{abelian\hfill}\psi_{\widehat{\gt{e}(7)}_\sfk}\bigl(
z^n,z^{n'},z^{n''}\bigr)=(-1)^{\sfk\,n\,n'\,n''}\cr
&&\hbox{3-cocycle}\cr
&&\hbox to 4cm{\hfill}\Om_{\widehat{\gt{e}(7)}_\sfk}\bigl(z^n,
z^{n'}\bigr)=\erm^{-\pi\sfi\,\sfk\,n\,n'\,/\,2}\cr\cr
% &&\rule{37em}{.5pt}\cr\cr
\nonumber
\qqq

\appendix

\sect{Appendix}

\subsection{Some background on group cohomology}\label{app:cohom}

In general, group cohomology is defined for a group $\,S\,$ and an
$S$-module $\,A$,\ see, e.g., \cite[chap.\ I, \S2]{Neukirch:2000}.
We shall only need the case of a finite group $\,S\,$ and the
$S$-module given either by $\,A={\rm U}(1)$,\ understood as an
$S$-module with trivial $S$-action, or by
$\,A=\check{C}^{p,r}(\Oc)$,\ understood as an $S$-module with an
$S$-action by pullback,
\qq
(g.\om)_{i_1 i_2\ldots i_{p+1}}=\bigl(g^{-1}\bigr)^*\om_{g^{-1}.i_1
\ g^{-1}.i_2\ \ldots\ g^{-1}.i_{p+1}}\,,
\qqq
where we have assumed the cover $\,\Oc\,$ to be $S$-invariant as in
\eqref{eq:ZG-acts-on-cech}.

An $n$-{\em cochain} on $\,S\,$ is a function $\,S^n
\rightarrow A$,\ and the set of $n$-cochains is denoted as
$\,C^n(S,A)$.\ The coboundary operator $\,\delta_{(n)}\,$ is a map
$\,C^n(S,A)\rightarrow C^{n+1}(S,A)\,$ which obeys $\,\delta_{(n+1)}
\circ \delta_{(n)} = 1$.\ For $\,n=1,2,3,4$, it is given by the
formul\ae
\be\begin{array}{l}
(\d_{(0)}\psi_{(0)})(a)=a.\psi_{(0)}-\psi_{(0)}\,,\qquad(\d_{(1)}
\psi_{(1)})(a,b)=a.\psi_{(1)}(b)-\psi_{(1)}(a\cdot
b)+\psi_{(1)}(a)\,,\cr\cr
(\d_{(2)}\psi_{(2)})(a,b,c)=a.\psi_{(2)}(b,c)-\psi_{(2)}(a\cdot
b,c)+\psi_{(2)}(a,b\cdot c)-\psi_{(2)}(a,b)\,,\cr\cr
(\d_{(3)}\psi_{(3)})(a,b,c,d)=a.\psi_{(3)}(b,c,d)-\psi_{(3)}(a\cdot
b,c,d)+\psi_{(3)}(a,b\cdot c,d)\cr\cr
\hspace{3.7cm}-\psi_{(3)}(a,b,c\cdot d)+\psi_{(3)}(a,b,c)
\end{array}
\ee
in the additive notation (e.g., for $\,A=\check{C}^{p,r}(\Oc)\,$
with $\,r>0$), and by the formul\ae
\be\begin{array}{l}
(\d_{(0)}\psi_{(0)})(a)=\frac{a.\psi_{(0)}}{\psi_{(0)}}\,,\qquad
\quad(\d_{(1)}\psi_{(1)})(a,b)=\frac{a.\psi_{(1)}(b)\,\psi_{(1
)}(a)}{\psi_{(1)}(a\cdot b)}\,,\cr\cr
(\d_{(2)}\psi_{(2)})(a,b,c)=\frac{a.\psi_{(2)}(b,c)\,\psi_{(2)}
(a,b\cdot c)}{\psi_{(2)}(a\cdot b,c)\,\psi_{(2)}(a,b)}\,,\qquad
\quad(\d_{(3)}\psi_{(3)})(a,b,c,d)=\frac{a.\psi_{(3)}(b,c,d)\,
\psi_{(3)}(a,b\cdot c,d)\,\psi_{(3)}(a,b,c)}{\psi_{(3)}(a\cdot
b,c,d)\,\psi_{(3)}(a,b,c\cdot d)}
\end{array}
\ee
in the multiplicative notation (e.g., for $\,A={\rm U}(1)\,$ or
$\,A= \check{C}^{p,0}(\Oc)$), all written for $\,\psi_{(n)}\in
C^n(S, A)\,$ and $\,a,b,c,d \in S$.\ The $n$-cocycles, the
$n$-coboundaries, and the $n$-th cohomology group are denoted as
\qq
Z^n(S,A)=\Ker\,\d_{(n)}\,,\qquad B^n(S,A)= \Im\,\d_{(n-1)}\,,\qquad
H^n(S,A)=\frac{Z^n(S,A)}{B^n(S,A)}\,,\cr
\qqq
respectively. We shall drop the subscript $n$ from the coboundary
operator henceforth, and we shall write $\,\d_S\,$ whenever we want
to emphasise that it is the coboundary operator for the cohomology
of $\,S$.
\medskip

For an abelian group $\,S$,\ one can introduce a different
cohomology, namely abelian-group cohomology
\cite{Eilenberg:1952,MacLane:1952}. We shall only need the third
abelian cohomology group of $\,S$,\ with values in the trivial
$S$-module $\,{\rm U}(1)$.

{\em Abelian $2$-cochains on $\,S\,$} are just ordinary 2-cochains
on the group, $\,C^2_\text{ab}(S,{\rm U}(1))=C^2(S,{\rm U}(1))$,\
and {\em abelian $3$-cochains} are defined as
\be
  C^3_\text{ab}(S,{\rm U}(1)) = \big\{\  (\psi,\Omega) \quad\big|\quad
  \psi \in C^3(S,{\rm U}(1))\,,\quad \Omega \in C^2(S,{\rm U}(1)) \ \big\}\,.
\ee
The set $\,C^3_\text{ab}(S,{\rm U}(1))\,$ is an abelian group under
element-wise multiplication. The coboundary operator
$\,\delta_\text{ab,(2)} :  C^2_\text{ab}(S,{\rm U}(1)) \rightarrow
C^3_\text{ab}(S,{\rm U}(1))\,$ is given by the formula
\be
  \delta_\text{ab,(2)} \varphi
  = \big(\, \delta_S \varphi \,,\, (a,b) \mapsto \varphi(a,b)/\varphi(b,a) \, \big)\,.
\labl{eq:def-cobound-ab}
The set of abelian 3-coboundaries $\,B^3_\text{ab}(S,{\rm U}(1))\,$
is the image of $\,\delta_\text{ab,(2)}$.\ An element
$\,(\psi,\Omega) \in C^3_\text{ab}(S,{\rm U}(1))\,$ is an {\em
abelian $3$-cocycle on $\,S\,$} if the following conditions are
satisfied for all $\,a,b,c,d \in S$,
\qq
\text{pentagon} \quad
&:&\quad\psi(b,c,d)\,\psi(a,b\cdot c,d)\,\psi(a,b,c)
=\psi(a\cdot b,c,d)\,\psi(a,b,c\cdot d)\,,\cr&&\hspace{5cm}\label{eq:penta-hexa}\\
\text{hexagon} \quad
&:&\quad\left\{\begin{array}{l}
\psi(c,a,b)\,\Omega(a\cdot
b,c)\,\psi(a,b,c)=\Omega(a,c)\,\psi(a,c,b)\,\Omega(b,c)
\cr\nonumber\cr
\psi(b,c,a)^{-1}\,\Omega(a,b\cdot
c)\,\psi(a,b,c)^{-1}=\Omega(a,c)\,\psi(b,a,c)^{- 1}\,\Omega(a,b)
\end{array}\right.\,.
\qqq
(In the notation used in \cite{MacLane:1952},
$\,\psi(a,b,c)=f(a,b,c)\,$ and $\,\Omega(a,b)=d(a\,|\,b)$,\ see
\cite[eqns.\,(17)--(19)]{MacLane:1952}.) Note that the pentagon
condition just says that $\,\delta_S \psi \equiv 1$.\ The set of
abelian $3$-cocycles is denoted by $\,Z^3_\text{ab}(S,{\rm U}(1))$.

The third abelian cohomology group of the abelian group $\,S$,\ with
values in the trivial $S$-module $\,{\rm U}(1)\,$ is defined as
\be
H^3_\text{ab}(S,{\rm U}(1)) = Z^3_\text{ab}(S,{\rm U}(1)) /
B^3_\text{ab}(S,{\rm U}(1))\,.
\ee

The set $\,Q(S,{\rm U}(1))\,$ of {\em quadratic forms} on a group
$\,S$,\ with values in $\,{\rm U}(1)\,$ is composed of all elements
$\,q \in C^1(S,{\rm U}(1))\,$ such that $\,q(a) = q(a^{-1})\,$ and
$\,\delta_S q : S\ti S\rightarrow {\rm U}(1)\,$ is a bihomomorphism.
The product of two quadratic forms is again a quadratic form, as is
the function $\,q \equiv 1$,\ and so $\,Q(S,{\rm U}(1))\,$ is an
abelian group. It is proved in \cite{Eilenberg:1952} (see
\cite[thm.\,3]{MacLane:1952}) that the map
\be \begin{array}{lrcl}
\mathrm{EM}\,:\ & H^3_{\text{ab}}(S,{\rm U}(1)) &\rightarrow&
Q(S,{\rm U}(1))
  \\{}\\[-.9em]&
  [\psi,\Omega] &\mapsto& q_{\psi,\Omega}(a)=\Omega(a,a)
  \end{array}  \labl{eq:em-def}
is an {\em isomorphism} of abelian groups. In particular,
$\,q_{\psi,\Omega}\,$ depends only on the class $\,[\psi,\Omega]\,$
of the abelian 3-cocycle $\,(\psi,\Omega)$.\ Using this isomorphism,
it was demonstrated in \cite[Lem.\,2.17]{Fuchs:2004dz} that the
class $\,[\psi]\in H^3(S,{\rm U}(1))\,$ of the 3-cocycle component
of an abelian 3-cocycle $\,(\psi,\Om)\,$ is trivial iff the identity
\qq
\Om(a,a)^{N_a}=1
\qqq
holds for every element $\,a\in S$,\ with $\,N_a\,$ the order of
$\,a$.

\subsection{Field equations and defect
conditions}\label{app:def-glue}

In this appendix, we perform a detailed derivation of the field
equations and defect conditions in a generic non-linear sigma model
with a topological term defined -- as in section \ref{sec:holo-net}
-- on a world-sheet $\,\Si\,$ with an embedded defect network
$\,\G\,$. The defect conditions, which characterise the defect in
the very same manner as boundary conditions characterise a boundary
state, are always to be imposed on the fields of the model, both in
the classical r\'egime (extremal field configurations) and in the
quantum r\'egime (the definition of the path integral for a
world-sheet with a defect network on it).
\smallskip

Let us start by stating some conventions. We use the two-dimensional
Levi-Civita symbols $\,\ep_{ab}\,$ and $\,\ep^{ab}\,$ such that
$\,\ep_{12}=1=\ep^{12}\,$ and $\,\ep_{ab}\,\ep^{cb}=\d_a^{\ c}$.\ In
the component notation for differential forms, we use the following
basis
\qq
\sfd y^{\mu_1}\wedge\sfd y^{\mu_2}\wedge\cdots\wedge\sfd y^{\mu_p}=
\sum_{\si\in S_p}\,(-1)^{{\rm sgn}(\si)}\,\sfd y^{\mu_{\si(1)}}
\oti\sfd y^{\mu_{\si(2)}}\oti\cdots\oti\sfd y^{\mu_{\si(p)}}\,.
\qqq

The standard `kinetic' term written in terms of the intrinsic
world-sheet metric $\,\g$,\ the associated metric volume form
$\,\vol_{\Si,\g}=\sqrt{\det\g}\,\sfd \si^1\wedge\sfd\si^2\,$
($\si^a\,$ are local coordinates on $\,\Si$), and the target-space
metric $\,G\,$ reads
\qq\label{eq:sigma-kin}
S_{\rm kin}[X;\g]=\int_{\Si-\G}\,G_X(\sfd X\overset{\wedge}{,}
\star_\g\sfd X)=\int_{\Si-\G}\,\vol_{\Si,\g}\,\bigl(\g^{-1}
\bigr)^{ab}\,G_{\mu \nu}(X)\,\p_a X^\mu\,\p_b X^\nu\,.
\qqq
The world-sheet metric defines, in particular, the Hodge operator
$\,\star_\g\,$ on $\,\Om^\bullet (\Si)\,$ as per
\qq
\star_\g 1=\vol_{\Si,\g}\,,\qquad\qquad\star_\g\sfd\si^a=\sqrt{\det
\g}\,\bigl(\g^{-1}\bigr)^{ab}\,\eps_{bc}\,\sfd\si^c\,,\qquad\qquad
\star_\g \vol_{\Si,\g}=1\,.
\qqq
We have also used the notation $\,\sfd X=\p_a X^\mu\,\sfd\si^a\oti
\p_\mu\in\sfT^*_\si\Si\oti\sfT_{X(\si)}M$,\ hence the familiar local
form of $\,S_{\rm kin}[X;\g]$.\ The integral in \eqref{eq:sigma-kin}
splits into contributions from the patches into which the
world-sheet is partitioned by the embedded defect network $\,\G$.\
Whenever a functional variation of the integral produces a
contribution from a component $\,e\,$ of the boundary of the patch,
we should use in the integrand the appropriate local extension
$\,X_{|\a}\,$ described in section \ref{sec:circle-hol}, with the
choice of $\,\a\in\{1,2\} \,$ depending on the relative orientation
of $\,e\,$ and that of the defect line covering $\,e$.

The variation of \eqref{eq:sigma-kin} in the direction of $\,X\,$
reads
\qq
\d_X S_{\rm
kin}[X;\g]&=&\,\int_{\Si-\G}\,\vol_{\Si,\g}\,\bigl(\g^{-1}
\bigr)^{ab}\,\bigl(2G_{\mu\nu}(X)\,\p_a\d X^\mu\,\p_b X^\nu+\d X^\la
\,\p_\la G_{\mu\nu}(X)\,\p_a X^\mu\,\p_b X^\nu\bigr)\cr\cr
&=&-2\,\int_{\Si-\G}\,G_X\bigl(\d X,\vol_{\Si,\g}\,\D_{(2)}X+
\G_{\textrm{L-C}}(\sfd X\overset{\wedge}{,}\star_\g\sfd X)\bigr)
\label{eq:var-kin}\\\cr
&&+2\,\int_{E_{\G}}\,\bigl(G_{X_{|1}}(\d X_{|1},\star_\g\sfd
X_{|1})-G_{X_{|2}}(\d X_{|2}, \star_\g\sfd X_{|2})\bigr)\,,
\nonumber
\qqq
where $\,\d X=\d X^\mu\,\p_\mu\,$ is the variation field, with the
one-sided (local) extensions $\,\d X_{|\a}\,$ to $\,\ovl U_\a$.\ We
have also used the notation
\qq
G_X\bigl(\d X,\vol_{\Si,\g}\,\D_{(2)}X\bigr)&=&\d X^\mu\,G_{\mu\nu}
(X)\,\bigl(\D_{(2)}X^\nu\bigr)\,\vol_{\Si,\g}\,,\cr&&\\
G_X\bigl(\d X,\G_{\textrm{L-C}}(\sfd X\overset{\wedge}{,}\star_\g
\sfd X)\bigr)&=&\d X^\mu\,G_{\mu\nu}(X)\,\bigl\{\begin{smallmatrix}
\nu \\ \rho\si\end{smallmatrix}\bigr\}(X)\,\sfd\,X^\rho\wedge
\star_\g\sfd X^\si\,,\nonumber
\qqq
with
\qq
\D_{(2)}=\tfrac{1}{\sqrt{\det\g}}\,\p_a\,\bigl(\sqrt{\det\g}\,
\bigl(\g^{-1}\bigr)^{ab}\,\p_b\bigr)
\qqq
the world-sheet Laplacian, and
\qq
\bigl\{\begin{smallmatrix} \nu \\ \rho\si\end{smallmatrix}\bigr\}=
\tfrac{1}{2}\,\bigl(G^{-1}\bigr)^{\nu\la}\,\bigl(\p_\rho G_{\si\la}
+\p_\si G_{\rho\la}-\p_\la G_{\rho\si}\bigr)
\qqq
the Christoffel symbols of the target-space metric $\,G$.\, As
usual, the boundary term in \eqref{eq:var-kin} comes from
integration by parts and the application of Stokes' theorem. Its
geometric interpretation becomes manifest upon introducing a
coordinate $\,t\in\mathbb{R}\,$ along an edge $\,e\in E_{\G}\,$ of
$\,\G$,\ together with the attendant normalised tangent vector field
$\,\widehat t=\frac{1}{\sqrt{\g(\p_t,\p_t)}}\,\p_t$.\ It is then
straightforward to show that the two normalised vector fields
$\,\widehat n_\a\,,\ \a=1,2\,$ normal to that edge which were
described in section \ref{sec:circle-hol} are given by $\,\widehat
n_\a=(-1)^\a\,(\widehat t\con\star_\g\sfd\si^a)\,\p_a$, \ and so the
variation of the `kinetic' term rewrites as
\qq
\d_X S_{\rm kin}[X;\g]&=&-2\,\int_{\Si-\G}\,G_X\bigl(\d X,\vol_\Si\,
\D_{(2)}X+\G_{\textrm{L-C}} (\sfd X\overset{\wedge}{,}\star_\g\sfd
X)\bigr)\cr&&\\
&&-2\,\int_{E_{\G}}\,\vol_{E_{\G},\g}\,\bigl(G_{X_{|1}}\bigl(\d X_{|1},
X_{|1*}\widehat n_1\bigr)+G_{X_{|2}}\bigl(\d X_{|2},X_{|2*}\widehat
n_2\bigr)\bigr)\,,\nonumber
\qqq
where $\,\vol_{E_\G,\g}\,$ is the volume form for $\,E_\G\,$
(locally given by $\,\sqrt{\g(\p_t,\p_t)}\,\sfd t$) and $\,X_{|\a*}
: \sfT\ovl U_\a \rightarrow \sfT M\,$ are the tangent maps for
$\,X_{|\a}$.

Passing, next, to the topological term
\qq
S_{\rm top}[X]&=&\sum_{t\in\triangle(\Sigma)}\left[\sfi\,\int_t\,
\widehat B_t+\sum_{e\subset t}\left(\sfi\,\int_e\,\widehat A_{te}
+\sum_{v\in e}\,\log\widehat g_{tev}(v)\right)\right]\cr
&+&\sum_{e\in\triangle(E_{\G})}\,\left(\sfi\,\int_e\,\widehat P_e
+\sum_{v\in e}\,\log\widehat K_{ev}(v)\right)\\
&+&\sum_{v\in V_{\G}}\,\log\widehat f_v(v)\,,\nonumber
\qqq
in which all triangulations have been correlated as discussed in
section \ref{sec:holo-net}, we find
\qq
&&\tfrac{1}{\sfi}\,\d_X S_{\rm top}[X]=\int_{\Si-\G}\,X^*\bigl(\d X\con
H\bigr)\cr\cr
&&+\sum_{t\in\triangle(\Si)}\,\sum_{e\subset t}\,\biggl[\int_e\,X^*
\bigl(\d X\con(B_{i_t}+\sfd A_{i_t i_e})\bigr)+\sum_{v\in e}\,
\vep_{tev}\,X^*\bigl(\d X\con\bigl(A_{i_t i_e}-\sfi\,\sfd\log g_{i_t
i_e i_v} \bigr)\bigr)(v)\biggr]\cr\cr
&&+\sum_{e\in\triangle(E_{\G})}\,\biggl[\int_e\,X^*(\d X\con\sfd
P_{i_e})+\sum_{v\in e}\,\vep_{ev}\,X^*\bigl(\d X\con\bigl(P_{i_e}+
\sfi\,\sfd\log K_{i_e i_v}\bigr)\bigr)(v)\biggr]\cr\cr
&&-\sfi\,\sum_{v\in V_{\G}}\,X^*\bigl(\d X\con\sfd\log f_{i_v}
\bigr)(v)\cr\cr
&&=\int_{\Si-\G}\,X^*\bigl(\d X\con H\bigr)+\sum_{t\in\triangle(\Si)}\,
\sum_{e\subset t}\,\biggl[\int_e\,X^*(\d X\con B_{i_e})-\sum_{v\in
e}\,\vep_{tev}\,X^*\bigl(\d X\con A_{i_e i_v}\bigr)(v)\biggr]\cr\cr
&&+\sum_{e\in\triangle(E_{\G})}\,\biggl[\int_e\,X^*(\d X\con\sfd
P_{i_e})+\sum_{v\in e}\,\vep_{ev}\,X^*\bigl(\d X\con\bigl(P_{i_e}+
\sfi\,\sfd\log K_{i_e i_v}\bigr)\bigr)(v)\biggr]\cr\cr
&&-\sfi\,\sum_{v\in V_{\G}}\,X^*\bigl(\d X\con\sfd\log f_{i_v}
\bigr)(v)\,,
\qqq
where -- so far -- we have only used the defining relations of the
local data of the gerbe, cf.\ \eqref{eq:def-gerbe}, alongside the
trivial relation $\,\sum_{e\subset t}\,\sum_{v\in e}\,\vep_{tev}\,\d
X\con A_{i_t i_v}=0$.\ The first line integral in the above formula
reduces to a contribution from the embedded defect network, and so
-- upon recalling the indexing conventions of section
\ref{sec:holo-net} and the definition \eqref{eq:stable-iso-local} of
the $\Gc$-bi-brane curvature $\,\om\,$ -- we readily see that it
combines with the other line integral as
\qq
&&\sum_{t\in\triangle(\Si)}\,\sum_{e\subset t}\,\int_e\,X^*(\d X\con
B_{i_e})+\sum_{e\in\triangle(E_{\G})}\,\int_e\,X^*(\d X\con\sfd
P_{i_e})\cr\cr
&=&\sum_{e\in\triangle(E_{\G})}\,\int_e\,\bigl[X_{|1}^*(\d X_{|1}\con
B_{\phi_1(i_e)})-X^*_{|2}(\d X_{|2}\con B_{\phi_2(i_e)})+X^*(\d
X\con\sfd P_{i_e})\bigr]\cr\cr
&=&\sum_{e\in\triangle(E_{\G})}\,\int_e\,X^*(\d X\con\om_{i_e})\equiv
\int_{E_{\G}}\,X^*(\d X\con\om)\,,
\qqq
where we have used that
\qq\label{eq:var-patch-edge}
\d X_{|\a}\vert_{E_\G}=\iota_{\a*}\d X\vert_{E_\G}\,,
\qqq
which holds by (L2).

Next, we turn to the vertex contributions. The one coming from the
internal vertices, $\,v\in\Si-V_{\G}$,\ is easily checked to vanish,
\qq
&&\sum_{e\in\triangle(E_{\G})}\,\sum_{v\in e-V_{\G}}\,
\vep_{ev}\,X^*\bigl(\d X\con\bigl(P_{i_e}+\sfi\,\sfd\log K_{i_e
i_v}\bigr)\bigr) (v)-\sum_{\substack{t\in\triangle(\Si)\\ e\subset
t}}\,\sum_{v\in e-V_{\G}}\,\vep_{tev}\,X^*\bigl(\d X\con A_{i_e
i_v}\bigr)(v)\cr\cr
&=&\sum_{e\in\triangle(E_{\G})}\,\sum_{v\in e-V_{\G}}\,\vep_{ev}\,
\bigl[X_{|2}^*\bigl(\d X_{|2}\con A_{\phi_2(i_e)\phi_2(i_v)}\bigr)
-X_{|1}^*\bigl(\d X_{|1}\con A_{\phi_1(i_e)\phi_1(i_v)}\bigr)\\\cr
&&\hspace{3cm}+X^*\bigl(\d X\con\bigl(P_{i_e}-P_{i_v}+\sfi\,\sfd
\log K_{i_ei_v}\bigr)\bigr)\bigr](v)=0\,,\nonumber
\qqq
by virtue of \eqref{eq:stable-iso-local} and
\eqref{eq:var-patch-edge}. The one coming from the vertices of the
defect network, on the other hand, does not vanish identically. At a
given vertex $\,v\in V_\G\,$ of, say, valence $\,n_v$,\ it splits
into a sum of terms sourced by the defect lines converging at the
vertex, completed with the vertex insertion of the 2-morphism data.
We shall first focus on the defect-line terms, further separating
the case of $\,\vep^{k,k+1}_{n_v}=+1\,$ from that of
$\,\vep^{k,k+1}_{n_v}=-1$.\ In the former case, we obtain
\qq
&&X_{k+1}^*\bigl(\d X_{k+1}\con A_{\phi_2(i_e)\psi^{k+1}_{n_v}(i_v)}
\bigr)(v)-X_k^*\bigl(\d X_k\con A_{\phi_1(i_e)\psi^k_{n_v}(i_v)}
\bigr)(v)\cr\cr
&&+X_{k,k+1}^*\bigl(\d X_{k,k+1}\con\bigl(P_{i_e}+\sfi\,\sfd\,\log
K_{i_e\psi^{k,k+1}_{n_v}(i_v)}\bigr)\bigr)(v)\cr\cr
&=&X_{k+1}^*\bigl(\d X_{k+1}\con A_{\phi_2(i_e)\psi^{k+1}_{n_v}(i_v)}
\bigr)(v)-X_k^*\bigl(\d X_k\con A_{\phi_1(i_e)\psi^k_{n_v}(i_v)}
\bigr)(v)\cr\cr
&&+X_{k,k+1}^*\bigl(\d X_{k,k+1}\con\bigl(\iota_1^*A_{\phi_1(i_e)
\phi_1\circ\psi^{k,k+1}_{n_v}(i_v)}-\iota_2^*A_{\phi_2(i_e)\phi_2
\circ\psi^{k,k+1}_{n_v}(i_v)}+P_{\psi^{k,k+1}_{n_v}(i_v)}\bigr)
\bigr)(v)\cr\cr
&=&X_{k,k+1}^*\bigl(\d X_{k,k+1}\con P_{\psi^{k,k+1}_{n_v}(i_v)}
\bigr)(v)\,,
\qqq
where we have used \eqref{eq:stable-iso-local} and a counterpart of
the consistency condition \eqref{eq:var-patch-edge} for the vertex
\qq\label{eq:var-patch-vertex}
\d X_k\vert_{V_{\G}}=\iota_{1*}^{\vep^{k,k+1}_{n_v}}\d X_{k,k+1}
\vert_{V_{\G}}\,,\qquad\quad\iota_{1*}^{\vep^{k,k+1}_{n_v}}\d
X_{k,k+1} \vert_{V_{\G}}=\iota_{2*}^{\vep^{k-1,k}_{n_v}}\d X_{k-1,k}
\vert_{V_{\G}}\,.
\qqq
Similarly in the second case, the defect-line terms reduce as
\qq
&&X_{k+1}^*\bigl(\d X_{k+1}\con A_{\phi_1(i_e)\psi^{k+1}_{n_v}(i_v)}
\bigr)(v)-X_k^*\bigl(\d X_k\con
A_{\phi_2(i_e)\psi^k_{n_v}(i_v)}\bigr) (v)\cr\cr
&&-X_{k,k+1}^*\bigl(\d X_{k,k+1}\con\bigl(P_{i_e}+\sfi\,\sfd\,\log
K_{i_e\psi^{k,k+1}_{n_v}(i_v)}\bigr)\bigr)(v)\cr\cr
&=&X_{k+1}^*\bigl(\d X_{k+1}\con A_{\phi_1(i_e)\psi^{k+1}_{n_v}(i_v)}
\bigr)(v)-X_k^*\bigl(\d X_k \con
A_{\phi_2(i_e)\psi^k_{n_v}(i_v)}\bigr) (v)\cr\cr
&&-X_{k,k+1}^*\bigl(\d X_{k,k+1}\con\bigl(\iota_1^*A_{\phi_1(i_e)
\phi_1\circ\psi^{k,k+1}_{n_v}(i_v)}-\iota_2^*A_{\phi_2(i_e)\phi_2\circ
\psi^{k,k+1}_{n_v}(i_v)}+P_{\psi^{k,k+1}_{n_v}(i_v)}\bigr)\bigr)(v)\cr\cr
&=&-X_{k,k+1}^*\bigl(\d X_{k,k+1}\con P_{\psi^{k,k+1}_{n_v}(i_v)}\bigr)
(v)
\qqq
Combining the two with the defect insertion and using the remaining
consistency condition
\qq
\d X_{k,k+1}\vert_{V_{\G}}=\pi^{k,k+1}_{n_v*}\d X
\qqq
for the vertex variations of the various maps involved yields
\be\begin{array}{l}
\sum_{k=1}^{n_v}\,\vep^{k,k+1}_{n_v}\,X_{k,k+1}^*\bigl(\d
X_{k,k+1}\con P_{\psi^{k,k+1}_{n_v}(i_v)}\bigr)(v)-\sfi\,X^*
\bigl(\d X\con\sfd\log f_{i_v}\bigr)(v)=X^*(\d X\con\th_{n_v})(v)
\,,
\end{array}
\ee
cf.\ \eqref{eq:2-morph-reder}.

Thus, at the end of the day, we find the neat result
\qq
\d_X S[X;\g]&=&-2\,\int_{\Si-\G}\,\Big[G_X\bigl(\d X,\vol_{\Si,\g}\,
\D_{(2)}X+\G_{\textrm{L-C}}(\sfd X\overset{\wedge}{,}\star_\g\sfd X)
\bigr)-\tfrac{\sfi}{2}\,X^*(\d X\con H)\Big]\cr\cr
&&-2\,\int_{E_{\G}}\,\vol_{E_{\G}}\,\Big[G_{X_{|1}}\bigl(
\iota_{1*}\d X,X_{|1*}\widehat n_1\bigr)+G_{X_{|2}}\bigl(\iota_{2*}
\d X, X_{|2*}\widehat n_2\bigr)-\tfrac{\sfi}{2}\,\om(\d X,X_*
\widehat t)\Big]\cr\cr
&&+\sfi\,\sum_{v\in V_{\G}}\,X^*(\d X\con\th_{n_v})(v)\,,
\label{eq:defect-eqcons}
\qqq
from which we read off the (dynamical) field equations
\qq
\D_{(2)}X^\la+\Big[\bigl\{\begin{smallmatrix} \la \\ \mu\nu
\end{smallmatrix}\bigr\}(X)\,\bigl(\g^{-1}\bigr)^{ab}-
\tfrac{3\sfi}{2\sqrt{\det\g}}\,\bigl(G^{-1}\bigr)^{\la\rho}(X)\,
H_{\rho\mu\nu}(X)\,\eps^{ab}\Big]\,\p_a X^\mu\,\p_b X^\nu=0\,,
\qqq
written in terms of the components of the curvature 3-form $\,H=
H_{\la\mu\nu}\,\sfd X^\la\wedge\sfd X^\mu\wedge\sfd X^\nu$,\ which
we take to be antisymmetric in their indices. The resulting defect
gluing conditions are
\be\begin{array}{l}
G_{X_{|1}}\bigl( \iota_{1*}\d X,X_{|1*}\widehat n_1\bigr)+G_{X_{|2}}
\bigl(\iota_{2*}\d X, X_{|2*}\widehat n_2\bigr)-\tfrac{\sfi}{2}\,
\om(\d X,X_*\widehat t)=0\qquad\textrm{at}\quad E_\G\,,\cr\cr X^*(\d
X\con\th_n)=0\qquad\textrm{at}\quad V_\G\,.
\end{array}
\ee
The latter of the two defect conditions forces us to set
\qq
\th_n=0\,,\qquad\qquad n\in\Zb_{>0}
\qqq
in the entire region of the $(\Gc,\Bc)$-inter-bi-brane world-volume
accessible to the string, and so it effectively eliminates
$\,\th_n\,$ from further analysis. This leaves us with only the
first of the defect gluing conditions as a non-trivial constraint of
the sigma-model dynamics.

\subsection{A homotopy move of the vertex}\label{app:assoc-move}

Our aim is to derive the variation of the sigma-model action
functional under a homotopy move of the defect network within the
world-sheet depicted in figure \ref{fig:homotopy-move}. To this end,
we demand that the initial network-field configuration $\,(\G,X)\,$
(for the drawing on the left-hand side) admit -- in a sense to be
specified below -- an extension which determines the final
network-field configuration $\,(\widetilde\G,\widetilde X)\,$ (for
the drawing on the right-hand side) and thereby defines the homotopy
move of the three-valent vertex of the embedded defect network along
the edge $\,e_3$.\

\begin{figure}[hb]
$$
   \raisebox{-32pt}{\begin{picture}(150,65)
   \put(0,0){\scalebox{0.65}{\includegraphics{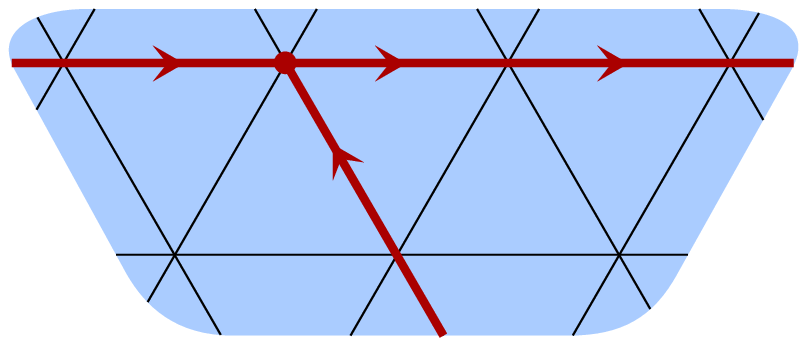}}}
   \put(0,0){
      \setlength{\unitlength}{.65pt}\put(-112,-87){
      \put(186,180)   {\scriptsize $ v_1 $}
      \put(249,180)   {\scriptsize $ v_2 $}
      \put(160,173)   {\scriptsize $ e_4 $}
      \put(224,173)   {\scriptsize $ e_3 $}
      \put(290,173)   {\scriptsize $ e_5 $}
      \put(192,140)   {\scriptsize $ e_1 $}
      \put(247,140)   {\scriptsize $ e_2 $}
      \put(157,146)   {\scriptsize $ t_1 $}
      \put(182,122)   {\scriptsize $ t_2 $}
      \put(259,122)   {\scriptsize $ t_3 $}
      \put(284,146)   {\scriptsize $ t_4 $}
      \put(221,146)   {\scriptsize $ t $}
      }\setlength{\unitlength}{1pt}}
   \end{picture}}
   ~~ \longrightarrow ~~
   \raisebox{-32pt}{\begin{picture}(150,65)
   \put(0,0){\scalebox{0.65}{\includegraphics{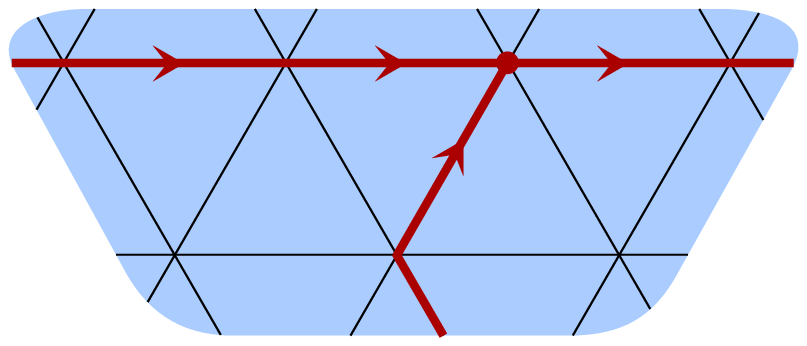}}}
   \put(0,0){
      \setlength{\unitlength}{.65pt}\put(-112,-87){
      \put(186,180)   {\scriptsize $ v_1 $}
      \put(249,180)   {\scriptsize $ v_2 $}
      \put(160,173)   {\scriptsize $ e_4 $}
      \put(224,173)   {\scriptsize $ e_3 $}
      \put(290,173)   {\scriptsize $ e_5 $}
      \put(192,140)   {\scriptsize $ e_1 $}
      \put(247,140)   {\scriptsize $ e_2 $}
      \put(157,146)   {\scriptsize $ t_1 $}
      \put(182,122)   {\scriptsize $ t_2 $}
      \put(259,122)   {\scriptsize $ t_3 $}
      \put(284,146)   {\scriptsize $ t_4 $}
      \put(221,146)   {\scriptsize $ t $}
      }\setlength{\unitlength}{1pt}}
   \end{picture}}
$$
\caption{A homotopy move of a three-valent vertex of the defect
network.} \label{fig:homotopy-move}
\end{figure}

We begin our description of the extension by choosing, for the sake
of simplicity, a sufficiently fine triangulation of the world-sheet,
so that the various embedding maps $\,X_k,X_{k,k+1}\,$ associated
with the vertex $\,v_1\,$ (as discussed in section
\ref{sec:holo-net}) are well-defined in the entire region of the
world-sheet shown in the left-hand side of figure
\ref{fig:homotopy-move}.\ Furthermore, we mark the defect edges
$\,e_4,e_1\,$ and $\,e_3\,$ converging at $\,v_1\,$ in the initial
defect network $\,\G\,$ as $\,e_{1,2},e_{2,3}\,$ and $\,e_{3,1}$,\
respectively. Similarly, the defect edges $\,e_3,e_2\,$ and
$\,e_5\,$ converging at $\,v_2\,$ in the final defect network
$\,\widetilde\G\,$ are marked as $\,e_{1,2},e_{2,3}\,$ and
$\,e_{3,1}$,\ respectively. We may now define the extension of the
map $\,X\,$ for $\,\G\,$ to be a pair of maps
\qq
\widehat X_{e_1}\ :\ t\rightarrow Q\,,\qquad\qquad\widehat X_{v_1}\
:\ e_3\rightarrow T_3
\qqq
such that the following compatibility conditions are satisfied
\be\begin{array}{c}
\widehat X_{e_1}\vert_{e_1}=X_{2,3}\vert_{e_1}\,,\qquad\qquad
\iota_2\circ\widehat X_{e_1}=X_3\vert_t\,,\cr\\
\widehat X_{v_1}\vert_{v_1}=X\vert_{v_1}\,,\qquad\qquad\pi^{3,1}_3
\circ\widehat X_{v_1}=X_{3,1}\vert_{e_3}\,,\qquad\qquad\pi^{2,3}_3
\circ\widehat X_{v_1}=\widehat X_{e_1}\vert_{e_3}\,,
\end{array}\label{eq:ext-homotopy-comp}
\ee
alongside the gluing condition
\be\begin{array}{l}
G_{\iota_1\circ\widehat X_{e_1}(p)}\bigl(\iota_{1*}v,\bigl(\iota_1
\circ\widehat X_{e_1}\bigr)_*\widehat u_2\bigr)-G_{\iota_2\circ
\widehat X_{e_1}(p)}\bigl(\iota_{2*}v,\bigl(\iota_2\circ\widehat
X_{e_1}\bigr)_*\widehat u_2\bigr)-\tfrac{\sfi}{2}\,\omega_{\widehat
X_{e_1}(p)}\bigl(v,\widehat X_{e_1*}\widehat u_1\bigr)=0\,,
\end{array}\label{eq:ext-homotopy-glue1}
\ee
to be satisfied at every $\,p\in t\,$ for all $\,v\in T_{\widehat
X_{e_1}(p)}Q\,$ and for any right-handed orthonormal basis
$\,\bigl(\widehat u_1,\widehat u_2\bigr)\,$ of $\,T_p \Si$,\ and the
gluing condition
\be\begin{array}{l}
G_{X_1(q)}\bigl(\iota_{1*}v,X_{1*}\widehat n_1\bigr)-G_{\iota_1
\circ\widehat X_{e_1}(q)}\bigl(\iota_{2*}v,\bigl(\iota_1\circ
\widehat X_{e_1}\bigr)_*\widehat n_2\bigr)-\tfrac{\sfi}{2}\,
\omega_{\pi^{1,2}_3\circ\widehat X_{v_1}(q)}\bigl(v,\bigl(\pi^{1,
2}_3\circ\widehat X_{v_1}\bigr)_*\widehat t\bigr)=0\,,
\end{array}\label{eq:ext-homotopy-glue2}
\ee
to be satisfied at every $\,q\in e_3\,$ for all $\,v\in T_{\pi^{1,
2}_3\circ\widehat X_{v_1}(q)}Q\,$ and for a triple of unit vectors
$\,\widehat t,\widehat n_1,\widehat n_2\in T_q\Si\,$ such that
$\,\widehat t\,$ is tangent to $\,e_3\,$ and points from $\,v_1\,$
to $\,v_2$,\ and $\,\widehat n_1\,$ (resp. $\,\widehat n_2$) is
normal to $\,e_3\,$ and points to the outside (resp. inside) of
$\,t$.\ The upper line in \eqref{eq:ext-homotopy-comp} in
conjunction with the gluing condition \eqref{eq:ext-homotopy-glue1}
identifies $\,\widehat X_{e_1}\,$ as an extension of $\,X\,$ to
$\,t\,$ across $\,e_1\,$ in the sense of section
\ref{sec:conf-top-def}. The bottom line in
\eqref{eq:ext-homotopy-comp}, on the other hand, is a
straightforward generalisation of the there defined notion of an
extension across a defect line to the setting of figure
\ref{fig:homotopy-move}, and \eqref{eq:ext-homotopy-glue2} ensures
that the defect gluing condition for the defect edge marked as
$\,e_{1,2}\,$ holds to the left of the three-valent defect vertex
all along the way as the latter gets shifted from $\,v_1\,$ to
$\,v_2$.

Keeping track of all the \v Cech indices involved quickly becomes
rather cumbersome, and so we make certain simplifying assumptions
which render our demonstration more tractable without any loss of
generality of the final result. Thus, we presuppose that $\,\iota_1
\circ\widehat X_{e_1}\,$ embeds $\,t\,$ in the same open set
$\,\Oc_{i_1}^M\,$ as the one into which the map $\,X\,$ sends the
adjacent triangles $\,t_1\,$ and $\,t_2$.\ Analogously, we assume
that all three triangles $\,t,t_3\,$ and $\,t_4\,$ are embedded in
the same set $\,\Oc_{i_2}^M\,$ by the original map $\,X$.\ The map
$\,\pi^{1,2}_3\circ\widehat X_{v_1}\,$ is taken to embed $\,e_3\,$
in a single set $\,\Oc_{i_3}^Q$,\ just as $\,X_{3,1}\,$ is taken to
embed $\,e_3\cup e_5\,$ in a single set $\,\Oc_{i_4}^Q$.\ Finally,
the map $\,\widehat X_{e_1}\,$ sends the entire triangle $\,t\,$
into $\,\Oc_{i_5}^Q$,\ which is also where $\,X\,$ sends $\,e_1$,\
and the map $\,\widehat X_{v_1}\,$ takes the entire edge $\,e_3\,$
into $\,\Oc_{i_6}^{T_3}$.\ We have the obvious compatibility
conditions for the index maps
\qq
&i_1=\phi_1(i_5)\,,\qquad i_2=\phi_2(i_5)\,,&\cr&&\\
&i_3=\psi^{1,2}_3(i_6)\,,\qquad i_4=\psi^{3,1}_3(i_6)\,,\qquad i_5=
\psi^{2,3}_3(i_6)\,.&\nonumber
\qqq

We may use $\,\widehat X_{e_1}\,$ and $\,\widehat X_{v_1}\,$ to
construct a new network-field configuration $\,(\widetilde\G,
\widetilde X)\,$ for the drawing on the right-hand side of figure
\ref{fig:homotopy-move} starting from the original one $\,(\G,X)$.\
This is achieved by setting
\be\begin{array}{ll}
\widetilde X\vert_{\Si-t}=X\vert_{\Si-t}\,,\qquad\quad&\widetilde X
\vert_{t-(e_2\cup e_3)}=\iota_1\circ\widehat X_{e_1}\vert_{t-(e_2
\cup e_3)}\,,\cr\cr \widetilde X\vert_{e_2-v_2}=\widehat
X_{e_1}\vert_{e_2-v_2}\,,\qquad\quad&\widetilde
X\vert_{e_3-v_2}=\pi^{1,2}_3\circ\widehat X_{v_1}\vert_{e_3-v_2}\,,
\cr\cr
\widetilde X\vert_{v_2}=\widehat X_{v_1}\vert_{v_2}\,.
\end{array}
\ee
We are now ready to compare the value of the holonomy for
$\,(\widetilde\G,\widetilde X)\,$ with that attained on $\,(\G,X)$.\
Upon rewriting \eqref{eq:hol-defect-network-1} in the simple setting
described and taking into account all the compatibility conditions
listed, alongside \eqref{eq:stable-iso-local} and
\eqref{eq:inter-bi-2morph-loc}, we obtain
\qq
\frac{1}{\sfi}\,\log\frac{\Hol(\til\G,\til X)}{\Hol(\G,X)}&=&\int_t
\,\bigl(\bigl(\iota_1\circ\widehat
X_{e_1}\bigr)^*B_{i_1}-X_3^*B_{i_2}\bigr)+\int_{e_2}\,\widehat
X_{e_1}^*P_{i_5}-\int_{e_1}\,X_{2,3}^*P_{i_5}\cr\cr
&&+\int_{e_3}\,\bigl(\bigl(\pi^{1,2}_3\circ\widehat X_{v_1}\bigr)^*
P_{i_3}-X_{3,1}^*P_{i_4}\bigr)-\sfi\,\log f_{i_6}\bigl(\widehat
X_{v_1}(v_2)\bigr)+\sfi\,\log f_{i_6}\bigl(X(v_1)\bigr)\cr\cr
&\equiv&\int_t\,\widehat X_{e_1}^*\bigl(\iota_1^*B_{\phi_1(i_5)}-
\iota_2^*B_{\phi_2(i_5)}\bigr)+\int_{e_2}\,\widehat X_{e_1}^*
P_{i_5}-\int_{e_1}\,\widehat X_{e_1}^*P_{i_5}\cr\cr
&&+\int_{e_3}\,\widehat X_{v_1}^*\bigl(\bigl(\pi^{1,2}_3\bigr)^*
P_{\psi^{1,2}_3(i_6)}-\bigl(\pi^{3,1}_3\bigr)^*P_{\psi^{3,1}_3(i_6
)}\bigr)\cr\cr
&&-\sfi\,\log f_{i_6}\bigl(\widehat X_{v_1}(v_2)\bigr)+\sfi\,\log
f_{i_6}\bigl(\widehat X_{v_1}(v_1)\bigr)\cr\cr
&=&\int_t\,\widehat X_{e_1}^*\om+\int_{e_3}\,\widehat X_{v_1}^*
\bigl(\bigl(\pi^{1,2}_3\bigr)^*P_{\psi^{1,2}_3(i_6)}+\bigl(\pi^{2,
3}_3\bigr)^*P_{\psi^{2,3}_3(i_6)}-\bigl(\pi^{3,1}_3\bigr)^*
P_{\psi^{3,1}_3(i_6)}\bigr)\cr\cr
&&-\sfi\,\log f_{i_6}\bigl(\widehat X_{v_1}(v_2)\bigr)+\sfi\,\log
f_{i_6}\bigl(\widehat X_{v_1}(v_1)\bigr)\cr\cr
&=&\int_t\,\widehat X_{e_1}^*\om\,.
\qqq
A straightforward calculation of the difference of the kinetic terms
of the sigma-model action functional evaluated on the two
network-field configurations $\,(\til\G,\til X)\,$ and $\,(\G,X)\,$
completes the derivation, cf.\ \eqref{eq:shift-action-diff}.
Thus, as explained below \eqref{eq:shift-action-diff}, we see, using \eqref{eq:ext-homotopy-glue1}, that the action functional remains invariant under the vertex move,
$\,S[(\widetilde\G,\widetilde X);\gamma_0] = S[(\G,X);\gamma_0]$.

Note, in particular, that upon fixing the trivial defect condition at
$\,e_4\,$ (whereby the relevant 2-morphism $\,f_{i_6}\,$ reduces to
the trivial death 2-isomorphism), we recover a result on the change
of the holonomy under a homotopy move of the vertex-free segment of
the defect network. It is unaffected by the presence of the defect
vertex due to the equality $\,\th_n=0$,\ imposed on the basis of the
analysis of appendix \ref{app:def-glue}.

\end{document}